\documentclass[fleqn,10pt]{wlscirep}
\usepackage[utf8]{inputenc}
\usepackage[T1]{fontenc}
\usepackage{float}
\usepackage{bm}
\usepackage{amsmath}
\usepackage{bigstrut}
\usepackage[ruled,vlined]{algorithm2e}
\usepackage{siunitx}
\usepackage{hyperref}
\usepackage{epstopdf}
\usepackage{subfigure}
\usepackage{lineno}

\title{\textcolor{black}{Geomechanical simulation of energy storage in salt formations}}

\author[1]{Kishan Ramesh Kumar}
\author[1]{Artur Makhmutov}
\author[2]{Christopher J Spiers}
\author[1,*]{Hadi Hajibeygi}
\affil[1]{Faculty of Civil Engineering and GeoSciences, Delft University of Technology, Delft, 2628 CD, The Netherlands}
\affil[2]{Faculty of Geosciences, Utrecht University, Utrecht, 3584 CS, The Netherlands}

\affil[*]{h.hajibeygi@tudelft.nl}

\begin{abstract}
A promising option for storing large-scale quantities of \textcolor{black}{green gases (e.g., hydrogen)} is in subsurface rock salt caverns. The mechanical performance of salt caverns utilized for long-term subsurface energy storage plays a significant role in long-term stability and serviceability. However, rock salt undergoes non-linear creep deformation due to long-term loading caused by subsurface storage. Salt caverns have complex geometries and the geological domain surrounding salt caverns has a vast amount of material heterogeneity.  To safely store gases in caverns, a thorough analysis of the geological domain becomes crucial. To date, few studies have attempted to analyze the influence of geometrical and material heterogeneity on the state of stress in salt caverns subjected to long-term loading. In this work, we present a rigorous and systematic modeling study to quantify the impact of heterogeneity on the deformation of salt caverns and quantify the state of stress around the caverns. A 2D finite element simulator was developed to consistently account for the non-linear creep deformation and also to model tertiary creep. The computational scheme was benchmarked with the already existing experimental study. The impact of cyclic loading on the cavern was studied considering maximum and minimum pressure that depends on lithostatic pressure. The influence of geometric heterogeneity such as irregularly-shaped caverns and material heterogeneity, which involves different elastic and creep properties of the different materials in the geological domain, is rigorously studied and quantified. Moreover, multi-cavern simulations are conducted to investigate the influence of a cavern on the adjacent caverns. An elaborate sensitivity analysis of parameters involved with creep and damage constitutive laws is performed to understand the influence of creep and damage on deformation and stress evolution around the salt cavern configurations. The simulator developed in this work is publicly available at https://gitlab.tudelft.nl/ADMIRE\_Public/Salt\_Cavern.

\end{abstract}
\begin{document}

%\linenumbers

\flushbottom
\maketitle
\thispagestyle{empty}

\section*{Introduction}
\label{intro}
Storage of green gases (eg. hydrogen) in salt caverns offers a promising large-scale energy storage option for combating intermittent supply of renewable energy, such as wind and solar energy.
Caverns are artificially created by a controlled dissolution mining process within the host rock formation \cite{DONADEI2016113}. Caverns typically hold volumes of about 300,000-500,000 m\textsuperscript{3}, with much larger outliers in the order of million cubic meters \cite{Laban_2020}. Salt caverns provide swift deliverability of the stored energy, i.e., excellent injection and production characteristics compared with porous rocks, and have strong sealing properties on time scales relevant for gas storage \cite{Lord2014,Caglayan2020}.

\begin{figure*}
     \centering
      \includegraphics[scale=0.4]{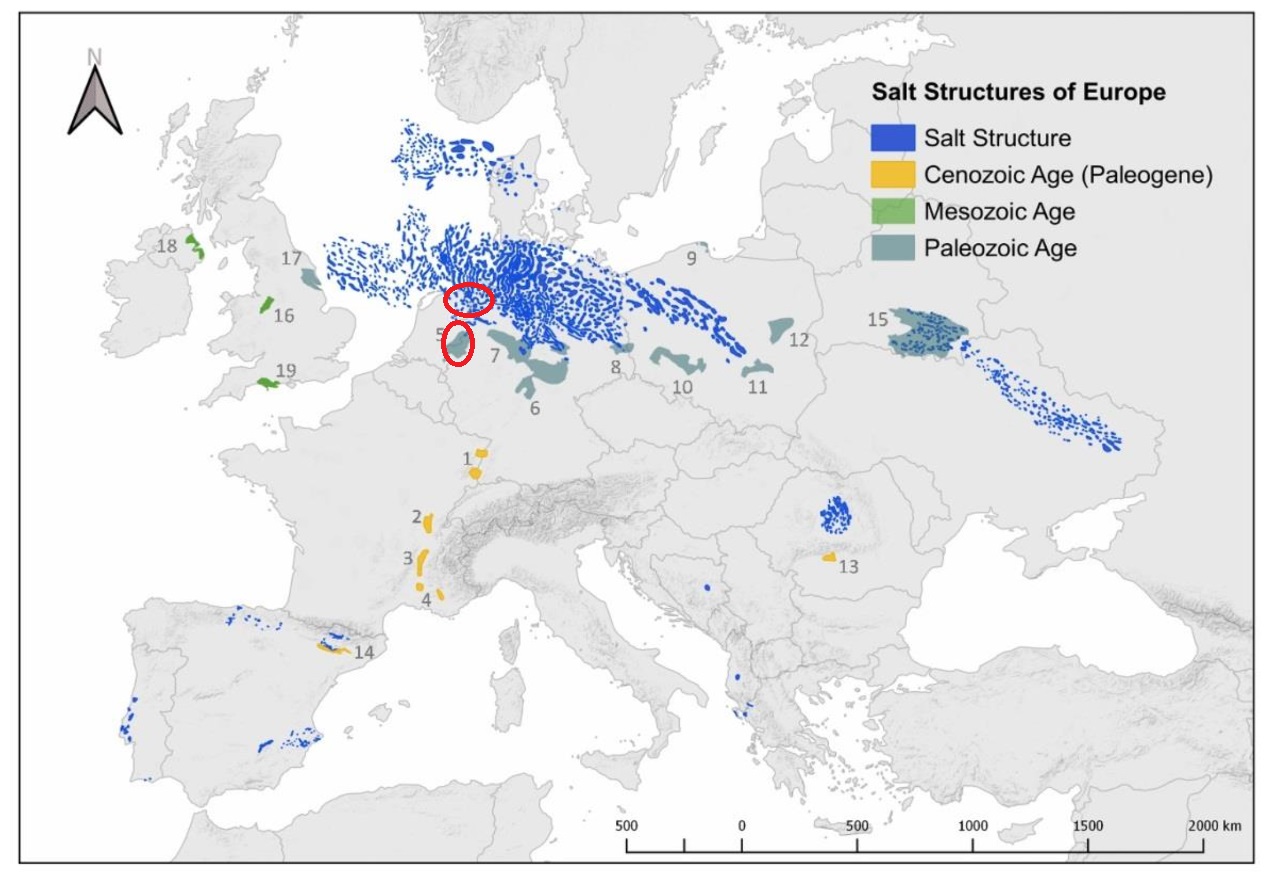}
	\caption{Illustration of the map of European salt deposits and salt structures as a result of suitability assessment, taken from the literature \cite{Caglayan2020}. Salt deposits in the Netherlands are marked in red circles.} \label{map}
\end{figure*}

Salt cavern construction focuses primarily on salt diapirs or domes, formed when part of a thick bed of salt migrates vertically into denser surrounding rock strata in response to buoyancy forces over geological time scales \cite{Jackson1986,Jeremic1994}. The primary depth target for salt cavern construction lies at 1,000-1,500 m, as within this depth range, the rock salt material behaves reasonably stably, and long-lived caverns can be constructed \cite{DONADEI2016113}. The plastic behavior of the salt defines the engineering limits at different depths, the operating pressures, and cycling characteristics, and the material characteristics of the rock salt \cite{DONADEI2016113}. Storage caverns usually have an elongated cylindrical shape because of their good stability, with a height from tens to a few hundred meters \cite{Caglayan2020,DONADEI2016113}.
Spherical, pear-shaped caverns and bell-shaped caverns have also been built in the past \cite{DONADEI2016113}.\\
Figure \autoref{map} shows a map of salt deposits in Europe which could be used for H2 storage \cite{Caglayan2020}. For salt cavern storage, a minimum salt thickness of 200 m and the depth range of 500 m to 2000 m was chosen.  In the Netherlands, the rock salt layers in the subsurface occur mainly in strata of the Permian Zechstein Group (laid down between approximately 251 and 260 million years ago) and the Triassic Röt Formation (laid down between around 238 and 244 million years ago) \cite{tno_2012}. As seen from Figure \autoref{map}, the majority of the salt deposits of the Netherlands, suitable for developing subsurface storage caverns, are located in the northern, north-eastern, and eastern parts of the country. The salt is originally accumulated in shallow, restricted salt lakes. Due to the evaporation of the seawater, salt crystals are precipitated to form today's solid rock salt layers, which in the case of the Zechstein salt in the Netherlands. The Netherlands has developed numerous domes, and pillow structures \cite{tno_2012}. 

\iffalse 
\begin{figure*}
	\centering
	\includegraphics[width=0.55\textwidth]{images/saltdepositionmap1.JPG}
	\caption{ This shows the map of European salt deposits and salt structures as a result of suitability assessment \cite{Caglayan2020}. Salt deposits in the Netherlands are marked in red circles. }
	\label{fig_002}
\end{figure*}
\fi 

Today's Dutch salt caverns are mainly the result of salt mining operations. However, a few are being used for the storage of natural gas, industrial gases, nitrogen, and compressed air \cite{tno_2012}. Near Zuidwending, for example, several caverns are currently being used for natural gas storage (Aardgasbuffer Zuidwending). Another example is the cavern near Heiligerlee (Stikstofbuffer Heiligerlee), which has been used for nitrogen storage since 2012 \cite{tno_2012}. Both purpose-built caverns, and existing caverns created during solution mining operations and preserved afterward, can potentially be used to store hydrogen or other gases. But first, it is necessary to conduct reliability and stability analyses to evaluate the safety of such structures.

%Modelling the state of the stress and deformation of the salt rocks, being the main focus of this work, has been the subject of research for a long time, and yet remains a challenging task. The main challenges are to model the complex nonlinear deformation behavior using constitutive formulae that are based on the physical processes that operate in salt under in-situ conditions while allowing for parameter uncertainties, heterogeneity in salt properties and structure, and complex cavern geometries. 

\begin{figure*}
	\centering
	\includegraphics[width=0.35\textwidth]{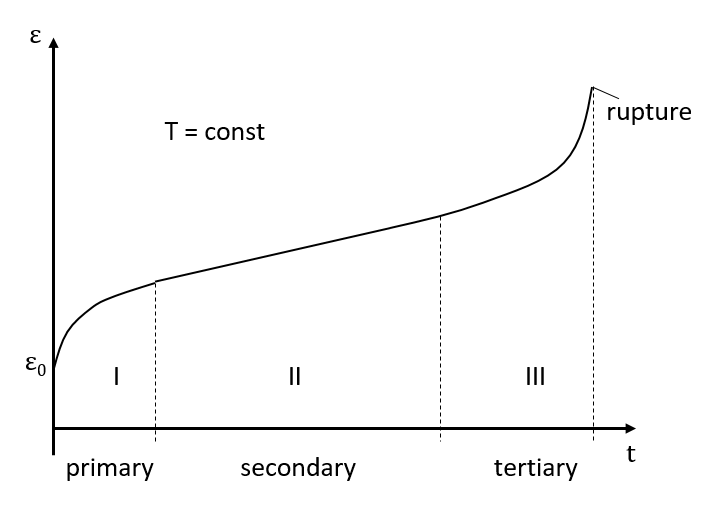}
	\caption{Illustration of the creep response of a material to uniaxial loading under conditions of low confinement that allow ultimate failure. Note the three stages of creep, starting from the initial elastic deformation $\varepsilon_0$. The illustration is modified after \cite{KonstantinNaumenko2007}.}
	\label{fig_creep_curve}
\end{figure*}

\begin{table*}[h]
	\small
	\caption{Constitutive relations presented in the literature to express creep strain rate and creep strain for Rock salt} \label{creepformulationssalt}
	\centering
	\begin{tabular}{c c}
		\hline
		\textbf{Model}  &\textbf{ Formulation}  \\ 
		\hline
		Power law \cite{KonstantinNaumenko2007,Betten2008,Carter1993,Spiers1990,Heege2005,Hunsche1999,Hampel2017} & $\dot{\varepsilon_{cr}} = A \exp \Bigg( \frac{-Q}{RT}  \Bigg)\sigma^n $ \\
		
		Hou/Lux
		\cite{Xing2014}  &$ \dot{\varepsilon_{cr}} = \frac{3}{2}\Bigg[ \frac{1}{\eta_k(\sigma_v)}\Bigg(1-\frac{\varepsilon^{tr}}{\textnormal{max}(\varepsilon^{tr})}\Bigg)   + \frac{1}{\eta_m(\sigma_v,T)}    \Bigg]\times s_{ij}$ \\
		
		MD
		\cite{Park2018,Munson1982}  &$  A_1\exp^{ \frac{-Q_1}{RT}}\Bigg[  \frac{\sigma}{\mu(1-\omega)}     \Bigg]^{n_1} + A_2\exp^{\frac{-Q_2}{RT}}\Bigg[\frac{\sigma}{\mu (1-\omega)}\Bigg]^{n_2} + |H(\sigma-\sigma_0)|\Bigg( B_1\exp^{\frac{-Q_1}{RT}} + B_2e^{\frac{-Q_2}{RT}}\Bigg) \sinh \Bigg[\frac{q\Big(\frac{\sigma}{1-\omega} - \sigma_0\Big)}{\mu}\Bigg]         $ \\
		
		Hoek Brown  
		\cite{Yang2015,Ma2013ab,Ma2017}  & $\varepsilon_{cr} = \frac{\bm{s}-D_b\bm{s}^{cr}}{1-D_b}\Bigg(\frac{1}{2G_1} + \frac{1}{2G_2}(1-e^{\frac{-G_2t}{\eta_1}}) + \frac{t}{2\eta_2}\Bigg)$ \\
		\hline
	\end{tabular}%
	\label{tab:input}%
\end{table*}%

The primary nonlinear deformation physics of rock salts is associated with time-dependent, i.e. creep processes. Creep is a phenomenon whereby a solid material permanently deforms with time under the influence of persistent mechanical stress. Of critical importance, here, is to predict the resulting evolution of stress and deformation field over long times, to ensure the safety of the storage facility up to and into abandonment, i.e., potentially over periods of hundreds of years. As shown in Figure \ref{fig_creep_curve}, after applying an external load, the material consistently goes through three stages of creep. These are often called transient (primary or reduced), steady (secondary or stationary), and tertiary (accelerated) creep stages \cite{KonstantinNaumenko2007}. The primary creep stage is characterized by a monotonic decrease in the rate of the creep. Secondary creep is characterized by a constant creep rate. Deformation occurring during primary and secondary creep are similar to pure elastic deformations and can be significant in the secondary phase. The tertiary creep phase, seen when mean stresses are low enough to allow the formation of microscopic cracks, is characterized by microcrack damage evolution, hence creep acceleration, and ultimately brittle rupture of the material \cite{betten_2014}.
Creep in salt occurs at rates that are significant on short and long engineering time scales at temperatures in the range of 20-200 \si{\degree}C \cite{Li2016} and stresses as low as 0.2 MPa \cite{Li2016,Berest2019}. The deformation mechanics are governed by dislocation motion within the crystalline grains and by a range of grain boundary processes. 
As explained in the literature \cite{Urai2017}, two main deformation mechanisms have been investigated by employing laboratory tests and micro-structural analyses.
The first is the dislocation creep mechanism, which in the steady case is characterized by a power-law dependence of creep strain rate on deviatoric stress and an Arrhenius dependence on temperature. This type of behavior is grain size insensitive, and generally shows stress exponents in the range of 3.5-5.5 and activation energies around 60 kJ/mol \cite{Berest2012,Carter1993,Hunsche1999}. When small (natural) quantities of brine are present, creep strains are over 5-10 \%, and mean normal stresses are high enough to suppress micro-cracking, the process is accompanied by a growth of new crystals during deformation, which slightly enhances creep rate, but without changing the overall flow law significantly \cite{Peach2001,TerHeege2005,TerHeege2005a,Urai2017}. Dislocation creep is favored towards higher stresses and temperatures through its highly nonlinear power-law stress dependence and Arrhenius temperature dependence. 
The second creep mechanism is solution-precipitation creep or pressure solution. It is a linear viscous creep mechanism that involves stress-driven dissolution-precipitation transfer of salt around water-bearing grain boundaries and is favored in fine-grained materials at low stresses, and temperatures \cite{Spiers1990,Urai2017}. Various formulations used in the literature to model creep and dilatancy are presented in \autoref{creepformulationssalt}. They are the power-law, Hou/Lux, Munson and Dawson (MD), and lastly, Hoek Brown models. For further details on these constitutive models, refer to the references in the \autoref{creepformulationssalt}. Note \autoref{creepformulationssalt} does not include models that explicitly account for pressure solution. It should be mentioned that there are many other formulations, including the Hampel/Schulze  \cite{Hampel2017} one used here, and several that do explicitly account for pressure solution \cite{Marketos2016,Urai2017,Cornet2018,Cornet2018a}.

%\begin{figure*}[h]
%	\centering
%	\includegraphics[width=0.46\textwidth]{images/creep_mech.png}
%	\caption{Illustration of different creep and damage mechanisms observed in deformation experiments on undeformed rock salt (top left picture). Different shades of green represent differently oriented crystals. Modified after \cite{Urai2017}.}
%	\label{fig_creep_mechs001}
%\end{figure*}

The present work uses the Carter \cite{Carter1993} constitutive model constants to describe the material behavior. The model is based on the power-law creep relationship derived by quantifying the creep processes observed in the laboratory tests on natural salt at differential stresses > 5 MPa. Laws for natural salt fitted above about 5 MPa will not contain pressure solution effects as it only becomes dominant in natural salt at lower stresses  \cite{Hampel2017,Peach2001,Hunsche1999,Carter1993}. We do not include the pressure solution creep and recrystallization effects mentioned above, in the present analysis, due to the uncertainties in the ranges of stress and strain where these processes operate (see \cite{TerHeege2005,TerHeege2005a,Urai2008}) and because we anticipate that hydrogen penetration into cavern walls will suppress these effects of water, due to desiccation. In salt bodies with a large lateral extent, pressure solution may become critical in controlling the far-field (low stress) behavior.
Damage continuum mechanics is also incorporated in this work. Several cyclic loading experiments were conducted to study the damage characteristics, and fatigue of rock salt \cite{Song2013,He2019,Liu2014,Yin2019,Wang2017}.  The formulations employed previously to study damage evolution of rock salt are presented in \autoref{damageformulationssalt}. In this work, the damage evolution process during the tertiary creep stage introduced by the Kachanov law is incorporated by introducing the dual damage variable $D$ with the assumption that creep rate depends on the current damage state and the stress state of the system, according to the literature, \cite{Kachanov1986}. The damage law incorporated increases the effective stress which has been caused due to the dilatancy of rocksalt occurring from the damage of existing microcracks or the creation of new microcracks. It can cause a change in the permeability of rock salt from $10^{-21}$ m$^2$ to $10^{-17}$ m$^2,$ \cite{Peach1996}. In this work, only geomechanical modeling of rock salt is performed without considering any change in permeability. Further research is needed to solve for poro-mechanics to incorporate the change in permeability of rock salt when damage laws are included.

%More precisely, we assumed that the damage process will only increase the effective stress, while no permeation of hydrogen into the rock salt nor subsequent creation of micro-cracks and larger failure will ever occur. Further research is required to quantify and include these risk potentials in the model. 

\begin{table*}[h]
	\caption{Constitutive relations presented in the literature to incorporate damage mechanics in creep models for rock salt} \label{damageformulationssalt}
	\centering
	\begin{tabular}{c c c}
		\hline
		\textbf{	Model}  & \textbf{ Formulation} & \textbf{Experiment} \\ 
		\hline
		Kachanov law \cite{Kachanov1986,Liu2017,Ma2013damage,Xu2018} & $\dot{D} = A \sigma^v (1-D)^{-v} $ & Uniaxial compression of rock salt\\ 
		
		Damage parameter
		\cite{Liu2014}  & $ D = 1 - \exp \Bigg[-B \frac{|Y-Y_0|}{Y^*}\Bigg]$ & Cyclic Uniaxial loading \\
	
		Damage parameter
		\cite{Hao2016}  & Mohr Coulomb criteria on three planes   & Yield criteria for triaxial expts. \\
		
		Modified Youngs modulus
		\cite{He2019}  & $ D = \frac{\varepsilon_d}{\varepsilon} \frac{\varepsilon -\varepsilon_0}{\varepsilon_d -\varepsilon_0}$ & Cyclic Uniaxial loading \\
		
		Damage \& Fatigue variable
		\cite{Zhao2021}  & $ D = k_1D_c^p + k_2D_F^{1-p}$ & Includes creep and fatigue damage    \\
		\hline
	\end{tabular}%
	\label{tab:input}%
\end{table*}%
In the past years, several numerical studies \cite{Chemia2008,Chemia2009,Chemia2008a,Koyi1996,Koyi1998,Li2009,Li2012,Poliakov1993,Schultz1993,Keken1993} were conducted to describe geo-mechanical rock salt deformation. More recently, multi-scale finite element schemes (MSFEM) where also developed for reservoir rock mechanics  \cite{Nicolai_MSFEM,Nicola_2019,Sokolova2019,Rameshkumar2020,Rameshkumar2021}, for both linear and non-linear elastic deformations. Single-scale finite element models \cite{Ma2013,Khaledi2014} were also developed to simulate salt caverns deformation under cyclic storage loading for homogeneous 2D caverns with simplified geometries using structured grids. However, only a few recent researchers have employed lab-scale constitutive models to reservoir scale analysis. One group of studies studied the creep behavior of Brazilian salt rocks using a multi-mechanism deformation creep model for symmetrical cavern shapes using commercial software ABAQUS \cite{Firme2016, Firme2019}. They conduct axis-symmetrical simulations with different elastic properties for heterogeneities such as shale to confirm the cavern tightness and integrity for energy storage. However, caverns with complex geometries and different governing creep mechanisms for heterogeneities such as carnallite were not considered in this work. Another group of studies \cite{Khaledi2016} conducted a stability analysis of symmetrical caverns subjected to cyclic loading for different salt mines using Perzyna's visco-plastic model in code bright finite element solver. Each salt mine was considered with different elastic and visco-plastic properties. However, individual mines have a different lithological composition comprising of different minerals and they can undergo plastic deformation with distinct governing equations due to varying crystal structure. To address the above challenges from the literature and distinctly account the heterogeneity in the lithology around salt caverns, we present an open-source python-based computational FE framework that addresses the influence of complex cavern shapes, varying elastic and creep properties, and suitable creep governing equations in reservoir scale on the state of stress and deformations around the caverns.  \\
The heterogeneous composition of the geological domain around salt caverns will affect the deformation of salt caverns. Elastic and creep properties of these rock formations vary along with the depth of the geological domain. Interlayers in salt formations are difficult to dissolve in water and pose many challenges in the process of designing caverns and during their operation \cite{Wang2015,Liang2007,Zhang2017, Liu2019}. Caverns within bedded salt formations are not so stable as those created within salt domes \cite{Lord2014}.  Heterogeneity of the rock salt may affect the solution mining process, which in turn will also affect the shape of the constructed cavern \cite{Jackson1986}. In field scale cases, salt properties are heterogeneous due to impurities such as anhydrite or potassium/magnesium salts, shale rocks, bischofite \cite{Jeremic1994,Ma2013damage,Nawaz2015}. Heterogeneity in the geological domain surrounding salt caverns involves many insoluble interlayers (anhydrites, potash salts, shale, gypsum, mudstone, etc.) \cite{Liu2014,Liang2007,Li2018}.
Few experimental researchers have attempted to study the impact of heterogeneity in the composition of rock salt \cite{Ma2013damage,Liang2007,Luangthip,Nawaz2015}. However, very few researchers have tried to study the impact of heterogeneity by considering elastic and creep properties of interlayers in the domain \cite{Ma2013damage}. Wang et al. \cite{Wang2015} showed the effect of heterogeneity by considering only elastic properties and by considering equivalent elastic modulus. Because of different crystal lattices and compositions, every material behaves differently under stress subjected to longer timescales. Hence, it becomes critical to incorporate creep properties of the heterogeneity. In this work, we study two approaches to study the impact of heterogeneity. The first method employs only elastic properties of the heterogeneity. The second method uses different elastic properties and creep constitutive relations to model creep aptly subjected for extended periods. One more aspect that becomes critical is studying the effect of the irregularly shaped cavern on the state of stress and deformation. This work attempts to study this effect allowing us to capture real field scenarios in the computational domain.\\
The computational framework employs an implicit simulation model on non-uniform fully-unstructured triangular mesh and constant strain triangular (CST) elements. The model is further expanded by incorporating Eulerian strains and considering the stored product (hydrogen) density and corresponding hydrostatic pressure. The rest of the paper is structured as follows. Firstly, creep and damage constitutive laws governing equations are presented. Numerical methodology is presented to account for creep consistently. Next, a comparison of numerical and experimental results is shown. Followed by, the influence of creep under monotonic and cyclic loading on salt caverns is discussed. Then the impact of complex, realistic cavern geometries and material heterogeneity inside the geological domain with different material properties and governing equations are elaborated. Then, the evolution of rock salt damage is studied, and a detailed sensitivity analysis of all the parameters in the chosen constitutive laws is presented. Lastly, the impact of multi-cavern simulations on stress is studied.   

\section{Governing equations}
In this work, conservation of momentum is employed to solve for drained solid with constant pore pressure,
\begin{equation}
	\nabla \cdot \bm{\sigma} = f^b,
	\label{eq:s000}
\end{equation}
where the $\bm{\sigma}$ and $f^b$ are the stress and body force terms. Note that the inertial terms are not included as for their minimal impacts. Through the theory of the linear elasticity, only the elastic part of the total strain is proportional to the stress field, i.e.,
\begin{equation}
	\bm{\sigma} = C:\bm{\varepsilon_{el}}.
	\label{eq:s001}
\end{equation}
Here, $C$ is the $4^{th}$ rank elasticity or stiffness tensor \cite{Sokolova2019}. For 2D isotropic homogeneous discrete elements, one can express it through the Lame's constants $\lambda$ and $\mu$ as 
\begin{equation}
	C = 
	\begin{bmatrix}
		\lambda+2\mu & \lambda & 0\\
		\lambda & \lambda+2\mu & 0\\
		0 & 0  & \mu
	\end{bmatrix},
	\label{eq:s002}
\end{equation}
where $\delta_{ij}$ is the Dirac delta.

When time dependent plastic flow (creep) deformation is considered, the elastic deformation is only part of the total strain \cite{KonstantinNaumenko2007}. In this case, the total strain can be expressed as
\begin{equation}
	\bm{\varepsilon} = \bm{\varepsilon_{el}} + \bm{\varepsilon_{cr}},
	\label{eq:003}
\end{equation}
where inelastic creep strain is expressed as $\varepsilon_{cr}$. 
Combining Equations \eqref{eq:s000}, \eqref{eq:s001} and \eqref{eq:003}, leads to 
\begin{equation}
	\nabla \cdot (C: \bm{\varepsilon_{el}} ) = \nabla \cdot (C: (\bm{\varepsilon} - \bm{\varepsilon_{cr}}) ) = f^b.
\end{equation}
One can state the total strain as a function of the gradient symmetric $\nabla^s$ of the displacement as
\begin{equation}
	\bm{\varepsilon} = \nabla^s u,
\end{equation}
where $\nabla^s = \frac{1}{2}(\nabla u + \nabla u^T)$ \cite{Nicolai_MSFEM}. Note that $u^T$ is the transpose of the displacement vector. In addition, in this work, derivatives of displacements are taken with respect to the deformed discrete geometries. Finally, the governing equation for displacement is found as
\begin{equation}
	\nabla \cdot (C:\nabla^s u) = 
	\nabla \cdot( C: \bm{\varepsilon_{cr}})  + f^b.
	\label{eq:004}
\end{equation}

For Equation \eqref{eq:004} to be well-posed, a closure term for the creep strain is necessary. It is elaborated in the next section.

%%%%%%%%----
\subsection{Creep formulation}
In this work, creep strain rate is expressed as Norton-Bailey power law using the parameters used for Halites by Carter law \cite{Carter1993}  as presented in \autoref{creepformulationssalt}. Using Carter law and under multiaxial stress condition the creep strain is expressed as \cite{Kraus1980,Xu2018},
\begin{equation}
	\dot{\bm{\varepsilon}}_{cr} = \frac{3}{2}ae^{-\frac{Q}{RT}} \bm{\sigma}_{vM}^{n-1} \bm{s},
	\label{eq:007}
\end{equation}
The above formulation incorporates temperature dependency as expressed by the Arrhenius law. here $s$ is deviatoric part of the stress tensor, and  $Q$, $R$ and $T$ denote the activation energy, Boltzmann's constant and temperature, respectively \cite{KonstantinNaumenko2007,TerHeege2005,Hunsche1999}. In this work it is also assumed that this flow rule is still valid for small volume changes that can accommodate damage law.

\iffalse 
Creep strain rate which is a function of equivalent stress $\sigma_{eq}$ and the temperature $T$ \cite{KonstantinNaumenko2007} is given by,
\begin{equation}
	\dot{\varepsilon}_{cr} = g(\sigma_{eq}) \ f(T).
	\label{eq:ss0001}
\end{equation}
Here $\sigma_{eq}$
The literature indicates broad agreement between the experimental data and the power-law creep models for rock salts \cite{Marketos2016}. This power-law function can be written as
\begin{equation}
	g(\sigma_{eq}) = a \ \sigma_{eq}^{n},
	\label{eq:ss0002}
\end{equation}
where the $a$ and $n$ are the material constants. The temperature dependency is expressed by the Arrhenius law as 
\begin{equation}
	f(T) = e^{\frac{-Q}{RT}}.
	\label{eq:ss0003}
\end{equation}

Assuming the Norton-Bailey type potential and von Mises $\sigma_{vM}$ equivalent stress, the above equations can be written as 
\begin{equation}
	\dot{\varepsilon}_{cr} = \frac{3}{2} e^{-\frac{Q}{RT}} a \sigma_{vM}^{n-1} s,
	\label{eq:007a}
\end{equation}
where $s$ is deviatoric part of the stress tensor, and  $Q$, $R$ and $T$ denote the activation energy, Boltzmann's constant and temperature, respectively \cite{KonstantinNaumenko2007,TerHeege2005,Hunsche1999}.
\fi 

\subsection{Tertiary creep}\label{damagesection}
In this work, damage is only considered in the tertiary stage of creep. To study tertiary phase of the creep which might involve initiation and propagation of microcracks leading to rupture it is important to study the damage mechanics of the rock salt. Damage mechanics of rock salt is still an active field of research both on the experimental and numerical studies. \\
In this work, the Kachanov law \cite{Kachanov1986,Liu2017,Ma2013damage} is employed to incorporate damage in studying rock salt. The creep damage rate is a function of stress $\sigma$ and the current damage state $D$. The constitutive creep equation is therefore stated as
\begin{equation}
	\dot{\bm{\varepsilon}}_{cr} = \dot{\bm{\varepsilon}}_{cr}(\bm{\sigma}, D),
	\label{eq:00001}
\end{equation}
where the damage state variable is expressed through the evolution equation as 
\begin{equation}
	\dot{D} = \dot{D}(\bm{\sigma}, D), \quad D|_{t=0}=0, \quad D < D_*.
	\label{eq:00002}
\end{equation}
Here, $D_*$ is the critical value of the damage, at which the given material breaks, and $\dot{D}$ is the damage evolution rate, expressed as 

\begin{equation}
	\dot{D} = \frac{\bm{\sigma}^r}{{B(1-D)}^r}.
	\label{eq:00003a}
\end{equation}
Finally, the creep strain rate is give ny \cite{Liu2017,Ma2013damage}
\begin{equation}
	\dot{\bm{\varepsilon}}_{cr} = a\Big(\frac{\bm{\sigma}}{1-D}\Big)^n.
	\label{eq:00004}
\end{equation}

Variables $a, B, n, r$  in the Equations \eqref{eq:00003a} and \eqref{eq:00004} above represent the material dependent constants. It is also important to note that in the case of $D=0$, equations \eqref{eq:00004} represents the well known power-law constitutive equation. More generally, equation \eqref{eq:00004} can be written as,
\begin{equation}
	\dot{\bm{\varepsilon}}_{cr} = a\bm{\sigma}_{d}^n,
	\label{eq:00004v2}
\end{equation}
where $\bm{\sigma}_{d}$ is the dilatancy intensified stress given by $\bm{\sigma}_{d} = \bm{\sigma}/(1-D)$. \\
The material parameters to incorporate damage strain used in \cite{Ma2013} cannot be used here due to different time scales. The time period used to study failure criteria  of bedded rock salt in \cite{Ma2013} is less than 10 days. However, since the focus of our work is mainly for long term energy storage (months/years) the time scale is off by a magnitude of 10-100. The value of the parameters $B$ is the main parameter that is considered to vary in our work. Due to lack of available experimental data in large timescales, the value of the parameters $B$ is assumed to be 100 times the magnitude in the paper \cite{Ma2013damage} mainly due to the difference in timescales.

\section{Numerical methodology}
Equation \eqref{eq:004} together with Eq. \eqref{eq:00004} forms a well-posed system for nonlinear time-dependent deformation vector $\mathbf{u} = (u,v)$ of salt rock with elastic and inelastic deformation. Here, $u$ and $v$ stand for displacement in $x$ and $y$ directions, respectively. 

\begin{figure*}[h]
	\centering   		
	\subfigure[]{
		\label{triangularmesh}
		\includegraphics[scale = 0.8]{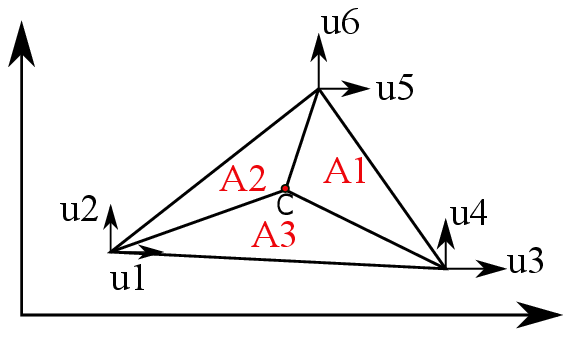}}
		\subfigure[]{
		\label{classicalcreep}
		\includegraphics[width=0.475\linewidth]{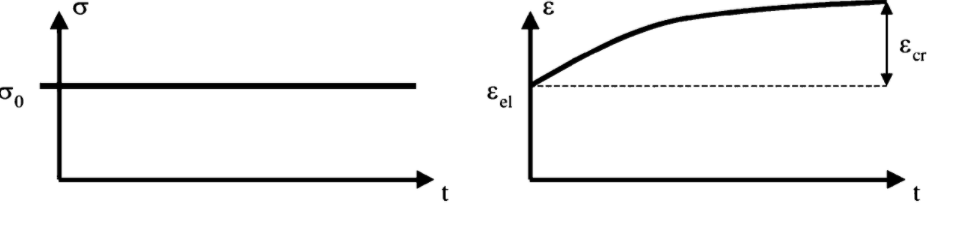}}
	\caption{ Figure \ref{triangularmesh} shows the 2D triangular mesh element with three nodes showing six deformations, 3 in each direction. Figure \ref{classicalcreep} shows the schematic diagram of classical creep \cite{Rust2015}. }
	\label{classictriangle}
\end{figure*}

The discrete system in space is found by using finite-element method (FEM), which can be stated as
\begin{equation}
	\mathbf{u} \ \approx \ \mathbf{N} \mathbf{u}^h,
	\label{eq:012}
\end{equation}
where $\mathbf{u}^h$ stands for the displacement vector at finite nodes, corresponding to the mesh resolution $h$. Here, $\mathbf{N}$ is the FE shape functions. In this work, bi-linear functions of the spatial coordinates are used as FE shape functions \cite{Nicolai_MSFEM}. Hereafter, since the solution is obtained on a single mesh resolution, the super-index $h$ is avoided for simplicity of the descriptions. In this work, 2D triangular mesh is used to employ finite element methodology. The deformation is stored in the nodes of the triangle and the stress is computed in the centre. Figure \autoref{triangularmesh} shows the mesh element with deformation stored at the corner nodes. The shape functions in the local coordinates system ($\xi,\eta$) are written as,
\begin{equation}
	N_1 = \xi \hspace{2cm} N_2 = \eta \hspace{2cm} N_3 = 1 - \xi - \eta 
\end{equation}
The shape functions are derived from the ratio of area of the triangular element opposite to the node i to total area. The areas can be seen in Figure \autoref{triangularmesh}. Using these shape functions in \autoref{eq:012} allows us to compute deformation and the same shape functions can be used to compute global coordinates of an interior point. The strain is computed using horizontal and vertical deformation ($u,v$),

\begin{equation}
	\label{strain}
	\bm{\varepsilon} =	\begin{bmatrix}
		\varepsilon_{xx} \\
		\varepsilon_{yy} \\
		\varepsilon_{xy}
	\end{bmatrix} = \begin{bmatrix}
		\frac{\partial u}{\partial x}\\
		\frac{\partial v}{\partial y}\\
		\frac{\partial u}{\partial y} + \frac{\partial v}{\partial x}
	\end{bmatrix}
\end{equation}
To compute differentials of displacements, shape functions and chain rule is employed considering local coordinates as shown \cite{KonstantinNaumenko2007}. 
\begin{equation}
	\begin{bmatrix}
		\frac{\partial u}{\partial x} \\
		\frac{\partial u}{\partial y}
	\end{bmatrix} = \begin{bmatrix}
	\frac{\partial x}{\partial \xi} & \frac{\partial y}{\partial \xi} \\
	\frac{\partial x}{\partial \eta} & \frac{\partial y}{\partial \eta}
\end{bmatrix}^{-1}
\begin{bmatrix}
	\frac{\partial u}{\partial \xi} \\
	\frac{\partial u}{\partial \eta}
\end{bmatrix}
\end{equation}
Stress is further computed from \autoref{eq:s001}.
\iffalse 
The above equation is employed to compute strain \autoref{strain}. The stress is computed further by, 
 \begin{equation}
 	\begin{bmatrix}
 		\sigma_{xx}\\
 		\sigma_{yy}\\
 		\sigma_{xy}
 	\end{bmatrix} = 
 \begin{bmatrix}
 	\lambda+2\mu & \lambda & 0\\
 	\lambda & \lambda+2\mu & 0\\
 	0 & 0  & \mu
 \end{bmatrix} \begin{bmatrix}
 	\varepsilon_{xx} \\
 	\varepsilon_{yy} \\
 	\varepsilon_{xy}
 \end{bmatrix}
 \end{equation}
\fi 
When heterogeneity is involved, each element will have different Lame parameters because of different elastic properties. Using the principle of virtual displacements \cite{KonstantinNaumenko2007} and Gauss divergence theorem, the final discrete equation for an element with unit thickness is obtained from \autoref{eq:004} is,
\begin{equation}
	\int_{\Omega_e}\bm{D}_e^T C \bm{D}_e \tilde{u}dA = \int_{\Omega_e} \bm{N}^T_efdA + \int_\Gamma \bm{N}_e^T t dS + \int_{\Omega_e} \bm{D}^T C \bm{\varepsilon}_{cr}dA
\end{equation}
The above area integrals are approximated using Gauss quadrature rule by transforming it from global to local coordinate system. Here, $\bm{D}_e = \bm{div}^T \bm{N}_e$, the applied stress on the boundary along the normal is $\bm{t}$ (Neumann bc) which is integrated over area $\Gamma$, the volumetric force is $\bm{f}$, integrated over an elemental volume $\Omega_e$. Here, fictitious forces $F_{cr} = \int_{\Omega_e} \bm{D}^T C \bm{\varepsilon}_{cr}dA$ are computed from creep strain. 

When the time dependent creep strain is included, two main time-integration approaches can be pursued: implicit (Euler backward) and explicit (Euler forward). In this work, both approaches are developed to allow for stability and efficiency analyses in a more comprehensive approach. In explicit formulation, the creep strain in continuum form is is obtained by integrating its rate functions, i.e.,
\begin{equation}
	\bm{\varepsilon_{cr}} = \int_0^t {\dot\varepsilon_{cr}} dt.
\end{equation}
In discrete form, it can be simply stated as
\begin{equation}
	\bm{\varepsilon}_{cr}^{n+1} = \bm{\varepsilon}_{cr}^{n} + f(T)g(\bm{\sigma}^n) \ \Delta t.
	\label{eq:t003}
\end{equation}
The above formula states that creep at new time step $(n+1)$ can be obtained by the already-available information at time step $n$. The explicit approach is summarised in Algorithm \ref{table_1}.

\begin{algorithm}[H]
	\SetAlgoLined
	\KwResult{Time-dependent displacement values}
	initialization: set $n = 0$, and $ \bm{\varepsilon}_{cr}^0 = 0$\ and compute elastic deformation ($F_{cr} = 0$)
	\While{$(\Delta t \cdot \ n) < T_{\text{simulation}}$}{
		Calculate creep strain
		$\bm{\varepsilon}_{cr}^{n+1} = \bm{\varepsilon}_{cr}^{n} + f(T)g(\bm{\sigma}^n) \ \Delta t$\;
		Calculate fictitious creep forces
		$F_{cr}^{n+1} = f(\varepsilon_{cr}^{n+1})$\;
		Solve for the displacement $\mathbf{u}^{n+1}$\;
		Calculate strain and stress\;
		$n \leftarrow (n + 1)$ \
	}
	\caption{Explicit displacement simulation}
	\label{table_1}
\end{algorithm}
Alternatively, one can simulate the nonlinear creep terms implicitly in time, i.e.,
\begin{align}
	\bm{\varepsilon}_{cr}^{n+1} &= \bm{\varepsilon}_{cr}^{n} + f(T) \ g(\bm{\sigma}^{n+1}) \ \Delta t.
	\label{eq:t004}
\end{align}
Here, the stress term in the creep model is obtained based on the new time step $(n+1)$. In this case, the Newton-Raphson iterative approach is needed to solve for the displacement iteratively. More precisely, the nonlinear residual reads
\begin{align}
	\Re ^ {n+1} = 
	\nabla \cdot( C: \bm{\varepsilon}_{cr}^{n+1})  + f^b - \nabla \cdot (C:\nabla^s \mathbf{u}^{n+1}).
\end{align}
The objective is to find $\mathbf{u}^{n+1}$, which also allows one to find $\bm{\varepsilon}^{n+1}$ and $\bm{\sigma}^{n+1}$. The linear form of this equation is then obtained using
\begin{align}
	\Re ^ {n+1} \approx \Re ^ {i+1} \approx \Re ^ {i} + \frac{\partial \Re}{\partial \mathbf{u}}\bigg|^{i} \delta \mathbf{u}^{i+1},
\end{align}
which is then solved iteratively until convergence is achieved, i.e. $\Re ^ {n+1} = 0$. Here, the Jacobian $J = \frac{\partial \Re}{\partial \mathbf{u}}\bigg|^{i}$ is computed based on the elastic part of the residual equation. The creep term is lagged by one iteration, i.e.,
\begin{align}
	\nabla \cdot (C:\nabla^s \mathbf{u}^{n+1}) \approx \nabla \cdot (C:\nabla^s \mathbf{u}^{i+1}),
\end{align}
and as such does not contribute in the Jacobian matrix. This allows for considering the creep term in the right-hand-side of the iterative linearised system, i.e.,

\begin{align}
	J^{i} \delta \mathbf{u}^{i+1}  = - \Re^{i}.
\end{align}
The iterations $i$ are repeated until the residual norm falls below the prescribed threshold, i.e., $||\Re^{i}||_2 < \epsilon_r$. This is called the convergence state. Algorithm \ref{table_2} provides an overview of the implicit time-integration scheme for non-linear displacement modelling. 
\begin{algorithm}[h]
	\SetAlgoLined
	\KwResult{Time-dependent displacement values}
	initialization: set $n = 0$, and $ \bm{\varepsilon}_{cr}^0 = 0$, compute elastic deformation ($F_{cr} = 0$) and set the Jacobian based on the elastic terms.\\
	\While{$(\Delta t \cdot \ n) < T_{\text{simulation}}$}{
		set iteration $i=0$\;
		set $F_{cr}^{n+1} = F_{cr}^{n}$\;
		Calculate the Residual
		$\Re^{i}$\;
		\While{$||\Re^{i}||_2 < \epsilon_r$ and $i< max(i)$}{
			Update the Residual
			$\Re^{i+1}$\;
			Update the displacement increment
			$\delta \mathbf{u}^{i+1} = - J^{i} \Re^{i}$\;
			Update the displacement $\mathbf{u}^{i+1} = \mathbf{u}^{i} + \delta \mathbf{u}^{i+1}$\;
			Calculate strain $\varepsilon^{i+1}$ and stress $\sigma^{i+1}$\;
			Calculate creep strain 
			$\bm{\varepsilon_{cr}}^{i+1} = \bm{\varepsilon}_{cr}^{n} + f(T) \ g(\sigma^{i+1}) \ \Delta t$\;
			Calculate fictitious creep forces
			$F_{cr}^{i+1} = f(\bm{\varepsilon}_{cr}^{i+1})$\;
			$i \leftarrow (i + 1)$ \
		}
		$n \leftarrow (n + 1)$ \
	}
	\caption{Implicit displacement simulation}
	\label{table_2}
\end{algorithm}

%%%% Hadi corrected up to here .. 21/Sep/2020, 00:21
This numerical framework accommodates both the types of creep which are classical and relaxation creep \cite{Rust2015}. Classical creep is a mechanism where the stress remains constant and the time dependent creep strain increases. Relaxation creep is when total strain is held constant and the increase in creep strain is compensated by decrease in the stress with respect to time. In this work, all the simulations are conducted with classical creep as shown in the schematic Fig. \autoref{classicalcreep}.

\section{Results}
This chapter presents numerical results of a series of 2D test cases beginning with the benchmarking of the developed simulator, and then to quantify the impact of nonlinear time-dependent creep physics. For this study, 8 test cases are considered: (1) linear elastic deformation under constant load, (2) consistency verification of linear elastic model, (3) nonlinear (creep) deformation under constant load, (4) nonlinear (creep) deformation under cyclic load, (5) complex heterogeneous model with creep deformation, (6) irregular cavern shape model with and without heterogeneity, (7) real field geometrical data, and (8) tertiary creep and material failure. The input parameters, used for the first 6 test cases, are summarized in Table \ref{tab:input}. The constants chosen here are obtained from lab-scale experimental data. These constants are used as a base test case for the remaining test cases. However, these constants can vary when energy is stored in the reservoir scale. Then accordingly, these constants have to be scaled considering the sensitivity of the parameter, loading conditions, and local heterogeneities beyond the scope of this work. 

\begin{table*}
	\centering
	\small
	\caption{Input parameters for simulation of base test case}
	\begin{tabular}{ l c l c }
		\hline
		\textbf{Parameter} & \textbf{Value} &  \textbf{Parameter} & \textbf{Value} \bigstrut\\
		\hline
		
		Rock salt density [kg/m$^3$] & 2200 \cite{Marketos2016} & Overburden rock density [kg/m$^3$] & 2200\bigstrut\\
	
		Rock salt Youngs modulus [GPa] & 35 \cite{Marketos2016} & Depth of the top of the salt layer [m] & 500 \bigstrut\\
	
		Rock salt Poisson ratio & 0.25 \cite{Marketos2016} & Cavern's radius [m] & 25 \bigstrut\\
	
		Roof thickness [m] & 200 \cite{Caglayan2020} & Cavern's height [m] & 250\bigstrut\\
	
		Floor thickness [m] & 200 \cite{Caglayan2020} & Time step size, [days] & 1.5\bigstrut\\
		
		Cavern volume [1000 m$^3$] & 670 & Damage evolution parameter r [-] & 2.5 \bigstrut\\
	
		Temperature gradient [\si{\degree}C/km] & 31.3 \cite{Bonte2012} & Damage evolution parameter l, [-] & 2.5\bigstrut\\
	
		Creep constant a [Pa$^n$] & 8.10E-27 \cite{Carter1993} & Damage evolution parameter B [-] & 4e4\bigstrut\\
	
		Creep exponent n [-] & 3.5 \cite{Carter1993} & Damage evolution parameter b [-] & 7E-22 \bigstrut\\
	
		Creep activation energy Q [J/mol] & 51600 \cite{Carter1993} \bigstrut  & Number of elements & 1960 \bigstrut\\
		\hline
	\end{tabular}%
	\label{tab:input}%
\end{table*}%
\begin{figure*}[h]
	\centering
	\includegraphics[scale=0.6,trim={0 18cm 0 0}]{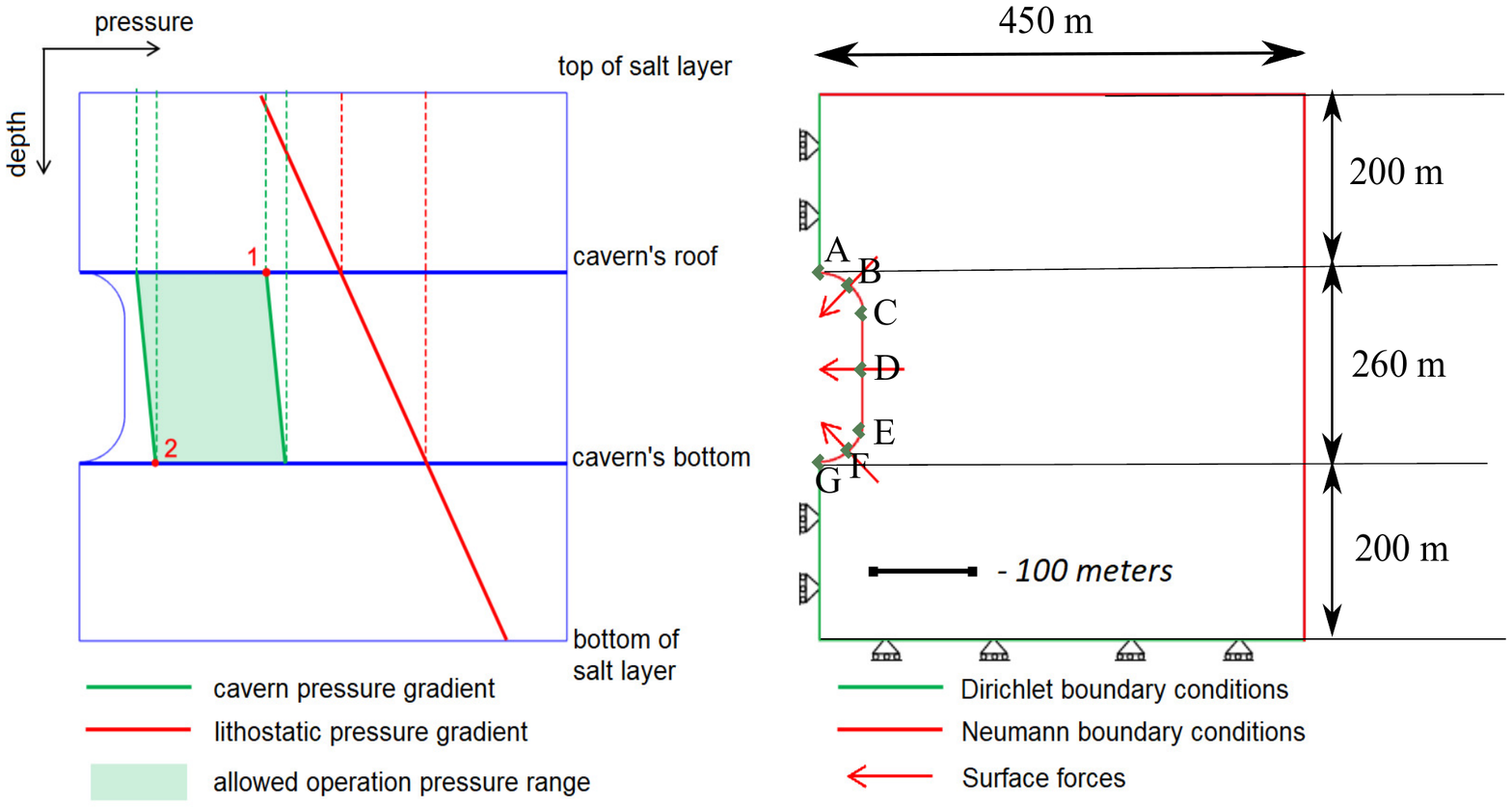}
	\caption{Pressure gradients and safe pressure range within the salt domain (left) and implied numerical model boundary conditions (right) for Test Cases 1-6. The dimensions of the geological domain is shown and the depth of the topmost salt layer from the ground is 600 m. Points (A,B,C,D,E,F,G) are located on the caverns wall are depicted to present the results from the numerical simulations.}
	\label{fig_grad}
\end{figure*}
\subsection{Defining the load for the boundary conditions}
To reflect real-field loading conditions, the cavern's fluid and lithostatic pressure values are set such that the fluid pressure does not exceed $24\%-80\%$ of lithostatic pressure \cite{Caglayan2020}. The minimum pressure difference between the cavern's fluid pressure and lithostatic pressure will be at the cavern's roof during the injection; and, the maximum will be at the cavern's bottom during production (see points 1 and 2 in the left Fig. \ref{fig_grad}). The cavern's fluid pressure, thus, is a function of the overburden rock density, rock salt density and depth of the roof and the bottom of the cavern. Roller boundary condition was imposed at the bottom and the face along the cavern. Traction free Neumann boundary conditions were imposed on the top and the far end face of the geological domain. This allows to observe any subsidence on the top face or any deformation that can cause in the far of the geological domain. 

The minimum and maximum cavern pressure and corresponding pressure difference are used to calculate the equivalent surface forces acting on the cavern's wall, as shown in Fig. \ref{fig_grad} (right). The density of hydrogen is employed to compute the forces on the cavern walls. These surface forces are then converted into equivalent nodal forces, which are finally used in the numerical model as input parameters. \\
Triangular mesh with refinement around the caverns was used for the simulations. A mesh convergence study was also conducted. Consistency and 2nd order accuracy of deformation in the implementation for the elastic domains are confirmed.

\subsection{Numerical results}
\subsubsection{Test case 1: Benchmarking with experiments}
\begin{figure*}[h]
	\centering   		
	\subfigure[]{
		\label{fig_exp_001}
		\includegraphics[width=0.25\linewidth]{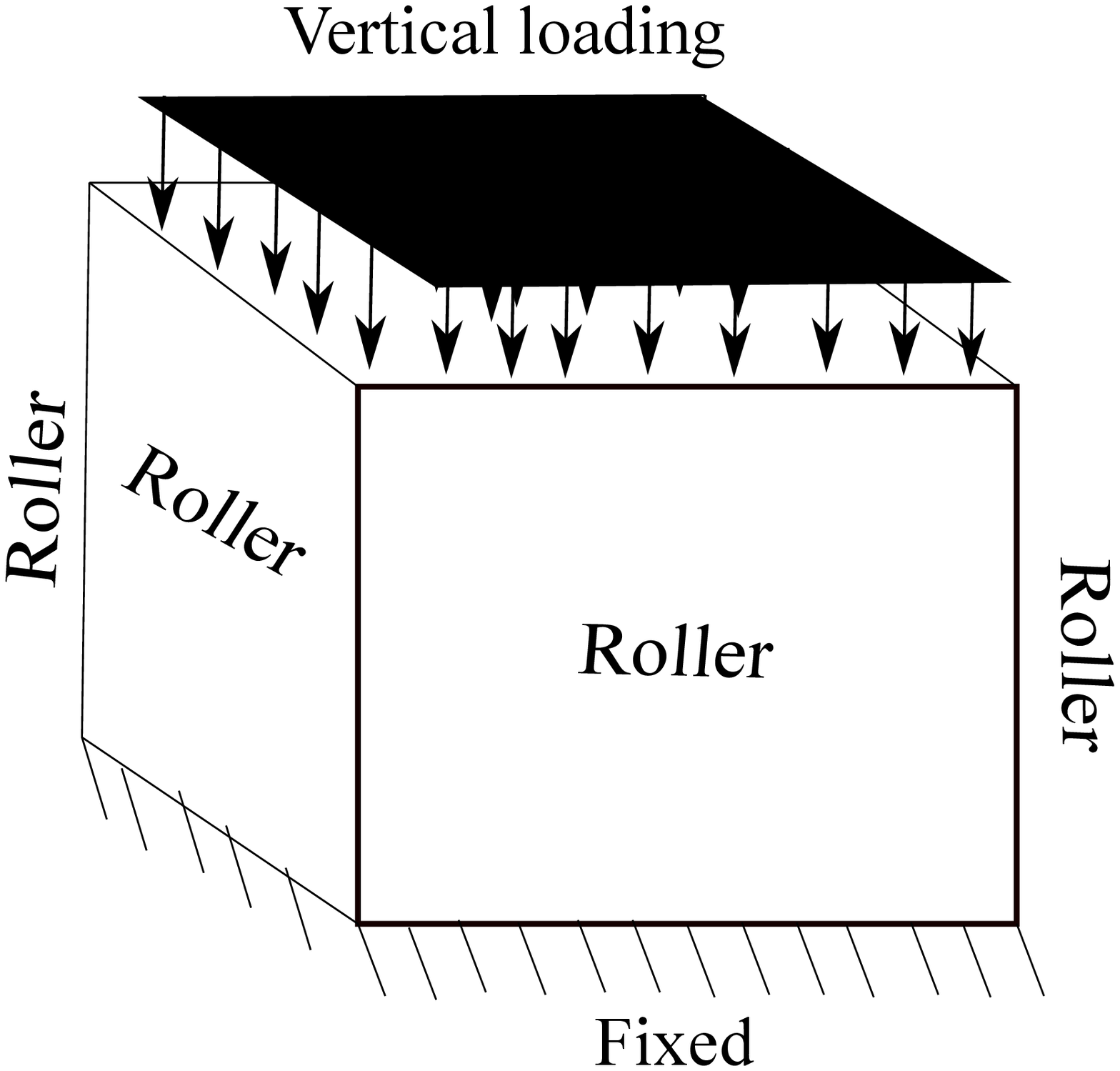}}
	\subfigure[]{
		\label{fig_exp_002}
		\includegraphics[width=0.6\linewidth]{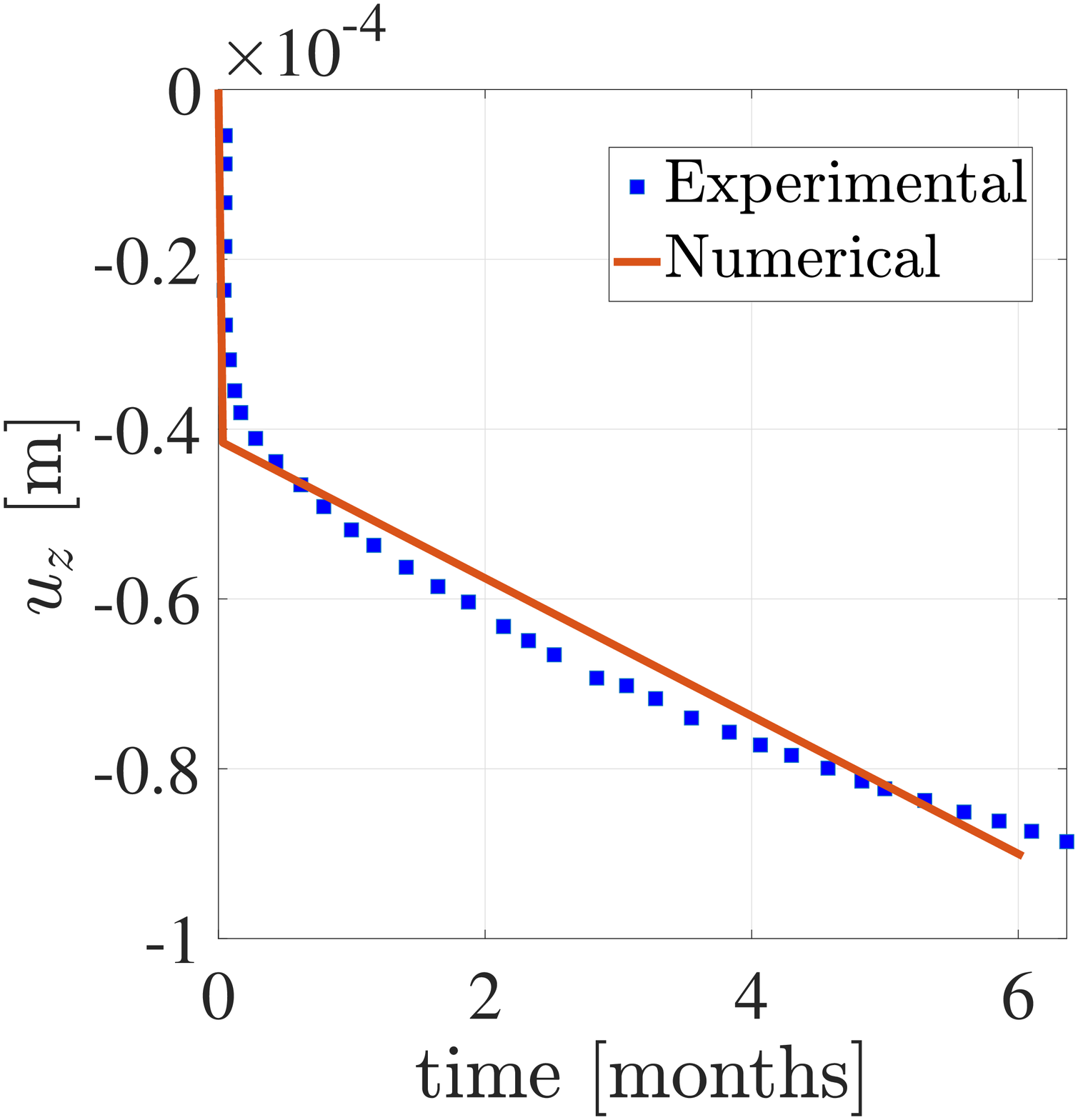}}
	\caption{Test case 1: Figure \ref{fig_exp_001} shows the schematic diagram of uniaxial compression experiment. Figure \ref{fig_exp_002} shows the variation of axial displacement $u_{z}$ with time for  numerical and experimental data of rock salt \cite{Berest2015}.}
	\label{fig_validation}
\end{figure*}

The computational framework which is developed here is compared with the experimental data.
The numerical results were compared with the uniaxial compression experiment for a  constant 20 MPa compressive load on the top face.  The computational schematic and the comparison of compressive axial deformation $u_z$ between experimental and numerical results are shown in \autoref{fig_validation}. The material parameters chosen in the creep constitutive law are shown in \autoref{tab:input}. Classical creep methodology is employed in this paper.

%Triaxial compression experiment was conducted for a constant deviatoric load of $10$ MPa on Rock salt \cite{Hunsche1999}. The confining stress is 15 MPa. The computational schematic and the comparison of compressive axial strain $\varepsilon_{zz}$ between experimental and numerical results are shown in \autoref{fig_validation}. The material parameters chosen in the creep constitutive law are shown in \autoref{tab:input}. Classical creep methodology is employed in this paper.  

The comparison between numerical and experimental results is satisfactory. The numerical results can be further optimised if necessary using optimization algorithms. However, in this work the chosen parameters for the creep constitutive law is obtained from the past literature \cite{Carter1993}. Accordingly using these constants, the numerical methodology is validated and further compared with the experiments. All the further numerical experiments are conducted using the parameters from this test case as tabulated in \autoref{tab:input}. 
 
\subsubsection{Test Case 2: Creep under monotonic loading}
In this test case, creep under monotonic constant load with respect to time is studied for the chosen parameters as described in Table \ref{tab:input}. \textcolor{black}{A constant fluid pressure of 20\% of lithostatic pressure with respect to time for 275 days is imposed on the caverns}. The time-dependent solutions are illustrated in Fig. \ref{fig_creepmodel_clusrer}, which shows the displacement evolution over time and von Mises stress distribution across the domain. A higher magnitude of displacement is observed around the cavern near larger curvatures at the end of 275 days. This is due to the reason of high stress accumulation near the curvature causing higher creep deformation. Due to the applied loading, the cavern volume shrinks.

\begin{figure*}[h]
	\centering   		
	\subfigure[]{
		\label{2a-case2}
		\includegraphics[scale=0.1234878997813971320]{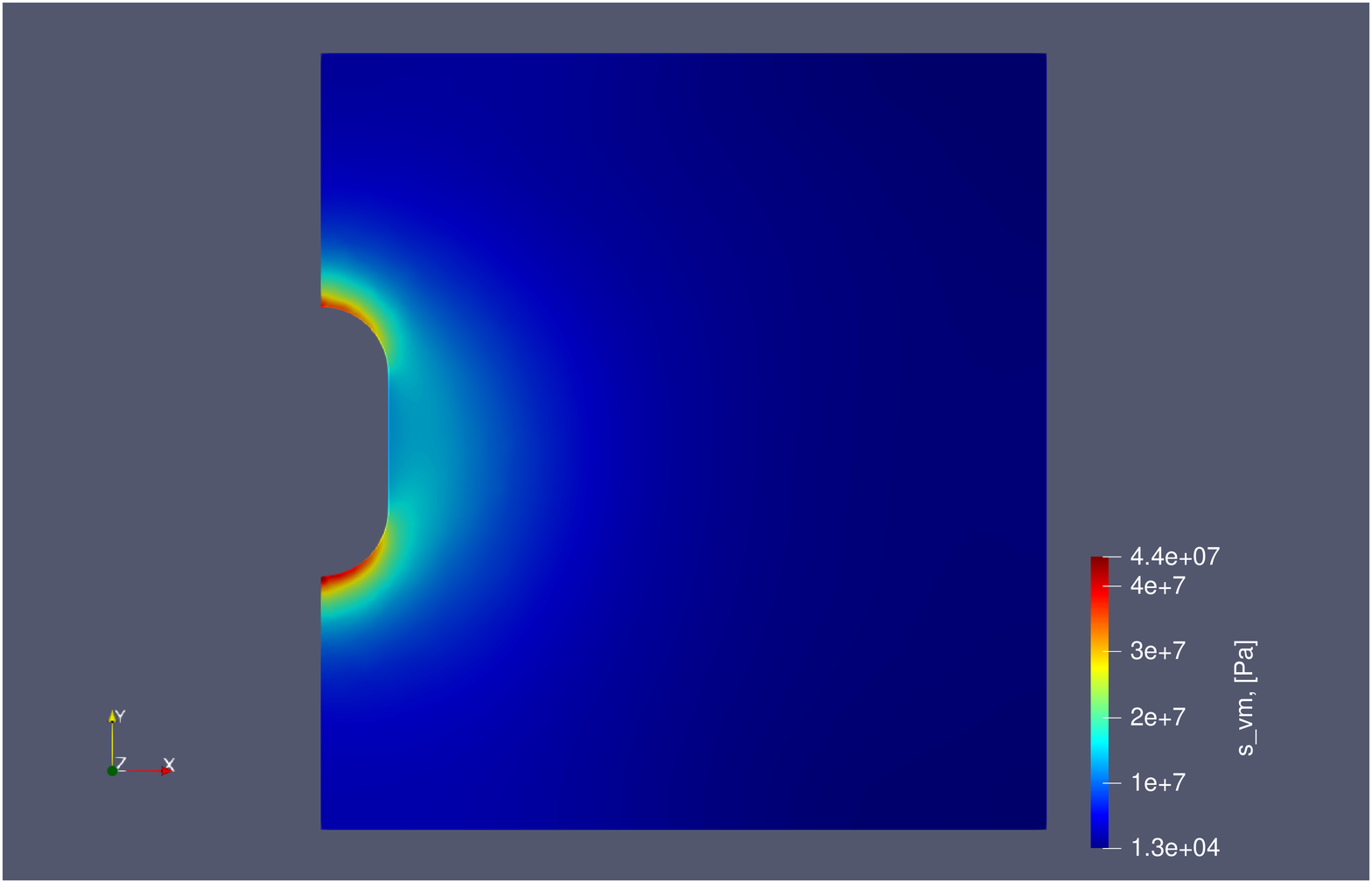}}
	\subfigure[]{
		\label{2b-case2}
		\includegraphics[scale=0.1234878997813971320]{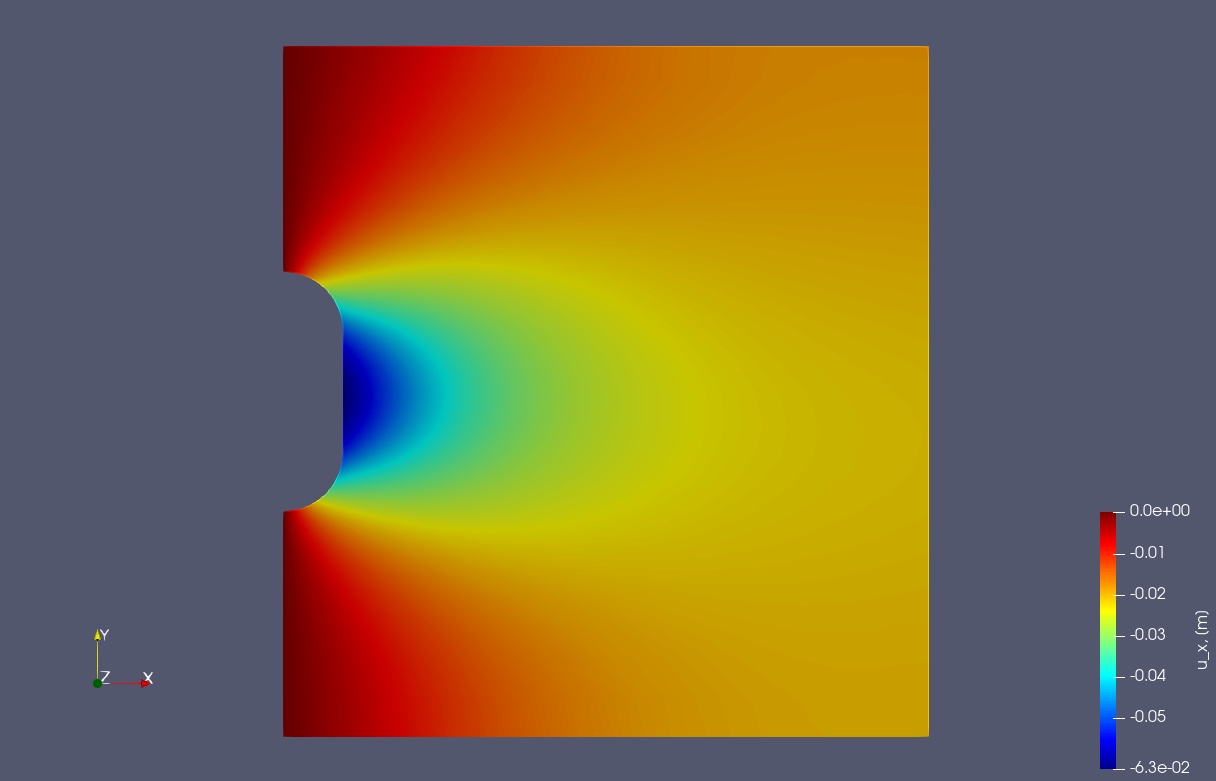}}
	\subfigure[]{
		\label{2c-case2}
		\includegraphics[scale=0.1234878997813971320]{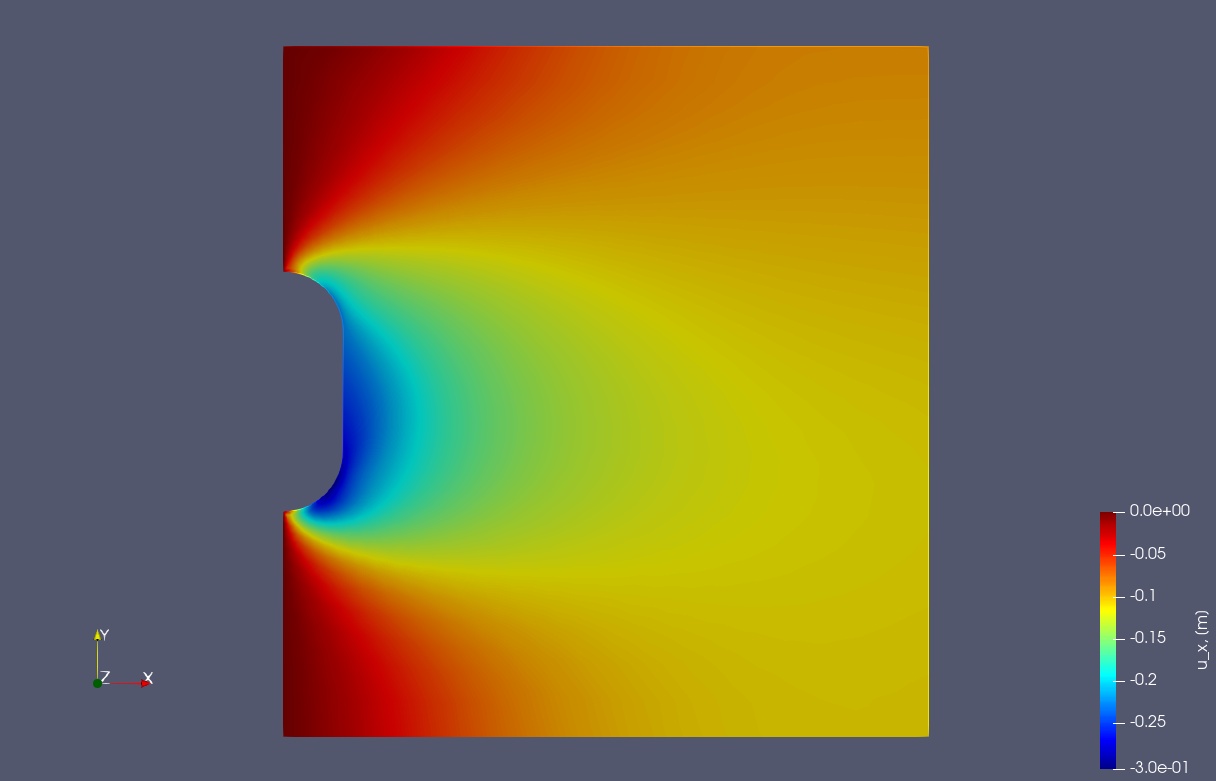}}
	\subfigure[]{
		\label{2d-case2}
		\includegraphics[scale=0.12348787813971320]{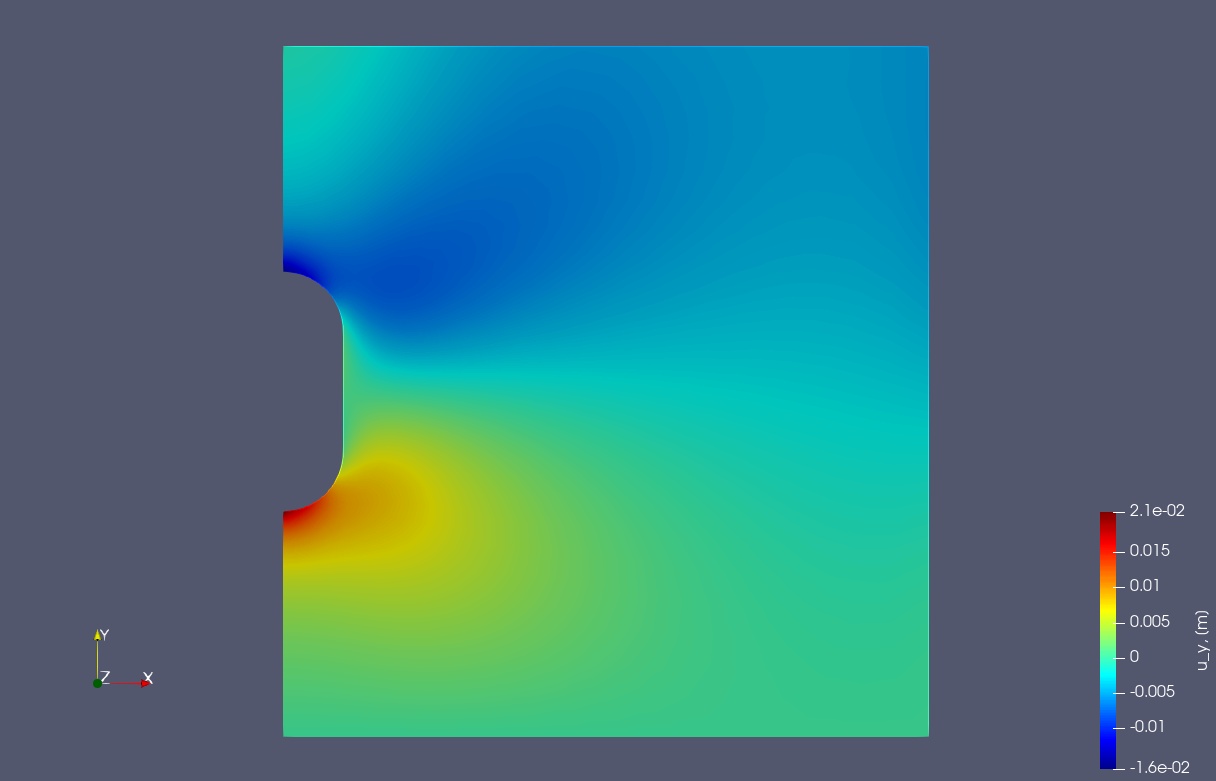}}
	\subfigure[]{
		\label{2e-case2}
		\includegraphics[scale=0.1234878813971320]{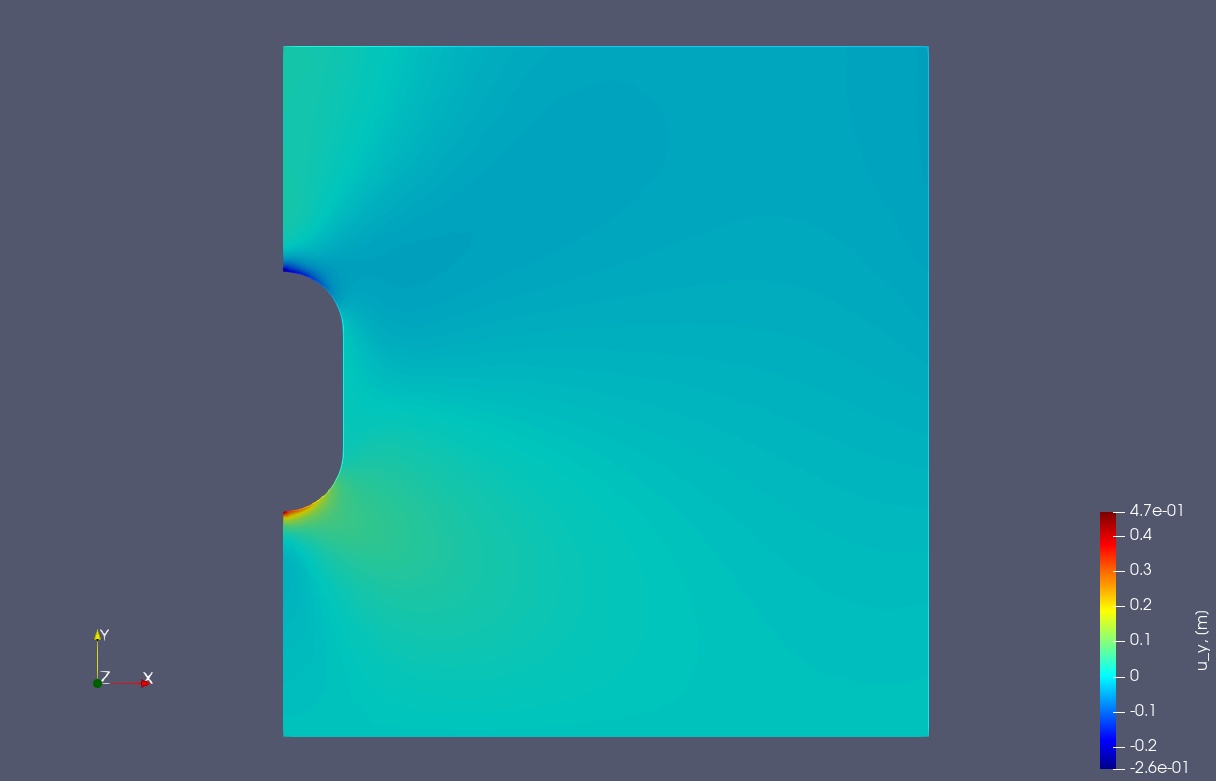}}
		\subfigure[]{
		\label{2f-case2}
		\includegraphics[scale=0.161234878813971320]{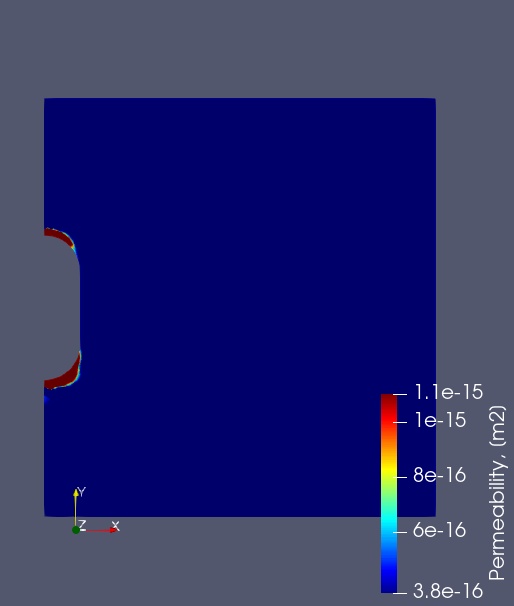}}
	\subfigure[]{
		\label{2aastrain}
		\includegraphics[width=0.32\linewidth]{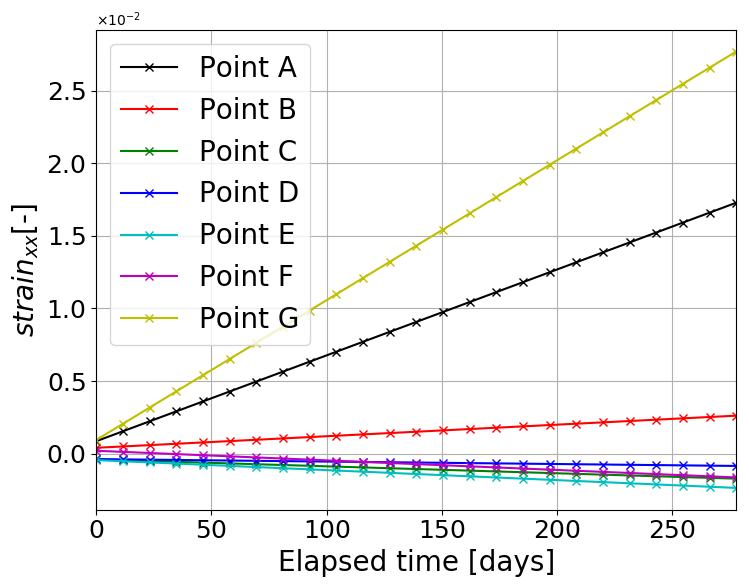}}
	\subfigure[]{
		\label{2bstrain}
		\includegraphics[width=0.32\linewidth]{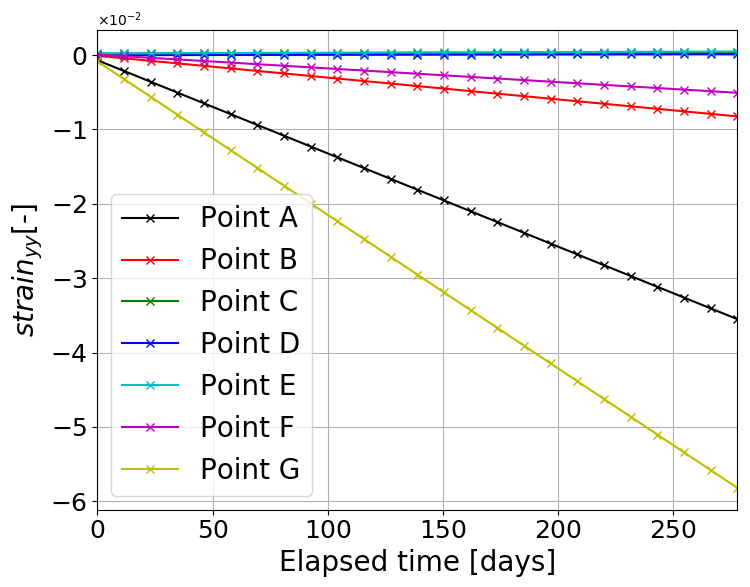}}
	\subfigure[]{
		\label{2cstress}
		\includegraphics[width=0.32\linewidth]{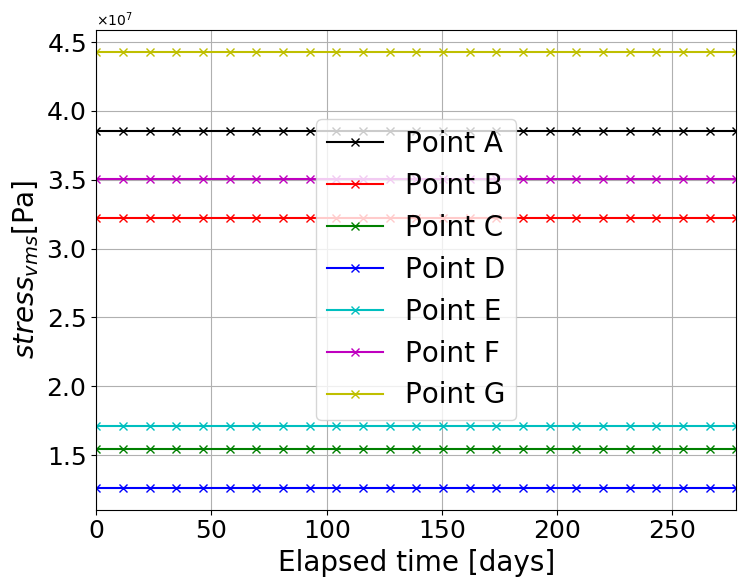}}
	\caption{\textbf{Test Case 2:} Fig. \ref{2a-case2} shows the Von Mises stress distribution. Fig. \ref{2b-case2} shows the horizontal displacement ($u_x$) at initial time step and Fig. \ref{2c-case2} shows the same after 275 days. Similarly  Fig. \ref{2d-case2} and Fig. \ref{2e-case2} show the vertical deformation at initial and final time step respectively. Fig. \ref{2f-case2} shows the permeability computed from the volumetric strain. Figure \ref{2aastrain}, Figure \ref{2bstrain} abd Figure \ref{2cstress} show the variation of strains ($\varepsilon_{xx}, \varepsilon_{yy}$) and von Mises stress respectively with time near the cavern at the points as shown in \autoref{fig_grad}.
	}
	\label{fig_creepmodel_clusrer}
\end{figure*} 

\iffalse 
\begin{figure*}
	\centering   		
	\subfigure[]{
		\label{2a-case2}
		\includegraphics[scale=0.197813971320]{images/creep_results/u_x0.pdf}}
	\subfigure[]{
		\label{2b-case2}
		\includegraphics[scale=0.197813971320]{images/creep_results/u_x3.pdf}}
	\subfigure[]{
		\label{2c-case2}
		\includegraphics[scale=0.197813971320]{images/creep_results/s_vm24.pdf}}
	\caption{\textbf{Test Case 2:} Fig. \ref{2a-case2} shows the horizontal displacement ($u_x$) at initial time step. Fig. \ref{2b-case2} shows the horizontal displacement ($u_x$) after 275 days. Fig. \ref{2c-case2} shows the initial von Mises stress distribution for salt rock.
	}
	\label{fig_creepmodel_clusrer}
\end{figure*} 
\fi 

\iffalse 
\begin{figure*}
	\begin{subfigure}{.3\textwidth}
		\centering
		\includegraphics[trim={7cm 0 0 0}, scale=0.17813971320]{images/creep_results/u_x0.pdf}
		%\caption{$u_x$}\label{fig_creepmodel00001}
	\end{subfigure}
	\begin{subfigure}{.3\textwidth}
		\centering
		\includegraphics[trim={5cm 0 0 0},scale=0.17813971320]{images/creep_results/u_x3.pdf}
		%\caption{$u_y$ }
		%\label{fig_creepmodel00002}
	\end{subfigure}
	\begin{subfigure}{.3\textwidth}
		\centering
		\includegraphics[trim={5cm 0 0 0},scale=0.17813971320]{images/creep_results/s_vm24.pdf}
		%\caption{$s_{\text{vms}}$}\label{fig_creepmodel00003}
	\end{subfigure}
	\centering
	\caption{Test Case 2: displacements after 0 (left) and 275 days (center), and the initial von Mises stress distribution for salt rock with creep behaviour.}
	\label{fig_creepmodel_clusrera}
\end{figure*}
\fi 

\autoref{fig_creepmodel_clusrer} also demonstrates the evolution of strain $\varepsilon_{xx}, \varepsilon_{yy}$ and stress $\sigma_{vM}$ over time, under constant load. Here the evolution of strain is linear because of the chosen formulation for the strain rate, i.e., Eq. \eqref{eq:007}. Higher rates of strains ( $\varepsilon_{xx}, \varepsilon_{yy}$) are observed for points A and B near the cavern roof and floor due to a high stress distribution caused by low surface area. \textcolor{black}{Points B and F show slightly lower strain rates. Lastly, the points close to the mid-plane of the domain (C, D and E) have very low strain rates compared to the rest of the labeled points.} This shows that if there are any cracks or heterogeneity near the curvature, there is a higher probability that it will lead to failure than near the mid plane region. The loss of volume of the cavern given by,
\begin{equation}
	\textnormal{Loss of volume} = \frac{V_f - V_0}{V_0}\times100 
\end{equation}
The change in the boundary of the cavern is not significant. In this case, the approximate loss of volume is around 3 \% after 275 days for a constant load. \\
Due to the low permeability of salt caverns, they are considered of best use in storage technology. However, due to volumetric strain permeability of salt rocks can increase, causing the stored gases to penetrate. Authors \cite{Peach1991, Khaledi2016} propose the permeability of salt caverns which is related to dilatant volumetric strain given by
\begin{equation}
	k =  \alpha_k \times \varepsilon^{\beta_k}_{vol}.
	\label{permeability}
\end{equation}
Here, constants $\alpha_k$ and $\beta_k$ are model parameters subjected to change due to the material properties and loading conditions. In this work $\alpha_k = 2.13e-8$ and $\beta_k = 3$ is employed as suggested by \cite{Peach1991}. Khaledi et al. \cite{Khaledi2016} showed the higher permeability near the roof and floor of the caverns. A similar observation is made here, as shown in Figure \autoref{2f-case2}. These zones also depict the potential fail zones around the cavern that can lead to gas leakages. However, it is important to see how the contours of permeability are observed in the geological domain with complex shapes as in the real field, which is not shown in the previous studies \cite{Khaledi2016}. This is further elaborated in the future sections.

\iffalse 
\begin{figure*}
	\centering   		
	\subfigure[]{
		\label{2astrain}
		\includegraphics[width=0.39\linewidth]{images/const_stress/strain_x_vs_disp.pdf}}
	\subfigure[]{
		\label{2bstress}
		\includegraphics[width=0.39\linewidth]{images/const_stress/stress_x_vs_disp.pdf}}
	\caption{Test Case 2: The above two plots show the variation of strain ($\varepsilon_{xx}$) and stress ($\sigma_{xx}$) respectively with time.}
	\label{cluster_creepgraphs1}
\end{figure*} 

\begin{figure*}
	\centering   		
	\subfigure[]{
		\label{2astrain}
		\includegraphics[width=0.39\linewidth]{images/const_stress/strain_x_vs_disp.pdf}}
	\subfigure[]{
		\label{2bstress}
		\includegraphics[width=0.39\linewidth]{images/const_stress/stress_x_vs_disp.pdf}}
	\caption{Test Case 2: The above two plots show the variation of strain ($\varepsilon_{xx}$) and stress ($\sigma_{xx}$) respectively with time.}
	\label{cluster_creepgraphs1}
\end{figure*} 
\fi 
\iffalse
\begin{figure*}
	\begin{subfigure}
		\centering
		\includegraphics[width=0.39\linewidth]{images/const_stress/strain_x_vs_disp.pdf}
		%\caption{Strain evolution.}
		%\label{fig_const001}
	\end{subfigure}
	\begin{subfigure}
		\centering
		\includegraphics[width=0.39\linewidth]{images/const_stress/stress_x_vs_disp.pdf}
		%\caption{Stress evolution.}
		%\label{fig_const002}
	\end{subfigure}
	\caption{Test Case 2: stress and strain history under constant load.}
	\label{cluster_creepgraphsa1}
\end{figure*}
\fi

\begin{figure*}[h]
	\centering
	\includegraphics[width=0.45\textwidth]{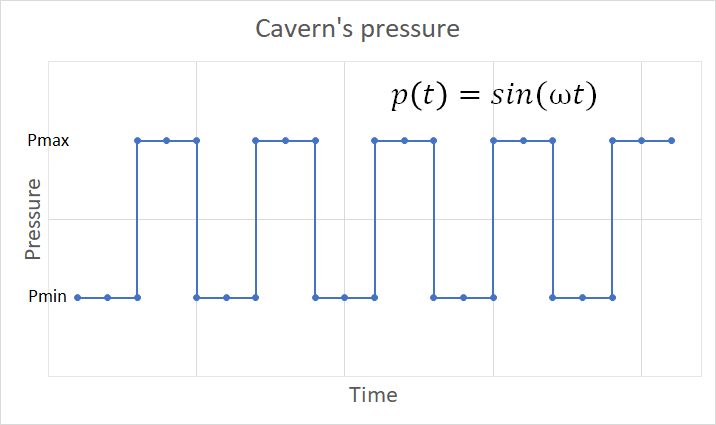}
	\caption{\textbf{Test Case 3}: This figure shows the step function used to impose time-dependent cyclic fluid pressure inside the cavern that varies between P$_{max} = 80\%$ and P$_{min} = 20\% $ of lithostatic pressure.}
	\label{fig_pres_cyc}
\end{figure*}
\begin{figure*}[h]
	\centering   		
	\subfigure[]{
		\label{2acyclic}
		\includegraphics[width=0.43025\linewidth]{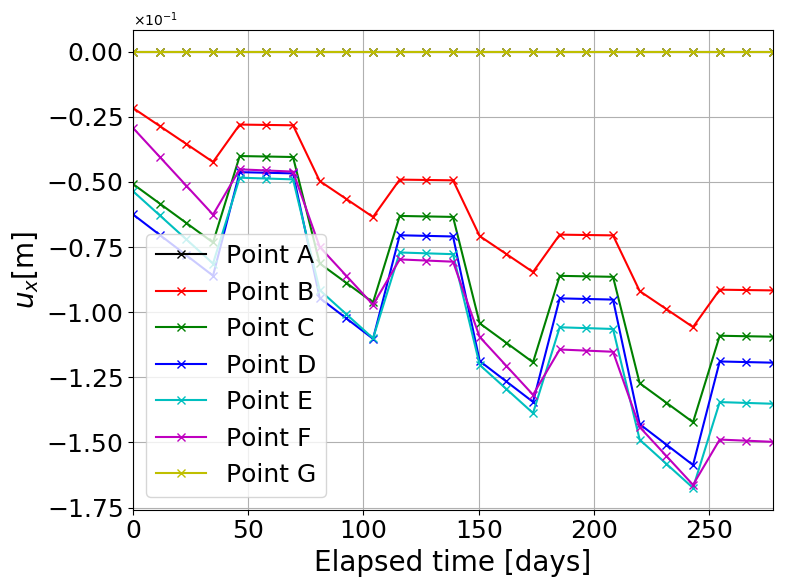}}
	\subfigure[]{
		\label{2bcyclic}
		\includegraphics[width=0.43025\linewidth]{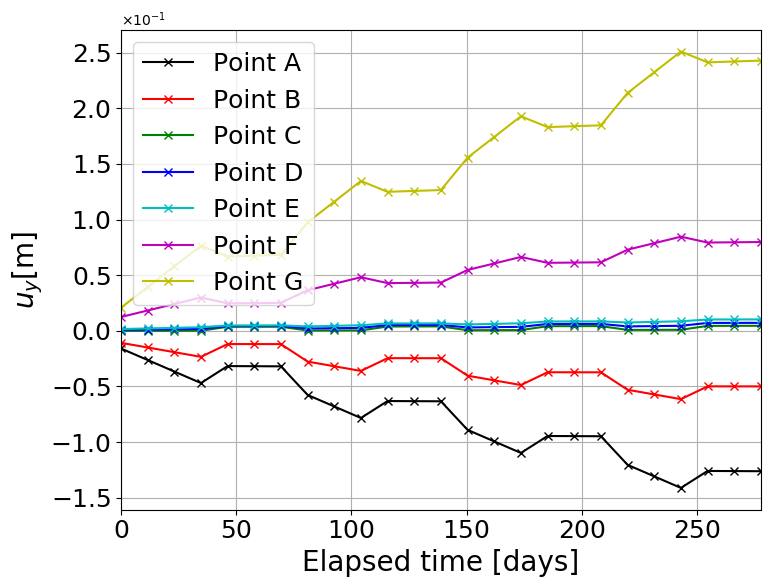}}
	
	\caption{\textbf{Test Case 3}: The above plots show the variation of deformation in x ($u_x$) and y-direction ($u_y$) over time for different points around the cavern as shown in \autoref{fig_grad}. }
	\label{Test3_creepgraphs}
\end{figure*} 
\subsubsection{Test Case 3: Creep under cyclic loading}
When excess renewable energy is produced, it is converted to green gas and stored in the subsurface. Depending on the supply and demand of the energy, the gas will be injected and produced. This will result in the cyclic loading of salt caverns. This test case addresses the important aspect of energy storage, i.e., the deformation under cyclic loading. For this reason, the cavern's fluid pressure is assumed to be a function of time. Here, a discrete step function is used as shown in Fig.  \ref{fig_pres_cyc}. The maximum and minimum pressure applied during cyclic loading depends on the lithostatic pressure. It  varies between P$_{max} = 80\%$ and P$_{min} = 20\% $ of lithostatic pressure.
Fig. \ref{Test3_creepgraphs} shows the variation of horizontal and vertical deformation with time for different points located around the cavern. Horizontal displacement is 0 for points A and G due to the imposed roller boundary conditions as shown in \autoref{fig_grad}. The full cycle of the cavern's loading and unloading in this simulation is evaluated for 6 days. \\
The high peak values are related to the instantaneous elastic response of the rock salt material, after which there is a short period of creep development with a linear slope representing the magnitude of the creep rate. The higher the load, the steeper the slope and the higher the creep rate. It can be seen that the creep deformation is insignificant at P$_{min}$ compared to P$_{max}$ (zero slope). \\
From Fig. \ref{2acyclic} it can be seen that the rate of deformation for every cycle is the highest point, F. Followed by the points E, D, C, and B. This trend is observed because of incorporating the lithostatic pressure in the geological domain. So accordingly, highest depth is observed at point F, causing the highest deformation. Similarly, from Fig. \ref{2bcyclic} it can be seen that the highest magnitude of vertical deformation is observed at point G, and the next highest magnitude is observed at point A. Although these points are symmetrically placed around the cavern along the x axis mid-plane, unequal vertical deformations are observed due to the lithostatic pressure. Accordingly, the magnitude of deformation reduces from highest stress to lowest stress (in the center point D) with the least curvature. In cyclic load, the loss of volume after 275 days is less than 1.5 \%.

\iffalse 
\begin{figure*}
	\centering   		
	\subfigure[]{
		\label{2a}
		\includegraphics[width=0.3025\linewidth]{images/cyclic/displacement_x_vs_disp.pdf}}
	\subfigure[]{
		\label{2b}
		\includegraphics[width=0.3025\linewidth]{images/cyclic/strain_x_vs_disp.pdf}}
	\subfigure[]{
		\label{2c}
		\includegraphics[width=0.3025\linewidth]{images/cyclic/creep_forces_x_vs_disp.pdf}}
	
	\caption{Test Case 3: The above plots show the variation of displacement in x direction ($u_x$), x component of strain ($\varepsilon_{xx}$), and x component of creep forces ($f_x$) over time for the point of the domain located at the roof of the cavern.}
	\label{Test3_creepgraphs}
\end{figure*} 
\fi

\begin{figure*}[h]
	\centering   		
	\subfigure[]{
		\label{2airreg}
		\includegraphics[scale=0.2197654397813971320]{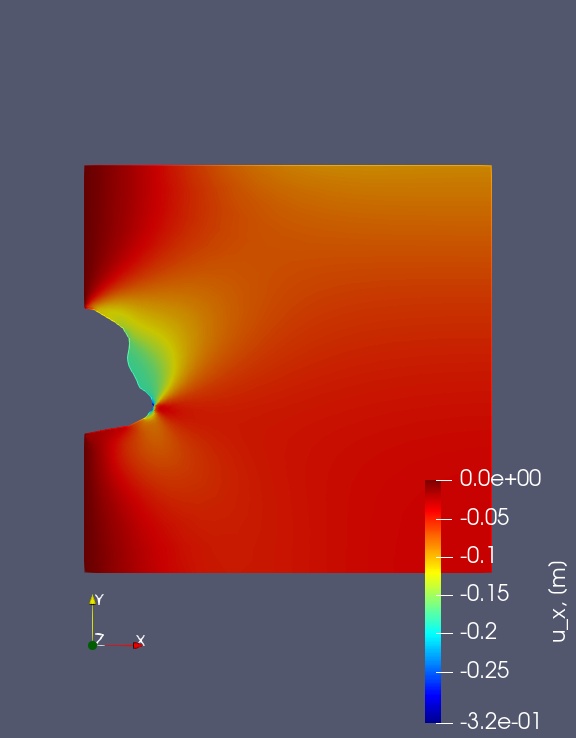}}
	\subfigure[]{
		\label{2birreg}
		\includegraphics[scale=0.2197654397813971320]{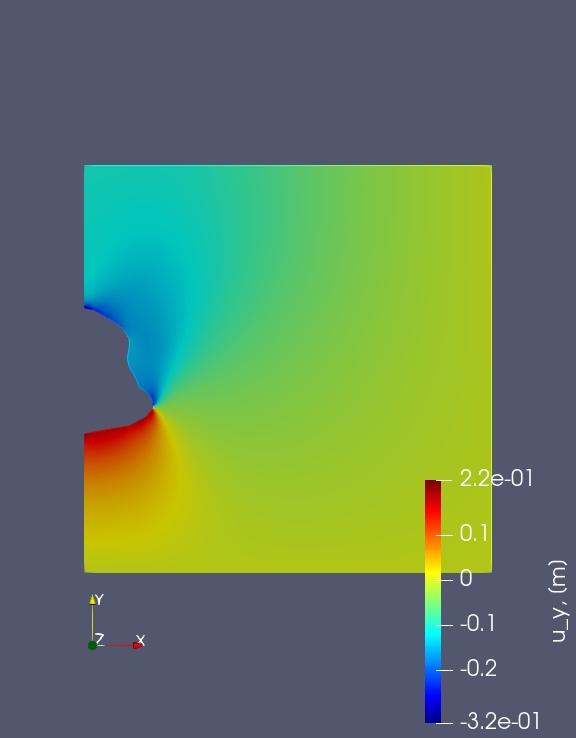}}
	\subfigure[]{
		\label{2cirreg}
		\includegraphics[scale=0.2197654397813971320]{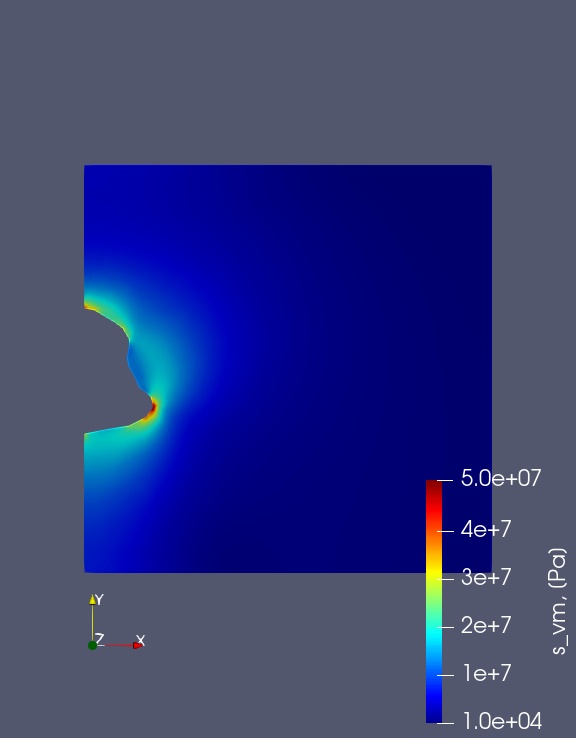}}
	\subfigure[]{
		\label{2dirreg}
		\includegraphics[scale=0.26397813971320]{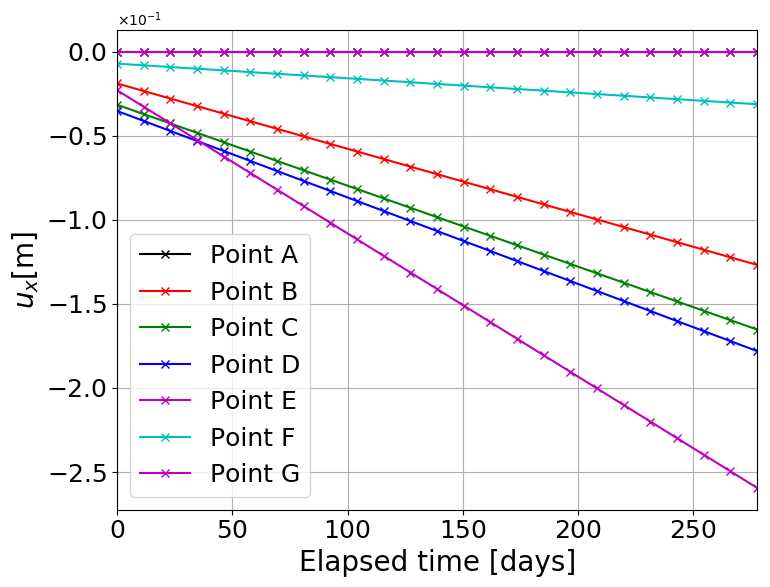}}
	\subfigure[]{
		\label{2eirreg}
		\includegraphics[scale=0.26397813971320]{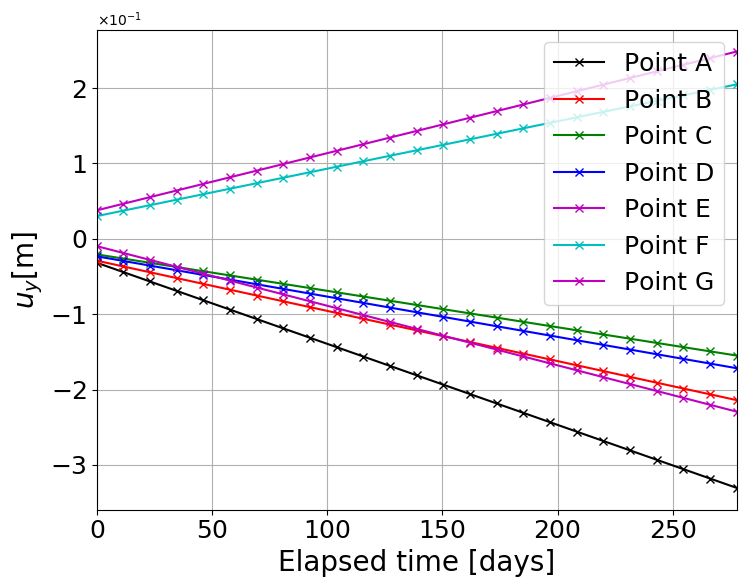}}
	\subfigure[]{
		\label{2firreg}
		\includegraphics[scale=0.2197654397813971320]{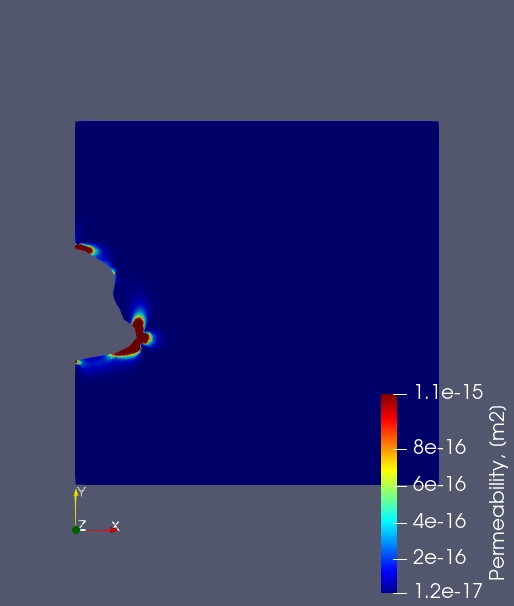}}
	\caption{\textbf{Test Case 4:} Illustration of the horizontal displacement $u_x$ (a), vertical displacement $u_y$ (b), von Mises stress $s_{vm}$ (c) and permeability (f), for irregular cavern shape and homogeneous properties. The variation of horizontal (d) and vertical (e) deformations around caverns are also shown, for the marked points in \autoref{creepgraphs_case4}.}
	\label{fig_irregular_cluster2homogeneous}
\end{figure*}

\subsubsection{Test Case 4: Caverns with irregular geometries}
In this test case, irregular-shaped caverns are studied for both homogeneous and heterogeneous domains. Fig. \ref{fig_irregular_cluster2homogeneous} shows the simulation results of the model with such an irregular cavern shape. From the figures, it can be seen that the maximum displacement is higher than that of the regular cylindrical-shaped cavern. Note that a larger surface is subject to the force exerted by the pore-fluid pressure fluctuations for irregular shapes. Also, it is noted that--for the shown geometry--the displacement is lower in the magnitude at the top and bottom of the cavern when compared with the cylinder-shaped cavern. Point E shows the maximum horizontal deformation in Fig. \autoref{2dirreg} at the end of 275 days and also has the highest rate of deformation. This is again due to large stresses at point E with high curvature. Fig. \autoref{2eirreg} shows the variation of vertical deformation with time. Since there is no boundary condition imposed on the vertical deformation highest magnitudes of the rate of vertical deformation is seen in Point A and Point G. The potential failure zones in the domain is point around E and the points near the floor and roof of the cavern. Due to the irregular shape of the caverns, the potential local failure zones are more important because of the high-stress zones, which might lead to micro-cracking or lead to damage. The above point is supported by high permeability values near high curvature, as shown in Figure. \ref{2firreg}. Compared to permeability observed in the cylindrical cavern, as shown in the previous test case, higher permeability is observed near point E than near the roof or floor of the cavern. Therefore, high permeability zones are potential failure zones that can reduce the cavern's tightness and lead to stored hydrogen or any other gas leakage. When heterogeneity is incorporated, different local properties or constitutive equations can cause different stress distributions in the geological domain leading to different potential failure zones. This is investigated in the following sections.
\begin{figure*}[h]
	\centering   		
	\subfigure[]{
		\includegraphics[width=0.345\textwidth]{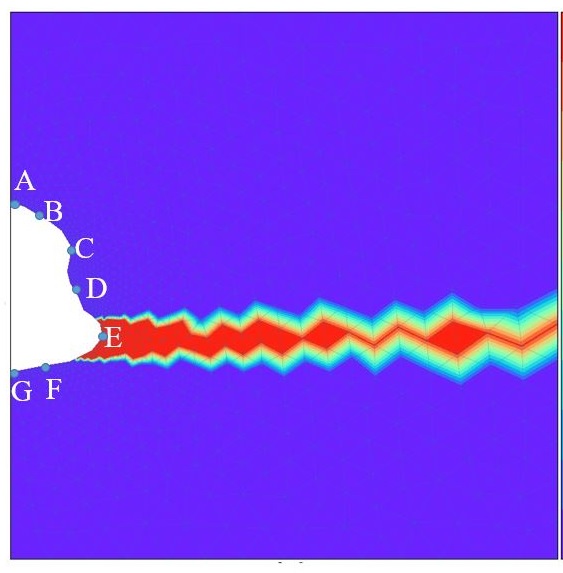}}
	\caption{\textbf{Test Case 5:} \textcolor{black}{Illustration of the heterogeneous cavern. An interlayer of 100 \% impurity (red colour) near the bottom of the cavern, consisting of potash salt. The rest of the geological domain is homogeneous (i.e.,100 \% pure halite). }  }
	\label{creepgraphs_case4}
\end{figure*}

\begin{figure*}[h]
	\centering   		
	\subfigure[]{
		\label{2ahetero}
		\includegraphics[scale=0.2317654397813971320]{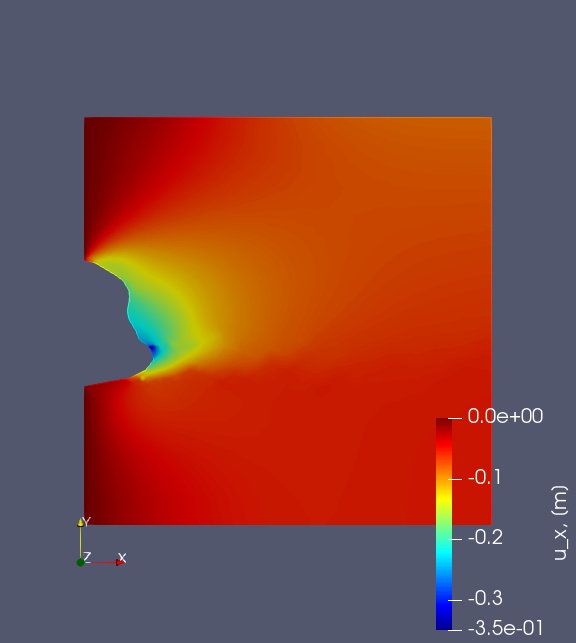}}
	\subfigure[]{
		\label{2bhetero}
		\includegraphics[scale=0.2317654397813971320]{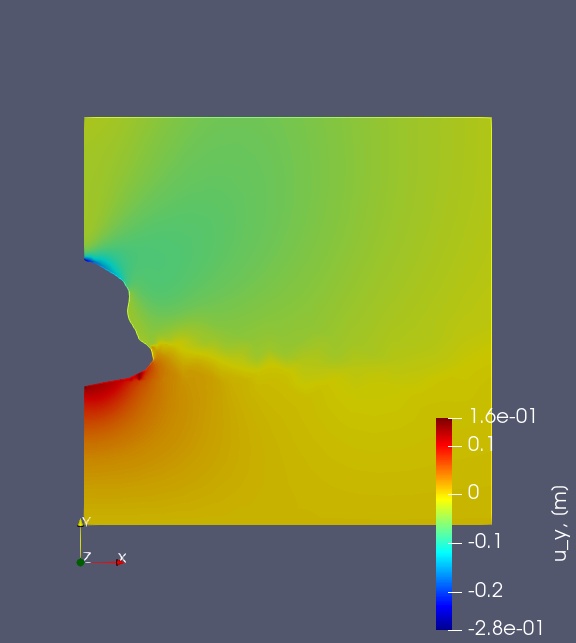}}
	\subfigure[]{
		\label{2chetero}
		\includegraphics[scale=0.2317654397813971320]{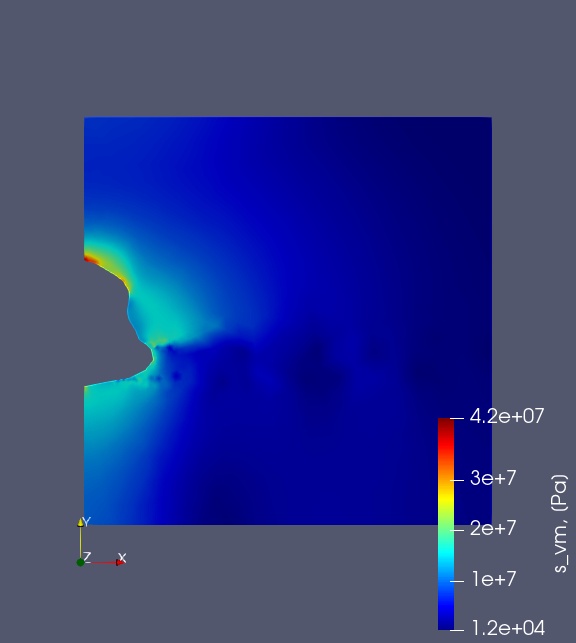}}
	\subfigure[]{
		\label{2dhetero}
		\includegraphics[scale=0.26197813971320]{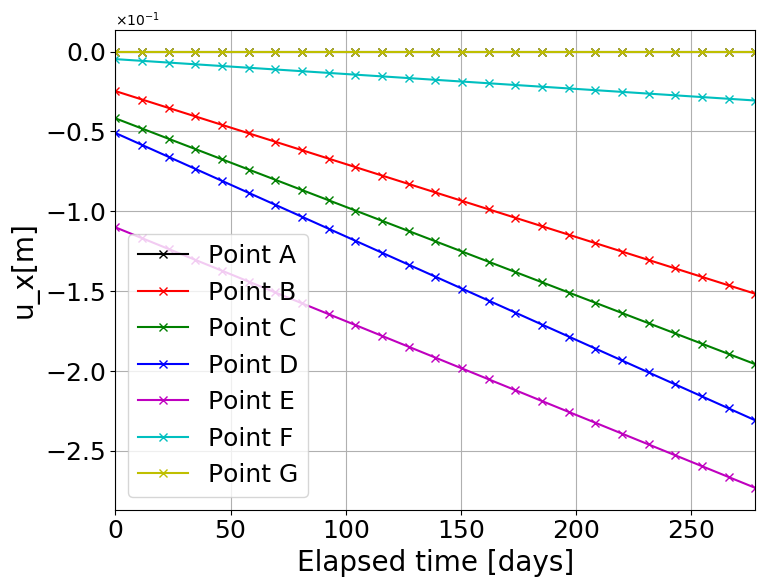}}
	\subfigure[]{
		\label{2ehetero}
		\includegraphics[scale=0.26197813971320]{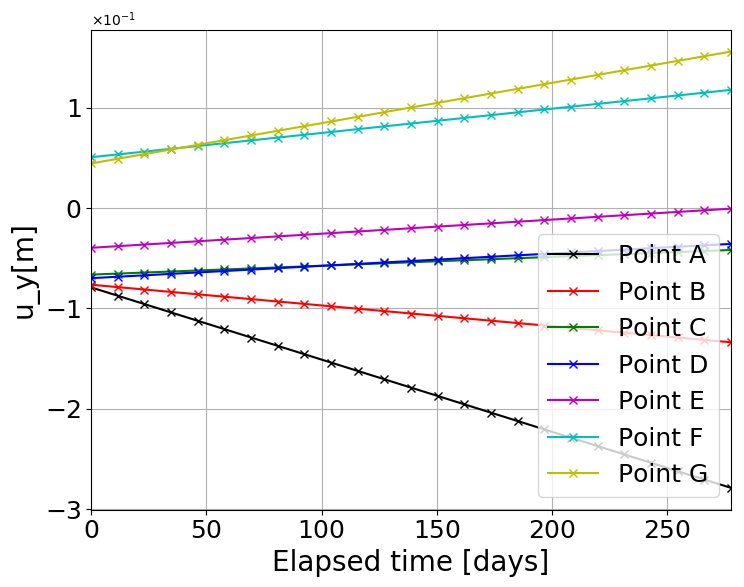}}
	\subfigure[]{
		\label{2fhetero}
		\includegraphics[scale=0.2197654397813971320]{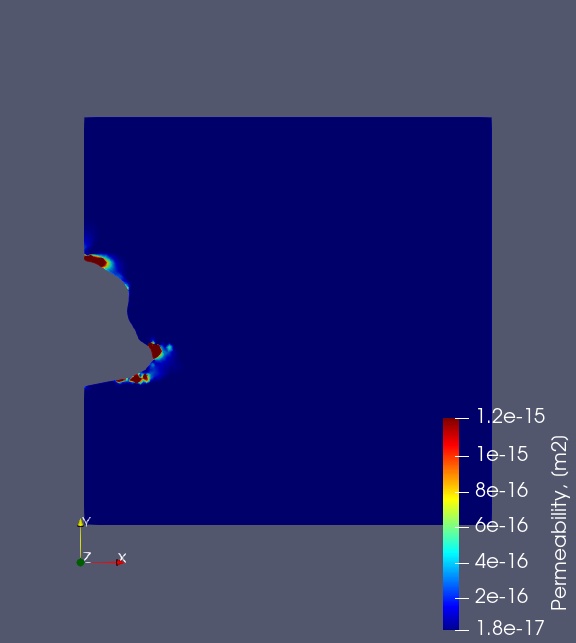}}
	
	\caption{\textbf{Test Case 5:} illustration of the horizontal displacement $u_x$ (a), vertical displacement $u_y$ (b), von Mises stress $s_{vm}$ (c) and permeability (f) for irregular cavern shape model with heterogeneous properties. The variation of horizontal (d) and vertical (e) deformations over time around caverns are also shown for the marked points in \autoref{creepgraphs_case4}.}
	\label{fig_irregular_cluster2heterogeneous}
\end{figure*}

 \iffalse 
 
 \begin{figure*}[h]
 	\centering   		
 		\subfigure[]{
 		\label{2adistribuition}
 		\includegraphics[width=0.345\textwidth]{images/download1.png}}
 	\subfigure[]{
 		\label{2ahetero}
 		\includegraphics[scale=0.2317654397813971320]{images/irreg/multipoint/withhetero/ux.jpeg}}
 	\subfigure[]{
 		\label{2bhetero}
 		\includegraphics[scale=0.2317654397813971320]{images/irreg/multipoint/withhetero/uy.jpeg}}
 	\subfigure[]{
 		\label{2chetero}
 		\includegraphics[scale=0.2317654397813971320]{images/irreg/multipoint/withhetero/svmwithhetero.jpeg}}
 	\subfigure[]{
 		\label{2dhetero}
 		\includegraphics[scale=0.26197813971320]{images/irreg/multipoint/withhetero/cylinder$u_x$85_vs_time.png}}
 	\subfigure[]{
 		\label{2ehetero}
 		\includegraphics[scale=0.26197813971320]{images/irreg/multipoint/withhetero/cylinder$u_y$85_vs_time.png}}
 	\subfigure[]{
 		\label{2fhetero}
 		\includegraphics[scale=0.2197654397813971320]{images/irreg/multipoint/withhetero/abspermwithhetrogeniety.jpeg}}
 	
 	\caption{\textbf{Test Case 5:} The above contours show the horizontal displacement $u_x$, vertical displacement $u_y$, permeability and von Mises stress $s_{vm}$ for irregular cavern shape model with heterogeneous domain. The variation of horizontal and vertical deformations with time around caverns are shown here for marked points as shown in \autoref{creepgraphs_case4}.}
 	\label{fig_irregular_cluster2heterogeneous}
 \end{figure*} 

\fi

\subsubsection{Test Case 5: Heterogeneity of elastic material properties}
\textcolor{black}{Heterogeneity is incorporated in this work by introducing a potash interlayer within a homogeneous halite rock salt deposit, as shown in Fig. \ref{creepgraphs_case4}. Firstly, in this Test Case 5, we allow only for elastic parameters to be heterogeneous, while plastic ones remain homogeneous. For this test case, potash Young's modulus and Poisson ratio are chosen to be 2.5 GPa and 0.35, respectively. To allow for significant deformation, these values are chosen in the lower limit of the reported values in the literature \cite{Nawaz2015,Duncan1993,XiaoXueying2011}. 
\\
The deformation and von Mises stress fields are shown in Fig. \ref{fig_irregular_cluster2heterogeneous}.  The magnitudes of deformation in x and y directions are comparable. Moreover, noticeable is that the introduction of potash interlayer has impacted the stress field and the deformation. For example, the point E within the interlayer (i.e., pink line in Fig. \ref{fig_irregular_cluster2heterogeneous}) shows higher deformation compared with the homogeneous test case. However, it shows slower creep rates for heterogeneous case, due to the lower local stress at the location of E. 
The maximum deformation change due to introducing the heterogeneous potash layer for this test case after 275 days is found to be approximately 9.5 \% and 38 \% for horizontal and vertical deformation, respectively. This is found by comparing the values of Fig. \ref{fig_irregular_cluster2homogeneous} and Fig. \ref{fig_irregular_cluster2heterogeneous}. Figure \ref{2fhetero} shows the variation of permeability in the domain, computed from the volumetric strain as stated in Eq. \ref{permeability}. It is observed that the permeability near the cavern wall within the heterogeneous potash is slightly smaller, compared with the homogeneous case. Note that here, only elastic parameters are considered to be heterogeneous. 
\\
These results motivate the next test case to investigate the impact of full heterogeneity in both elastic and plastic properties for the potash interlayer.}

\begin{figure*}
	\centering   		
	\subfigure[]{
		\label{midstrip}
		\includegraphics[scale=0.25211234246097813971320]{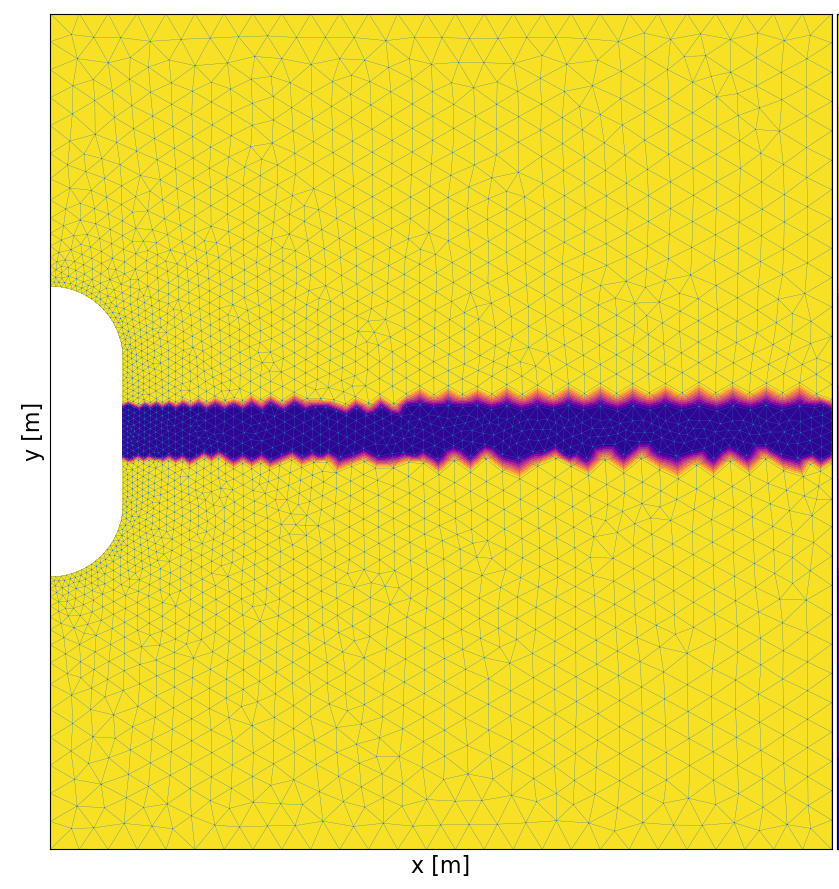}}
	\subfigure[]{
		\label{botstrip}
		\includegraphics[scale=0.25211234246097813971320]{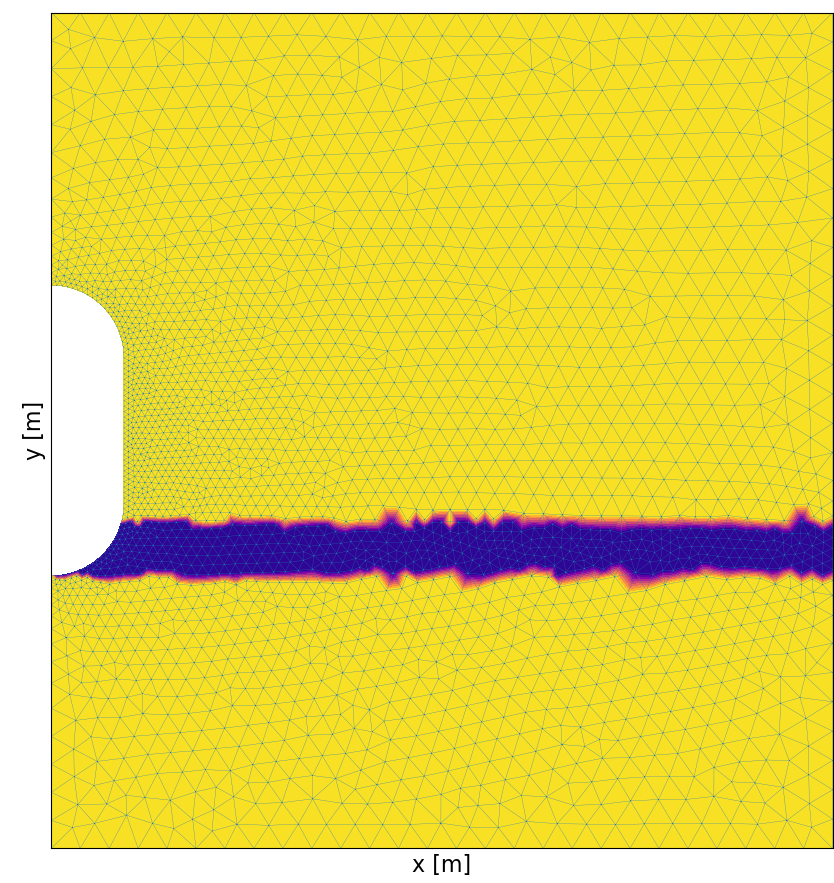}}
	\caption{\textbf{Test Case 6:} The above contours show the interlayers in geological domain around a cylindrical cavern. Interlayers are located in the mid strip as shown in Fig. \ref{midstrip} and  near the floor of the cavern as shown in Fig. \ref{botstrip}. Material properties chosen for interlayers are presented in \autoref{tab:inputinterlayers}.  }
	\label{figstrips}
\end{figure*}

\begin{table*}[h]
	\caption{\textbf{Test case 6:} Creep formulations of heterogeneity in rock salt \cite{Zhang2012,Nawaz2015,Duncan1993} } \label{creepformulationssalthetereogenietyformulaion}
	\centering
	\begin{tabular}{ c c c c c c }
		\hline
		\textbf{Material}  & \textbf{Formulation}  & \textbf{Youngs modulus} & \textbf{Poisson ratio} & \textbf{n} \\ 
		\hline
		Carnallite
		& 2.6804e-14 $\times \sigma^{5}$  & 17 GPa  & 0.33  &  5 \\

		Bischofite
		& 1.1e-9$\times \sigma^{4.6}$ & 18 GPa & 0.36  &  4.6  \\
		\hline
	\end{tabular}%
	\label{tab:inputinterlayers}%
\end{table*}%

\begin{figure*}
	\centering   		
	\subfigure[]{
		\label{midCarnallitea}
		\includegraphics[scale=0.19138352097813971320]{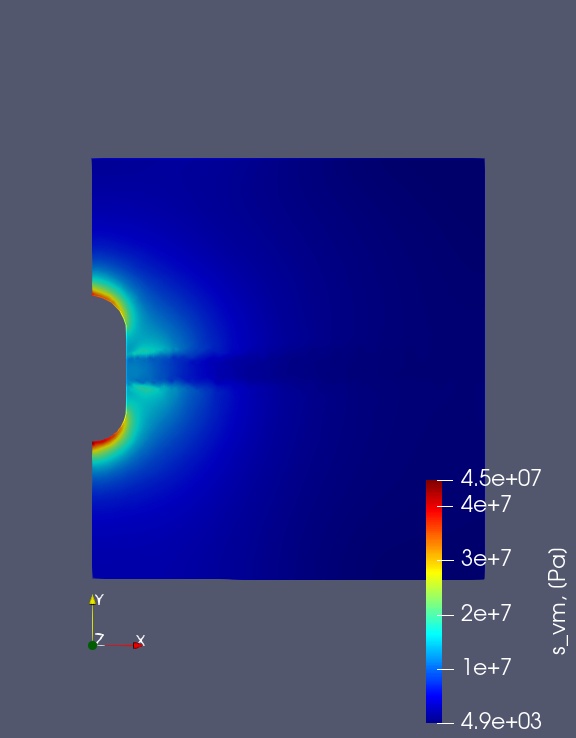}}
	\subfigure[]{
		\label{midCarnalliteb}
		\includegraphics[scale=0.191352097813971320]{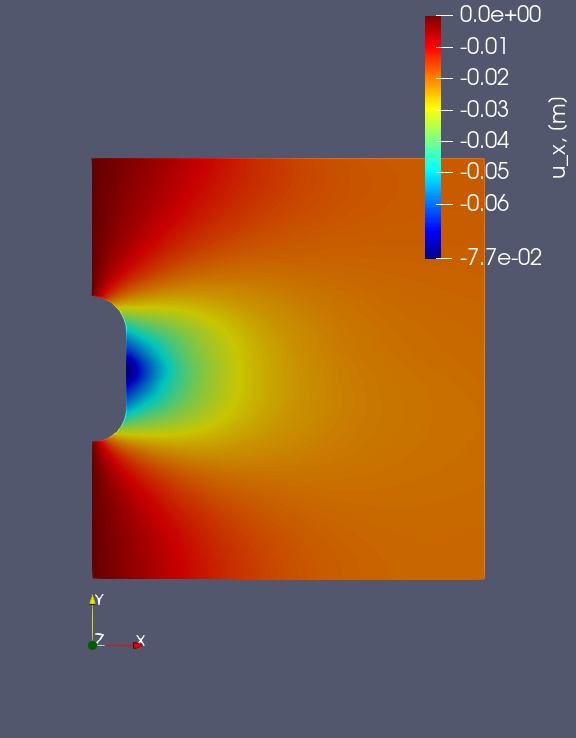}}
	\subfigure[]{
		\label{midCarnallitec}
		\includegraphics[scale=0.1913052097813971320]{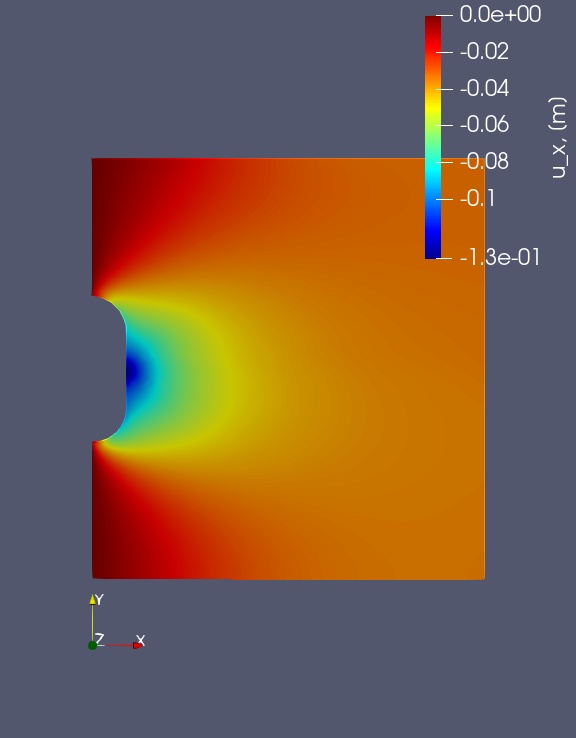}}\\
	\subfigure[]{
		\label{midCarnallited}
		\includegraphics[scale=0.26453052097813971320]{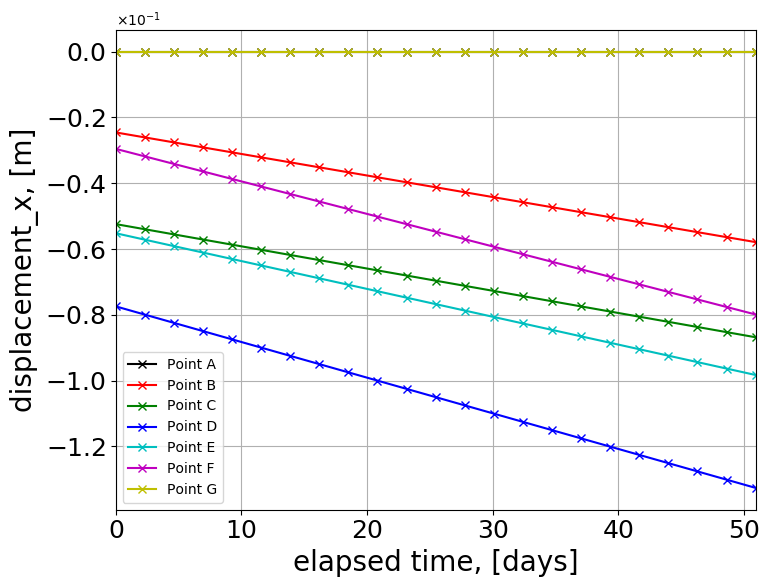}}
	\subfigure[]{
		\label{midCarnallitee}
		\includegraphics[scale=0.26453097813971320]{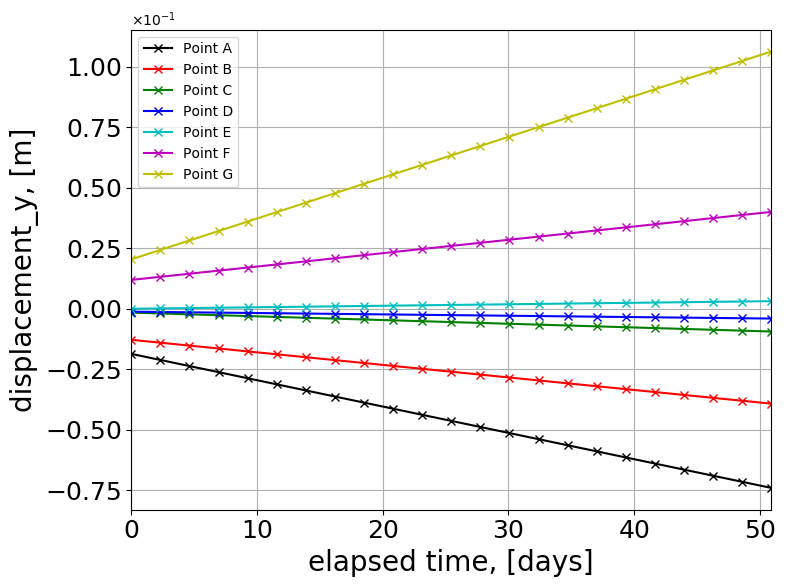}}
	\subfigure[]{
		\label{midCarnallitek}
		\includegraphics[scale=0.19453097813971320]{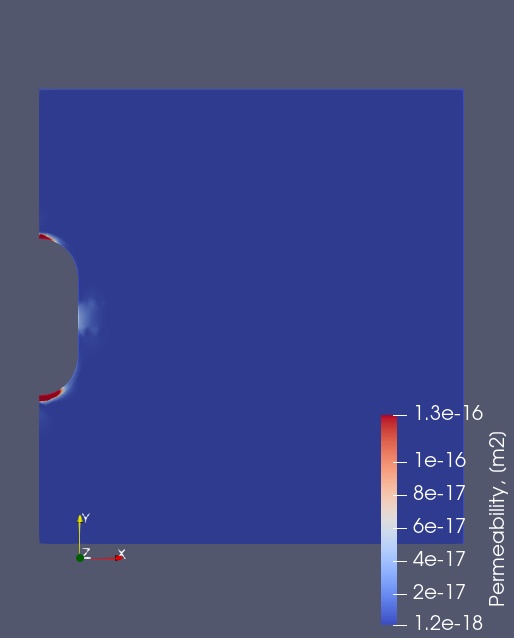}}
	\subfigure[]{
		\label{midCarnallitef}
		\includegraphics[scale=0.1933097813971320]{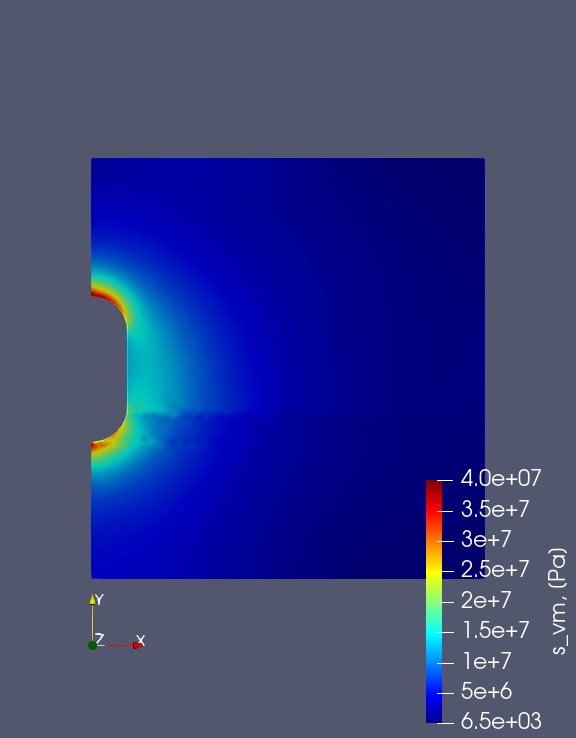}}
	\subfigure[]{
		\label{midCarnalliteg}
		\includegraphics[scale=0.193352097813971320]{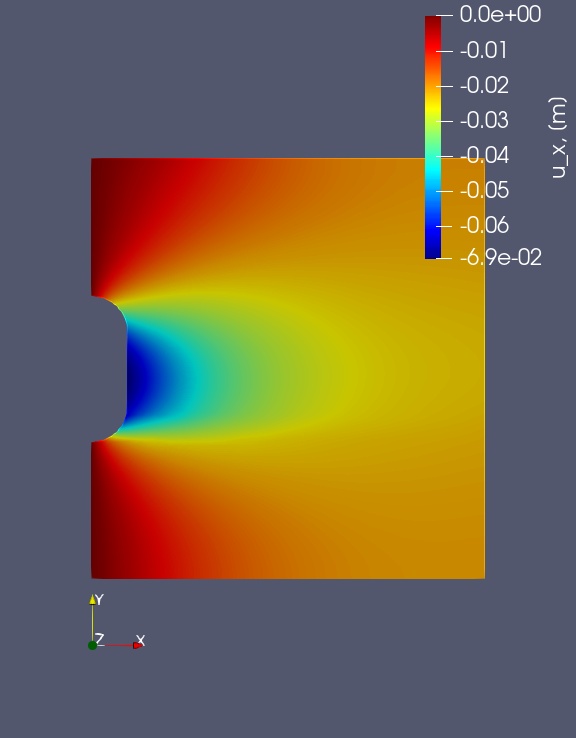}}
	\subfigure[]{
		\label{midCarnalliteh}
		\includegraphics[scale=0.193352097813971320]{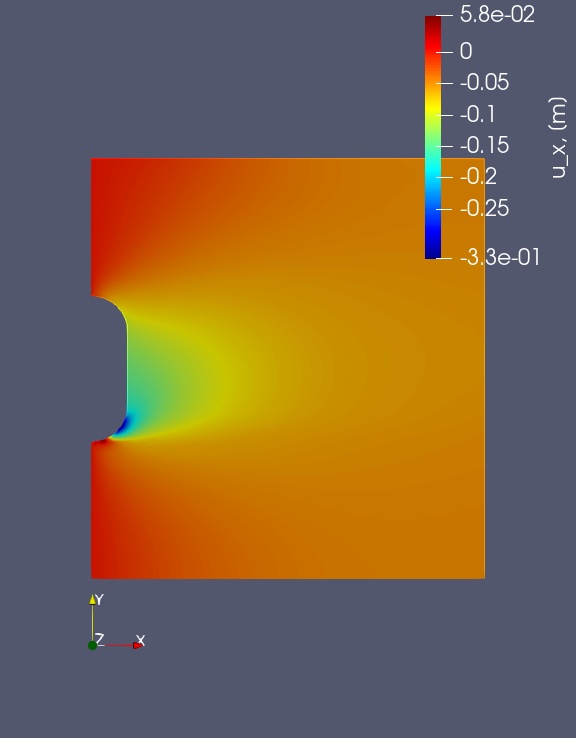}}\\
	\subfigure[]{
		\label{midCarnallitei}
		\includegraphics[scale=0.260352097813971320]{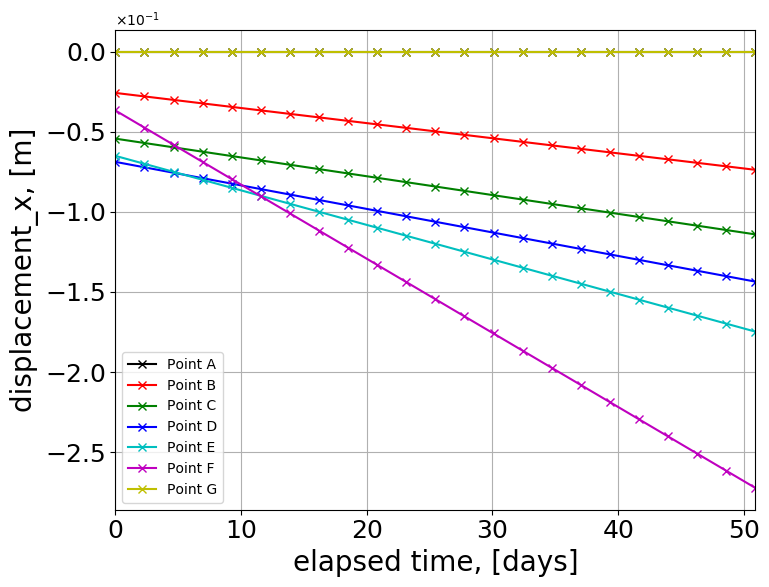}}
	\subfigure[]{
		\label{midCarnallitej}
		\includegraphics[scale=0.260352097813971320]{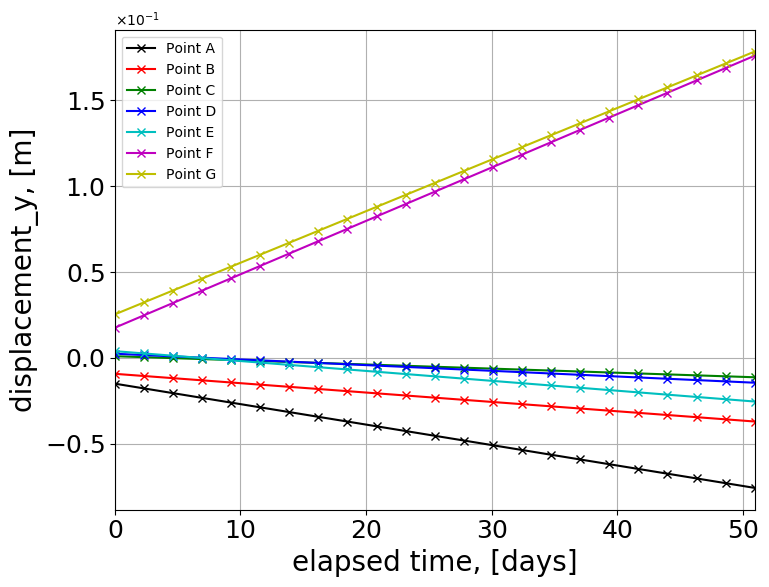}}
	\subfigure[]{
		\label{midCarnallitel}
		\includegraphics[scale=0.19]{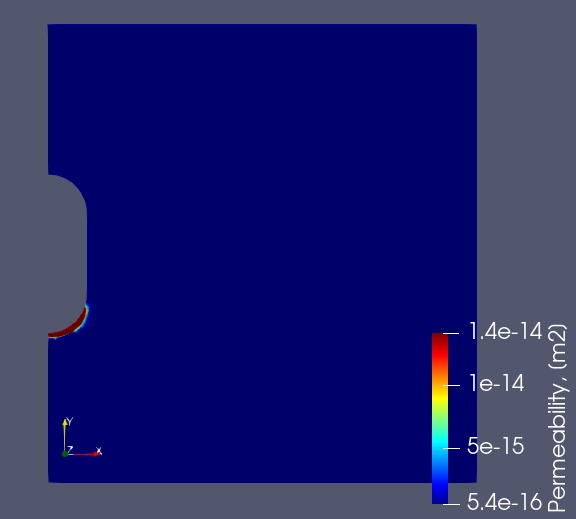}}
	\caption{\textbf{Test Case 6:} illustration of the contours for heterogeneous domain with an interlayer of {Carnallite}. Figure \ref{midCarnallitea} to Fig. \ref{midCarnallitee} show the contours when the interlayer is in the midgle of the vertical dimension. Figure \ref{midCarnallitef} to Fig. \ref{midCarnallitej} show the contours when the interlayer is near the floor of the cavern. Figure \ref{midCarnallitea} and Fig. \ref{midCarnallitef} show the von Mises stress distributions. Figure \ref{midCarnalliteb} and Fig. \ref{midCarnallitec} show the horizontal deformation distribution at initial and final time steps, respectively. Figure \ref{midCarnalliteg} and Fig. \ref{midCarnalliteh} show the horizontal deformation distribution at initial and final time steps, respectively. Figure \ref{midCarnallited}, Fig. \ref{midCarnallitee}, Fig. \ref{midCarnallitei} and Fig. \ref{midCarnallitej} show the variation of horizontal and vertical deformation with time for simulation period of 50 days.}
	\label{Carnallitegraphs}
\end{figure*}

\begin{figure*}
	\centering   		
	\subfigure[]{
		\label{midBischofitea}
		\includegraphics[scale=0.19352097813971320]{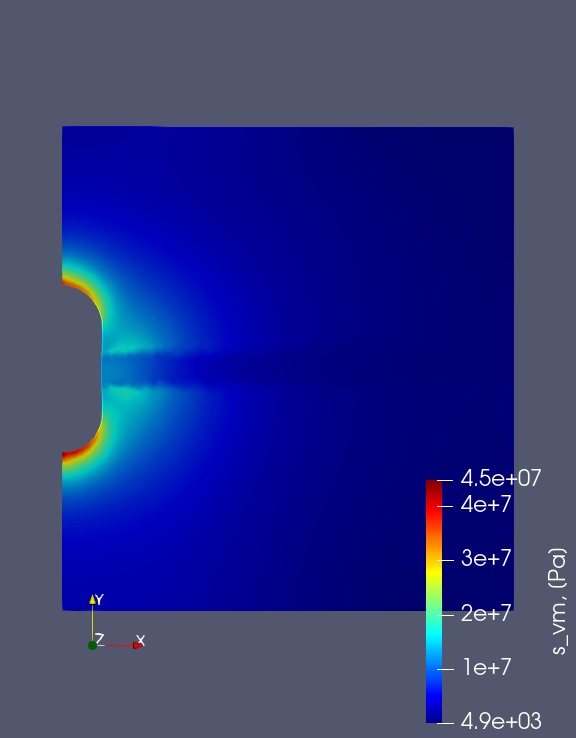}}
	\subfigure[]{
		\label{midBischofiteb}
		\includegraphics[scale=0.19352097813971320]{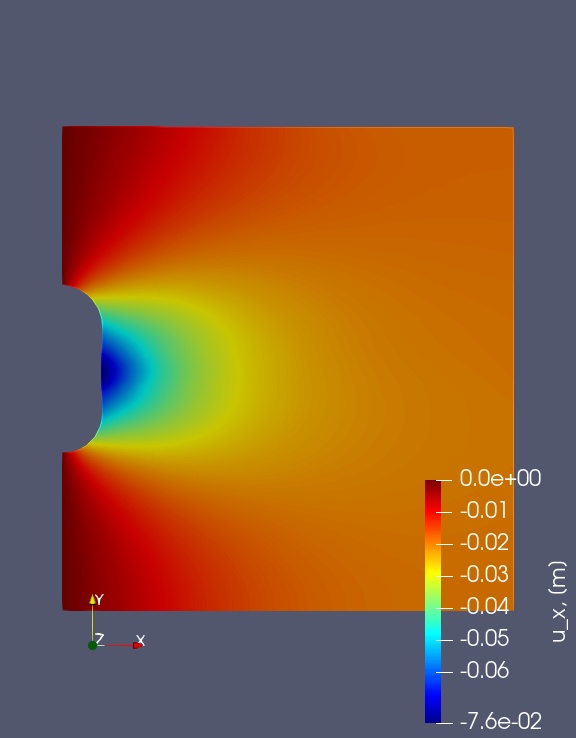}}
	\subfigure[]{
		\label{midBischofitec}
		\includegraphics[scale=0.193052097813971320]{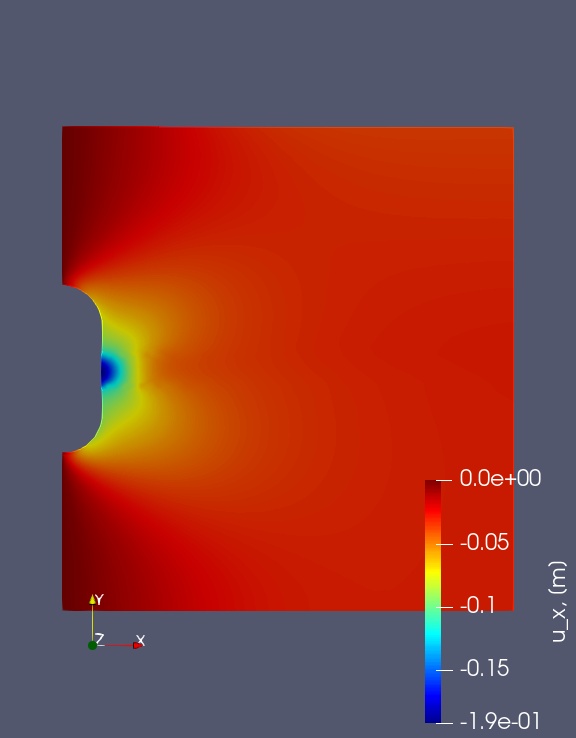}}\\
	\subfigure[]{
		\label{midBischofited}
		\includegraphics[scale=0.26453052097813971320]{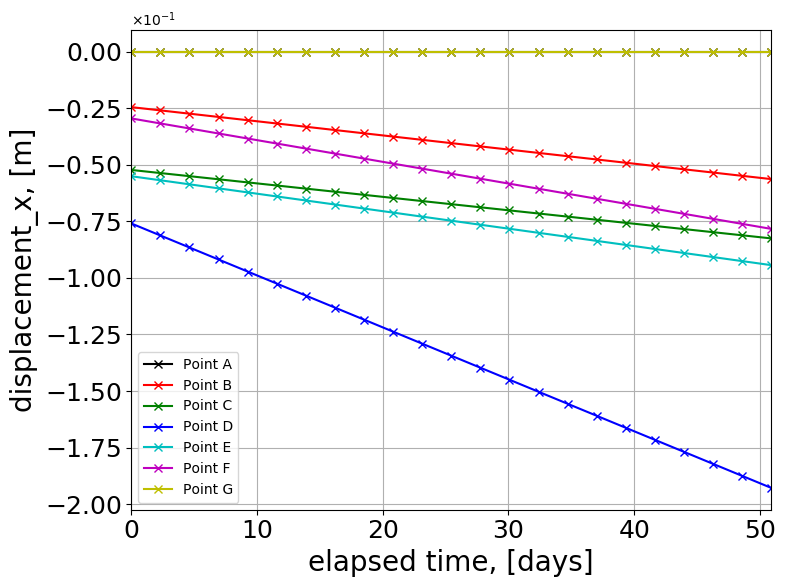}}
	\subfigure[]{
		\label{midBischofitee}
		\includegraphics[scale=0.26453097813971320]{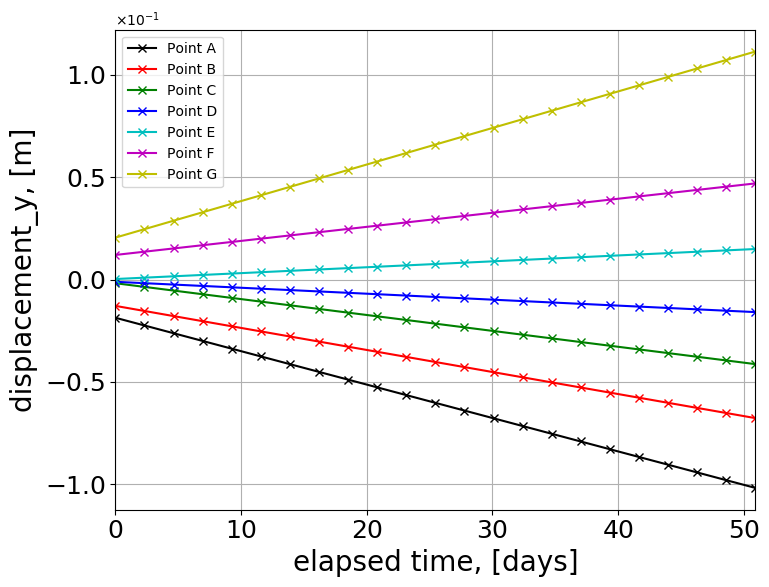}}
	\subfigure[]{
		\label{midBischofitek}
		\includegraphics[scale=0.19453097813971320]{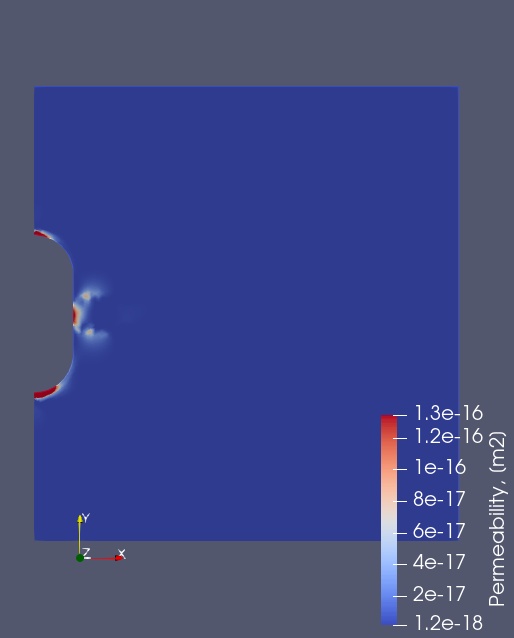}}
	\subfigure[]{
		\label{midBischofitef}
		\includegraphics[scale=0.1903097813971320]{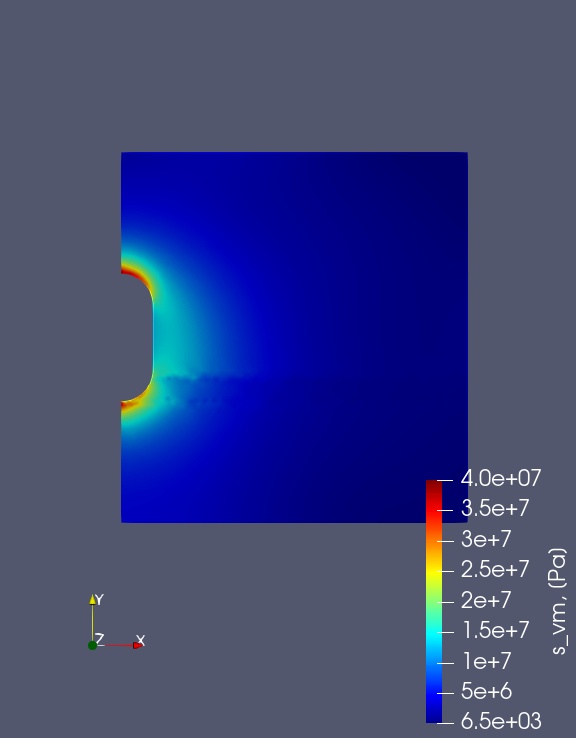}}
	\subfigure[]{
		\label{midBischofiteg}
		\includegraphics[scale=0.190352097813971320]{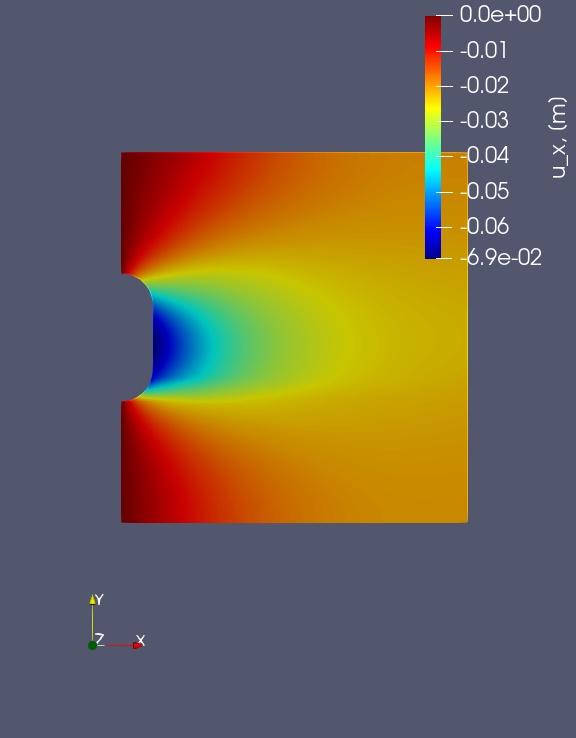}}
	\subfigure[]{
		\label{midBischofiteh}
		\includegraphics[scale=0.190352097813971320]{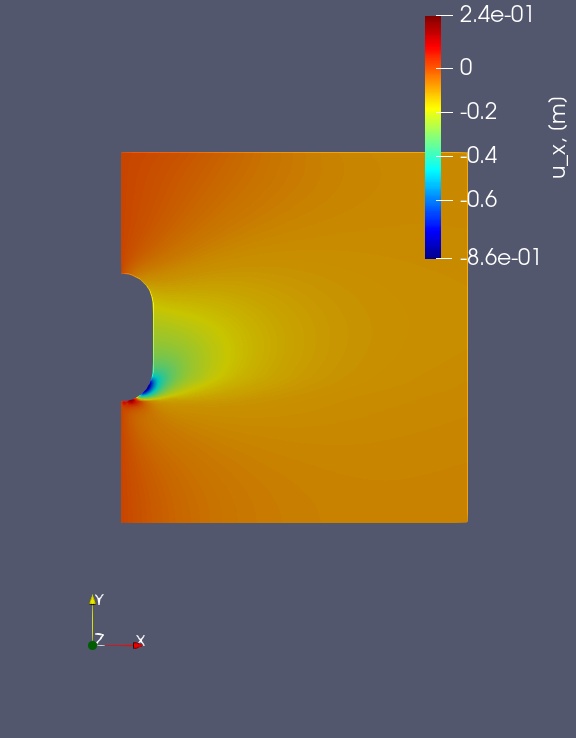}}\\
	\subfigure[]{
		\label{midBischofitei}
		\includegraphics[scale=0.240352097813971320]{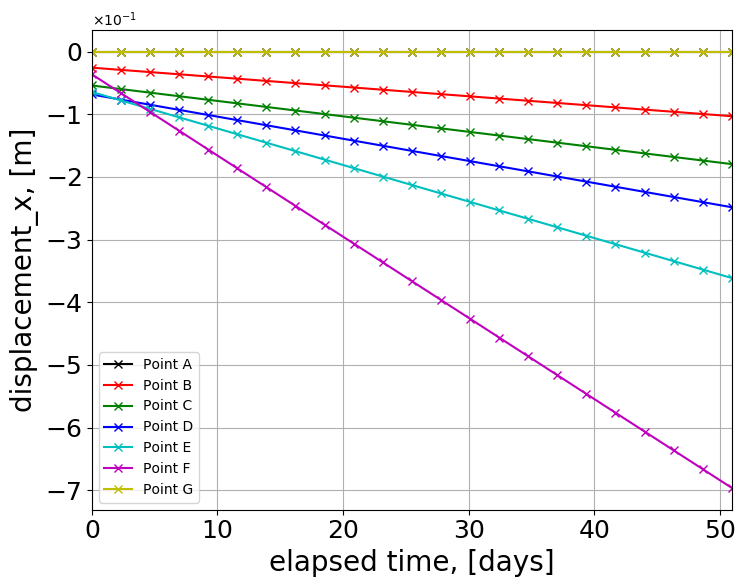}}
	\subfigure[]{
		\label{midBischofitej}
		\includegraphics[scale=0.240352097813971320]{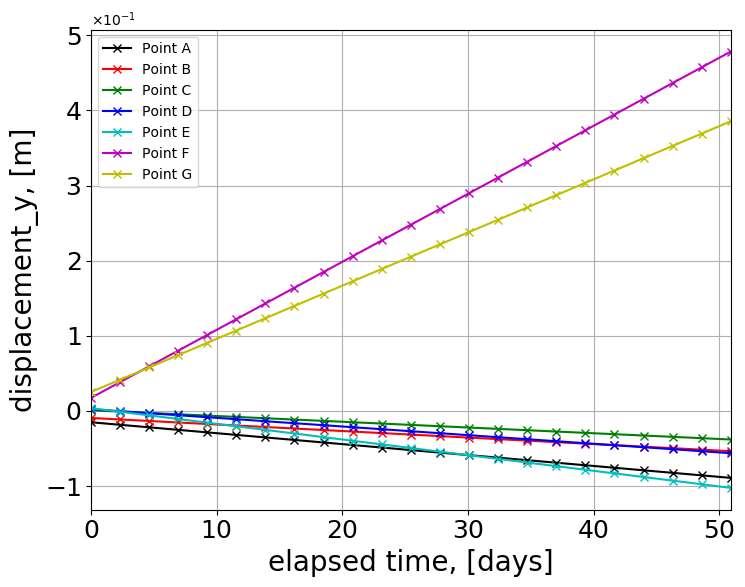}}
	\subfigure[]{
		\label{midBischofitel}
		\includegraphics[scale=0.19453097813971320]{images/heterogenietystrips/botstrip/Carnallite/absperm.jpeg}}
	\caption{\textbf{Test Case 6:} illustration of the contours for heterogeneous domain with an interlayer of {Bischofite}. Figure \ref{midBischofitea} to Fig. \ref{midBischofitee} show the contours when the interlayer is in the middle of the vertical dimension. Figure \ref{midBischofitef} to Fig. \ref{midBischofitej} show the contours when the interlayer is near the floor of the cavern. Figure \ref{midBischofitea} and Fig. \ref{midBischofitef} show the von Mises stress distribution. Figure \ref{midBischofiteb} and Fig. \ref{midBischofitec} show the horizontal deformation distribution at initial and final time-steps, respectively. Figure \ref{midBischofiteg} and Fig. \ref{midBischofiteh} show the horizontal deformation distribution at initial and final time-steps, respectively. Figure \ref{midBischofited}, Fig. \ref{midBischofitee}, Fig. \ref{midBischofitei} and Fig. \ref{midBischofitej} show the variation of horizontal and vertical deformation with time for simulation period of 50 days.}
	\label{Bischofitegraphs}
\end{figure*}

\subsubsection{Test Case 6: Heterogeneous interlayers}
The second, more realistic method is employing different insoluble layers and using their elastic and creep properties. To incorporate heterogeneity in the geological domain surrounding salt caverns, interlayers considering many insoluble impurities (anhydrites, potash salts, shale, gypsum, mudstone, etc.) \cite{Liu2014,Liang2007,Li2018} are included. The detailed composition of rock salt and interlayers can vary depending on the type of region. Experimental analysis needs to be conducted to understand the structure and the lithological composition of the geological domain \cite{Zhang2020,Liang2007}. In this work, interlayers of different materials are shown in \autoref{creepformulationssalthetereogenietyformulaion} are chosen as heterogeneity in the geological domain of halite. The table shows the constitutive parameters of creep formulations for different materials are taken from the literature. Considering i) different laboratory conditions, ii) timescales for conducting experiments, ii) water content in the rock samples, iv) different loading conditions, and v) purity of the samples chosen in the literature, these creep constitutive parameters can change depending on the type of application it is being used for instance, in this case, being energy storage. This study becomes very critical given right parameters of the interlayers are chosen. The experiments were conducted for roughly about 500 hrs, and at 70$^\circ $ C for both carnallite and bischofite \cite{Nawaz2015}. This work used these constants due to the lack of other suitable literature for energy storage applications. The creep constants are reduced by three orders of magnitude for both carnallite and bischofite because of the difference in temperature the experiments are conducted and the current case study because of their temperature dependence from the Arrhenius formulation as shown in \cite{Taheri2020}. \\
Fig. \ref{figstrips} shows the interlayers in the geological domain. The width of the interlayer is 30m. The interlayers are placed in the two regions near the salt cavern. Fig. \ref{midstrip} shows the interlayer around the midplane of the cavern. Locally around that interlayer, the curvature is minimum. Fig. \ref{botstrip} shows the interlayer near the roof of the cavern with the highest curvature and more lithostatic pressure compared to the midplane. The entire simulation was conducted for 50 days. Fig. \ref{Carnallitegraphs} and Fig. \ref{Bischofitegraphs} shows various graphs for carnallite and bischofite respectively. Von mises stress distribution, horizontal deformation at initial and final time step, and variation of horizontal and vertical deformation with time for points around the cavern as shown in the schematic Fig. \ref{fig_grad} are shown for both the interlayer distributions. The magnitudes of von Mises stress for both carnallite and bischofite materials are the same. However compared to the homogeneous test case of cylindrical cavern Fig. \ref{2a-case2}, the minimum value of stress has reduced and the distribution is not smooth anymore. This is due to the discontinuous distribution of Young's modulus and the magnitudes of Young's modulus of interlayers is lower than halite. Due to this, the magnitudes of horizontal deformation at t=0 are similar as shown in Fig. \ref{midCarnalliteb}, Fig. \ref{midCarnalliteg}, Fig. \ref{midBischofiteb} and Fig. \ref{midBischofiteg}. However, the magnitudes of these deformations are higher than the homogeneous test case as shown in Fig. \ref{2b-case2}. The deformation magnitude is higher for interlayers in the mid than the interlayer near the floor due to low stress in the central region and low curvature. \\
When creep is incorporated in these interlayers with separate properties, it's an additional nonlinear physics that depends on the stress distribution. Fig. \ref{midCarnallitec}, Fig. \ref{midBischofitec}, Fig. \ref{midCarnalliteh} and Fig. \ref{midBischofiteh} show the horizontal deformation distribution at end of simulation after 50 days when interlayer are incorporated comprising carnallite and bischofite respectively. Due to different stress distributions caused by lithostatic pressure and curvature, the accumulated creep deformation also varies. Fig. \ref{midCarnallitec} and Fig. \ref{midBischofitec} show a similar distribution of deformation due to the same location of heterogeneity; however, the bischofite shows higher deformation magnitudes compared to carnallite. This can be because even though the creep exponent for carnallite is higher than bischofite, the creep constant of bischofite is higher than carnallite by five orders of magnitude. The sensitivity of the creep constant and creep exponent is studied in the later section. Fig. \ref{midCarnalliteh} and Fig. \ref{midBischofiteh} show the horizontal deformation when the interlayer is located near the floor of the cavern. Qualitatively, they look similar; however, the interlayer with bischofite shows higher horizontal deformation than the carnallite test case and a homogeneous test case. \\
Fig. \ref{midCarnallited}, Fig. \ref{midCarnallitee}, Fig. \ref{midCarnallitei} and Fig. \ref{midCarnallitej} show the variation of horizontal and vertical deformation with time at different points as shown in the schematic for carnallite interlayers.  Fig. \ref{midBischofited}, Fig. \ref{midBischofitee}, Fig. \ref{midBischofitei} and Fig. \ref{midBischofitej} show the variation of horizontal and vertical deformation with time at different points as shown in the schematic for bischofite interlayers. Compared to the homogeneous test case as shown in Fig. \ref{fig_creepmodel_clusrer}, the horizontal deformation plots for each point are more widely spread when the interlayer is in the midplane. However, when the interlayer is near the floor, Point F shows a much steeper slope than the rest of the points. This shows that given a local heterogeneity in the domain, a large amount of deformation could occur, causing failure of the salt cavern. In the vertical deformation, the plots are magnified proportionately compared to homogeneous test cases where the magnitude is higher but the distribution looks similar when the interlayer is the midplane. When the interlayer is near the cavern floor, the magnitude of vertical deformation at point F with high curvature and heterogeneity has a higher rate of increase in deformation than a homogeneous test case. Permeability of salt caverns which is obtained from volumetric strain, can be seen in the graphs Figure \ref{midCarnallitek} and Figure \ref{midCarnallitel} for carnallite with interlayers in the mid strip and near the floor of the cavern.  Figure \ref{midBischofitek} and Figure \ref{midBischofitel} show the variation of permeability with interlayer as bischofite in the mid and bottom strip, respectively. When the interlayer is the midplane, we can see that the potential failure zones are the floor, roof (same as without interlayers), and the midplane section with the interlayer. This potential failure zone is created only because of the interlayer heterogeneity. When the interlayer heterogeneity is near the cavern floor, the main potential failure zone is near the floor around point E. Followed by the roof of the cavern location. Here the impact of material and geometrical heterogeneity has combined, resulting in a  very high potential failure zone. Also, the magnitude of bottom strip interlayer permeability is higher. Similar permeability distribution is observed for both carnallite and bischofite when the interlayers are in the mid or near floor, respectively.    
\\
The maximum increase in percentage for horizontal and vertical deformation in the geological domain when compared to homogeneous test case (fig. \ref{fig_creepmodel_clusrer}) after 50 days of simulation is shown in \autoref{Percentageincreasetable}. The percentages are higher for bischofite material compared to carnallite. When the interlayers are present near the cavern floor, a higher increase in deformation is observed than the mid-plane interlayer. Also, the percentage increased for both the deformations when the creep properties are incorporated to study heterogeneity is much higher than when only elastic properties are included, as shown in the previous test case. Considering the chosen creep parameters are in the realistic range, the results from this section clearly show that interlayers can fail the cavern locally around the cavern.

\begin{table*}
	\caption{Maximum percentage increase in deformation when interlayers are incorporated when compared to homogeneous test case after 50 days } 
	\centering
	\begin{tabular}{ c c c }
		\hline
		\textbf{Interlayer}  & $u_x$ & $ u_y $\\  
		\hline
		Midplane-carnallite 
		& 44.4\%  & 12\%  \\
		
		Floor-carnallite 
		&  200 \% & 75 \%  \\
		
		Midplane-bischofite
		&  111 \% & 20 \%  \\
		
		Floor-bischofite 
		& 677 \%  &  380 \% \\
		\hline
	\end{tabular}%
	\label{Percentageincreasetable}%
\end{table*}%

\subsubsection{Test Case 7: Real field test case}
Using the field data on cavern shape from echo logs of a salt cavern in Germany \cite{Laban_2020}, a cross-section of the modeled cavern was generated to be used in the developed simulator. The test case simulation results are shown in Fig. \ref{fig_real_cav}. The computed displacement distribution, the simulation output, has a similar distribution to the cylinder-shaped cavern. Due to the increased lithostatic pressure, the maximum horizontal deformation is just below the midplane of the cavern. While the maximum horizontal deformation appears at the cavern roof, lower deformation values are predicted due to the Dirichlet boundary conditions along the bottom boundary. This simulation was run for two years.

\begin{figure*}
	\centering   		
	\subfigure[]{
		\label{2areal}
		\includegraphics[scale=0.1797813971320]{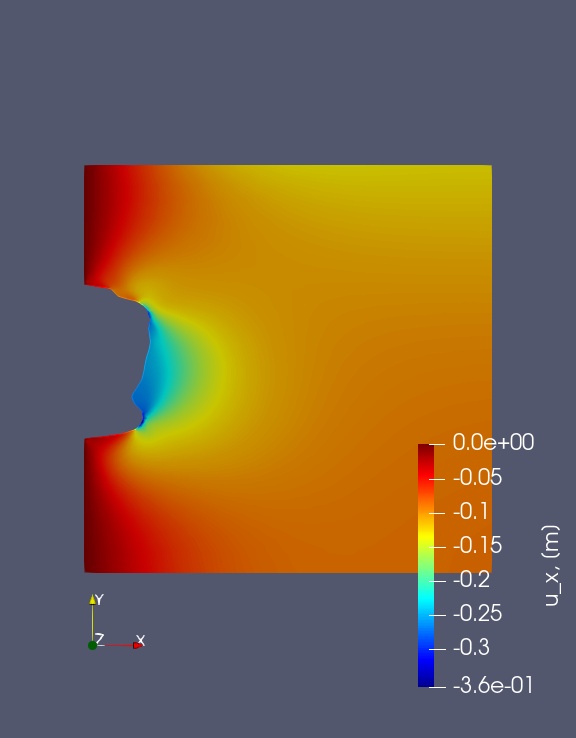}}
	\subfigure[]{
		\label{2breal}
		\includegraphics[scale=0.1797813971320]{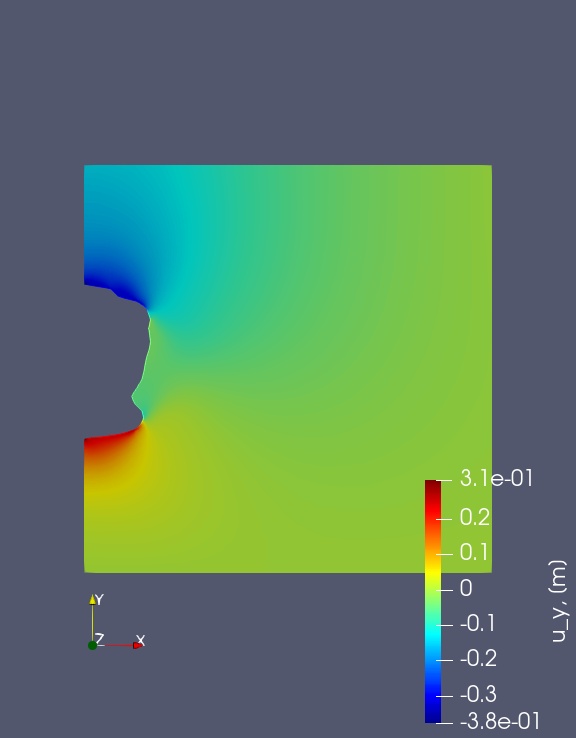}}
	\subfigure[]{
		\label{2creal}
		\includegraphics[scale=0.1797813971320]{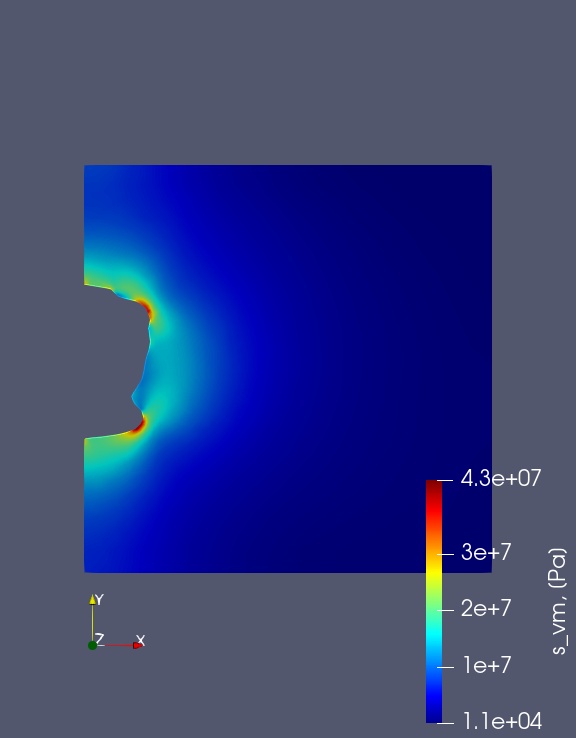}}
	\caption{\textbf{Test Case 7:} The above contours show the horizontal displacement (Fig. \ref{2areal}), vertical displacement (Fig. \ref{2breal}) and von Mises stress (Fig. \ref{2creal}) for real field cavern shape model. The simulation was run for a period of 2 years.}
	\label{fig_real_cav}
\end{figure*}

\iffalse 
\begin{figure*}
	\centering   		
	\subfigure[]{
		\label{2areal}
		\includegraphics[scale=0.197813971320]{images/h2cav/u_x0.pdf}}
	\subfigure[]{
		\label{2breal}
		\includegraphics[scale=0.197813971320]{images/h2cav/u_y0.pdf}}
	\subfigure[]{
		\label{2creal}
		\includegraphics[scale=0.197813971320]{images/h2cav/s_vm0.pdf}}
	\caption{\textbf{Test Case 5:} The above contours show the horizontal displacement (Fig. \ref{2areal}), vertical displacement (Fig. \ref{2breal}) and von Mises stress (Fig. \ref{2creal}) for real field cavern shape model. The simulation was run for a period of 3 years.}
	\label{fig_real_cav}
\end{figure*} 
\fi

\iffalse 
\begin{figure*}
	\centering
	\begin{subfigure}{.32\textwidth}
		\centering
		\includegraphics[trim={7cm 0 0 0}, scale=0.17813971320]{images/h2cav/u_x0.pdf}
	\end{subfigure}
	\begin{subfigure}{.32\textwidth}
		\centering
		\includegraphics[trim={6cm 0 0 0},scale=0.17813971320]{images/h2cav/u_y0.pdf}
	\end{subfigure}
	\begin{subfigure}{.32\textwidth}
		\centering
		\includegraphics[trim={5cm 0 0 0},scale=0.17813971320]{images/h2cav/s_vm0.pdf}
	\end{subfigure}
	\centering
	\caption{\textbf{Test Case 6}: displacement in x (left), y (center) and von Mises stress (right) for real field cavern shape model.}
	\label{fig_real_cava}
\end{figure*}
\fi

\subsubsection{Test Case 8: Tertiary creep and material failure}
\begin{figure*}
	\centering   		
	\subfigure[]{
		\label{fig_damage_001}
		\includegraphics[width=0.321\linewidth]{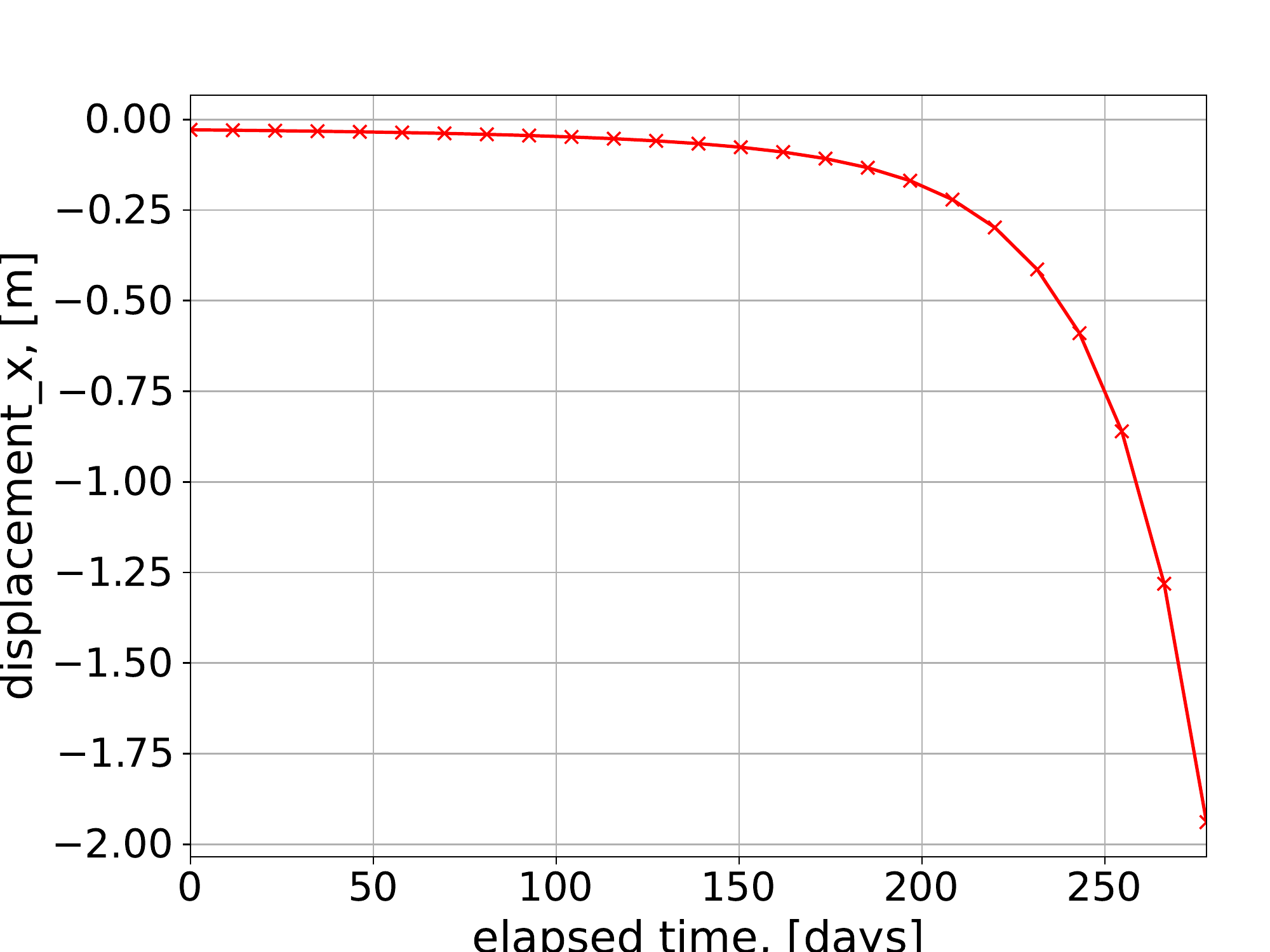}}
	\subfigure[]{
		\label{fig_damage_002}
		\includegraphics[width=0.321\linewidth]{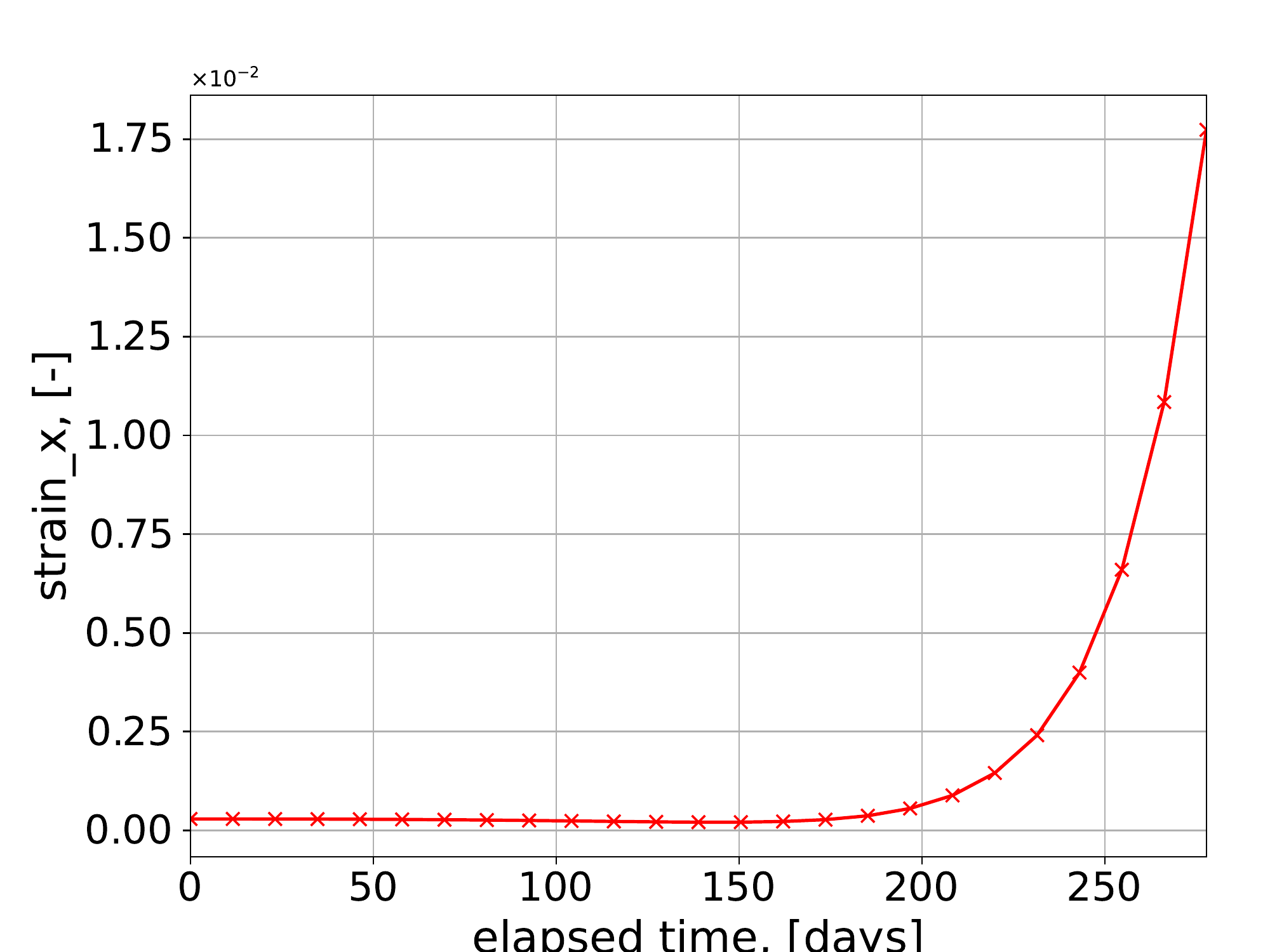}}
	\subfigure[]{
		\label{fig_damage_003}
		\includegraphics[width=0.321\linewidth]{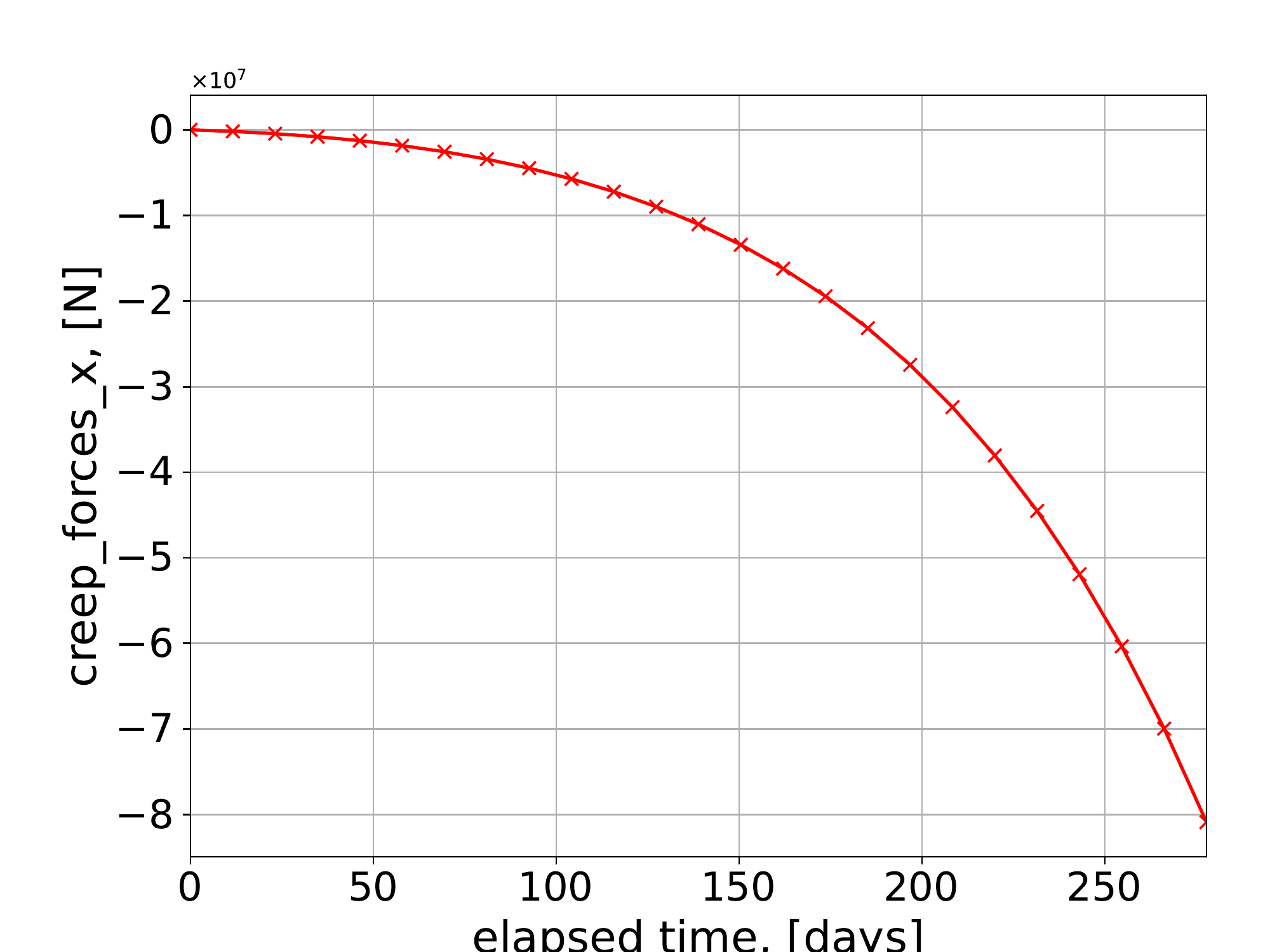}}
	\caption{\textbf{Test Case 8:} The above plots show the variation of horizontal displacement $u_x$, horizontal strain $\varepsilon_{xx}$ and horizontal component of creep forces $f_x$ over time respectively at the roof of the cavern.}
	\label{fig_creepgraphs}
\end{figure*}

\begin{figure*}[h]
	\centering   		
	\subfigure[]{
		\label{fig_sensB}
		\includegraphics[width=0.31\linewidth]{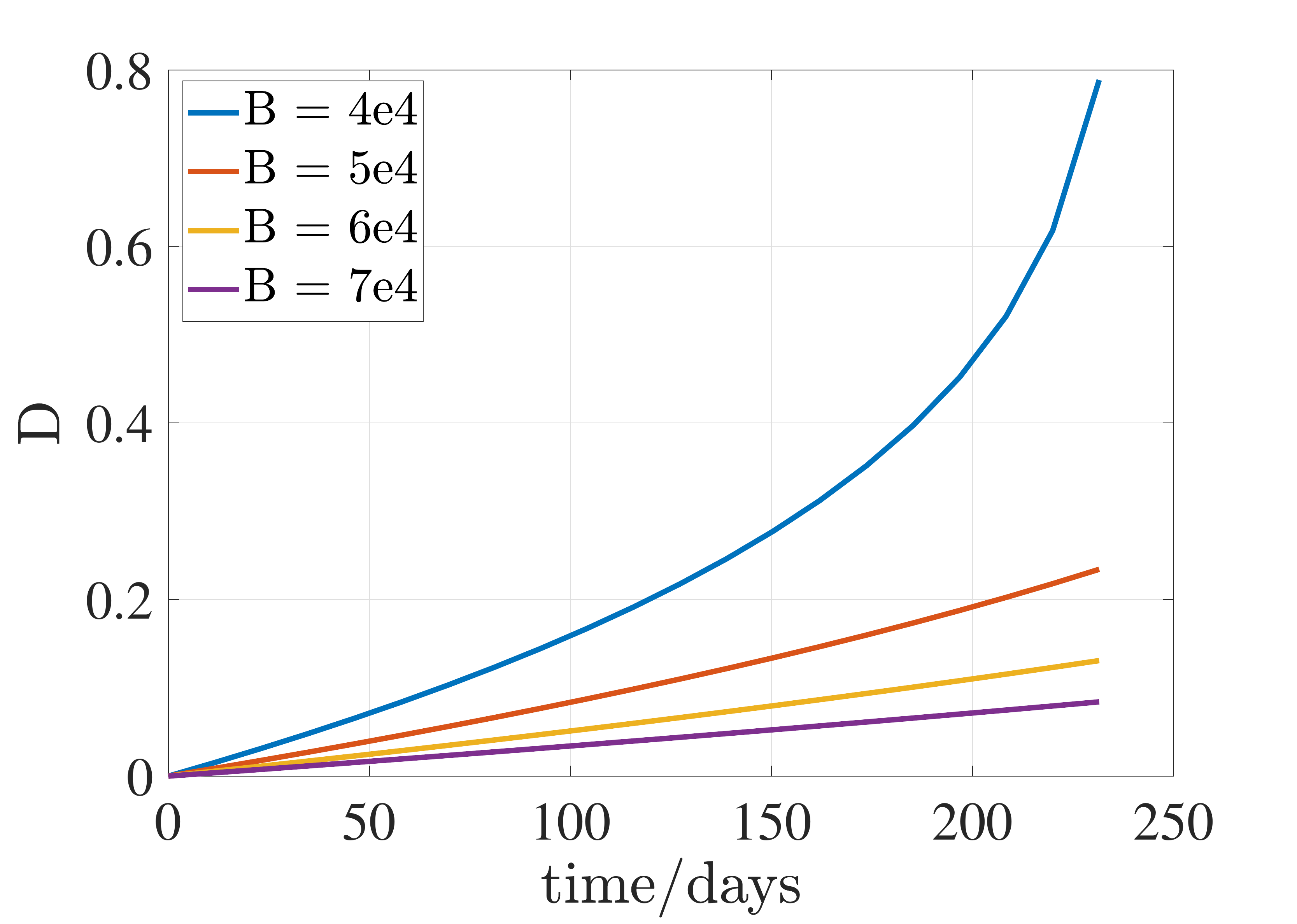}}
	\subfigure[]{
		\label{fig_sensr}
		\includegraphics[width=0.31\linewidth]{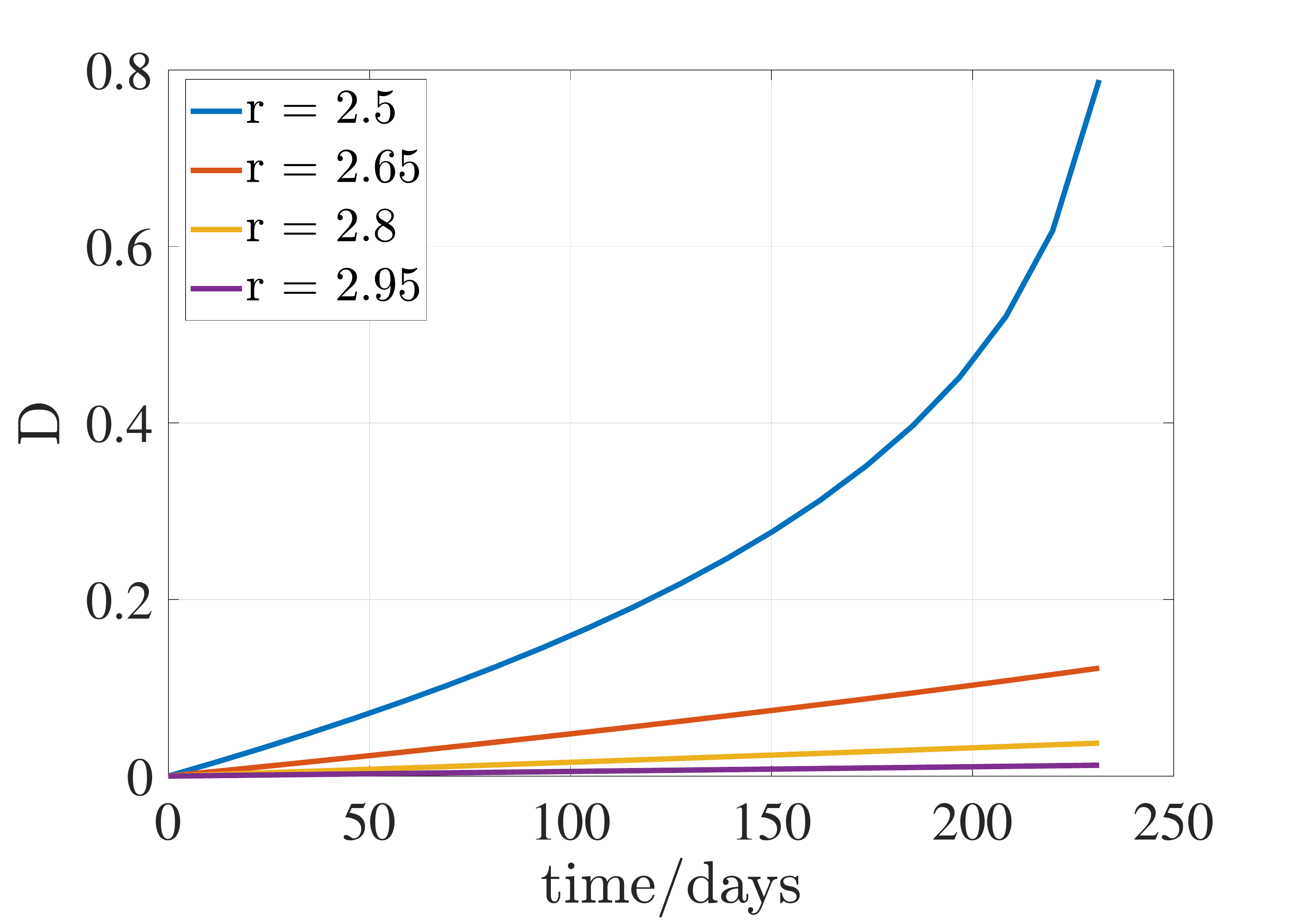}}
	\subfigure[]{
		\label{fig_senssigma}
		\includegraphics[width=0.31\linewidth]{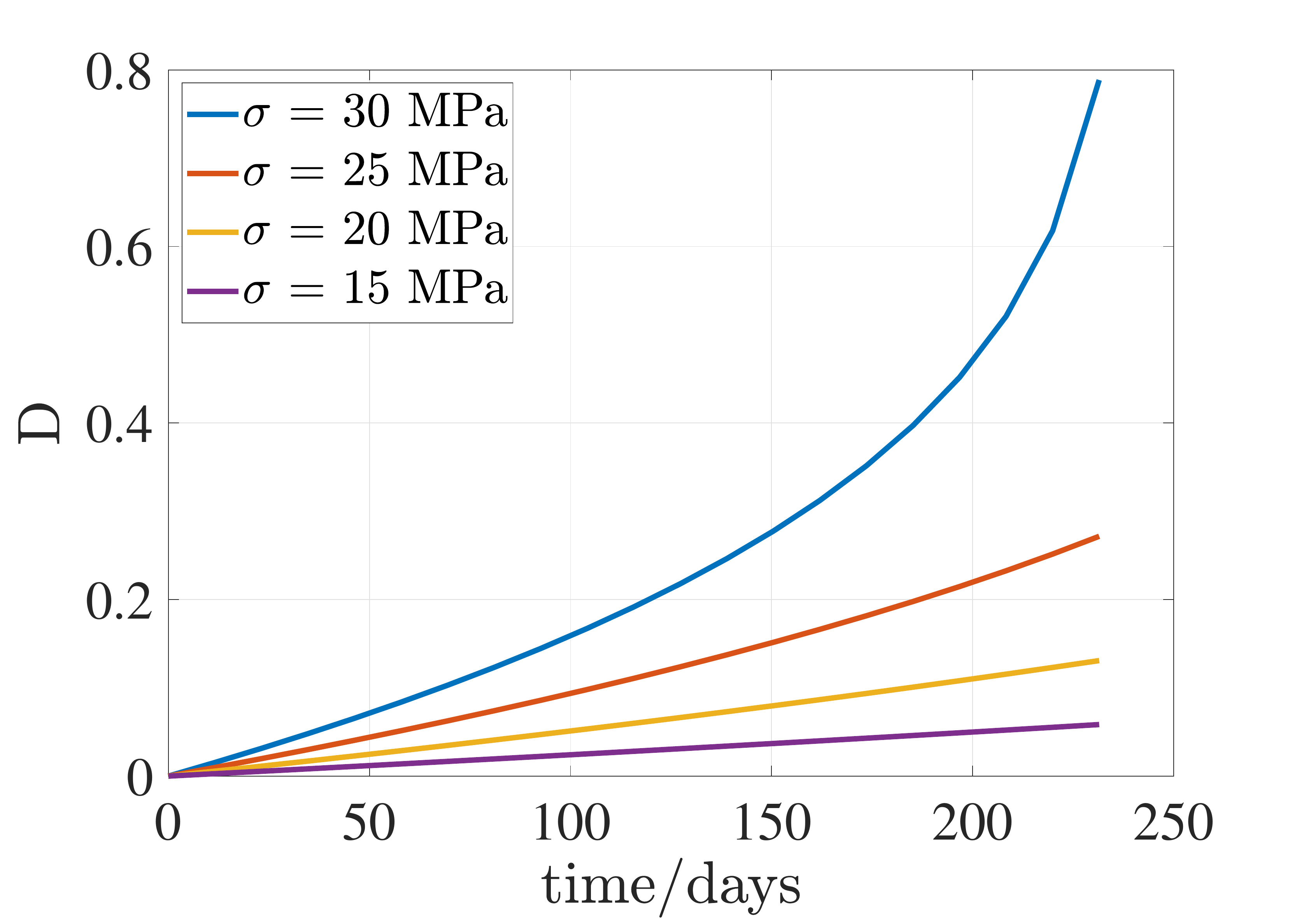}}
	\caption{Sensitivity analysis of damage constitutive law. The base test case is $r = 2.5, B = 4e4, \sigma = 30$ MPa. Accordingly the other dependent parameters are varied as shown in the legends.  }
	\label{fig_sensdamage}
\end{figure*}

In this test case, the evolution of damage within the material is studied for a given parameter set. Fig. \ref{fig_creepgraphs} show the transition from secondary creep to tertiary state and subsequent material failure. These graphs are plotted for a node inside the domain with a maximum strain, where the material failure would begin. In a homogeneous geological domain with rock salt, the location of that node is observed to be on the cavern's wall. This is because the cavern wall would undergo maximum stress in given time duration. From the results, it can be seen that as the creep forces increase over time, the strain rate becomes steeper, causing the magnitude of the deformation to increase drastically. Depending on the chosen damage parameters, the rate at which the material fails can be calibrated. \\
The damage constants used in the reference \cite{Ma2013damage} cannot be used in this formulation mainly due to different time scales and different rock salt properties. Fig. \autoref{fig_sensB} shows for different $B$ (constant parameter in the formulation) the damage parameter increases significantly. No literature is available to compare the numerical results with damage parameters incorporated with the timeline of 250 days. Hence the damage parameters had to be assumed by considering the same physics involved as explained in the \autoref{damagesection}. \\
Damage continuum analysis using Kachanov law involves three material constants $B,r,\sigma$. The variation of damage parameter $D$ with time for different $B,r,\sigma$ is shown in  \autoref{fig_sensdamage}. Fig. \autoref{fig_sensB} shows for different $B$ (constant parameter in the formulation) the damage parameter increases significantly. Lower the value of B higher the D. In \cite{Ma2013damage} the parameter B is lower by 100 in magnitude due to the time scale less than ten days. So from this analysis, it can be said that the parameter B depends on the timescale of the simulation significantly. Figure \autoref{fig_sensr} shows the variation of damage parameter for different $r$. Lower the value of $r$ higher is the damage parameter. Higher the stress higher is the expected damage can be seen in Figure \autoref{fig_senssigma}.

\iffalse 
\begin{figure*}
	\begin{subfigure}{.32\textwidth}
		\centering
		\includegraphics[width=1\linewidth]{images/damage/displacement_x_vs_disp.pdf}
		% \caption{$u_x$}
		\label{fig_damage_001}
	\end{subfigure}
	\begin{subfigure}{.32\textwidth}
		\centering
		\includegraphics[width=1\linewidth]{images/damage/strain_x_vs_disp.pdf}
		% \caption{$\varepsilon_{xx}$}
		\label{fig_damage_002}
	\end{subfigure}
	\begin{subfigure}{0.32\textwidth}
		\centering
		\includegraphics[width=1\linewidth]{images/damage/creep_forces_x_vs_disp.pdf}
		% \caption{Creep forces.}
		\label{fig_damage_003}
	\end{subfigure}
	\centering
	\caption{Test Case 6: variation of displacement in x direction, $\varepsilon_{xx}$ and x component of creep forces over time at the roof of the cavern.}
	\label{fig_creepgraphsa}
\end{figure*}
\fi 
\subsection{Multi-cavern simulations}

\begin{figure*}[h]
	\centering   		
	\subfigure[]{
		\label{fig_multi_001}
		\includegraphics[width=0.21\linewidth]{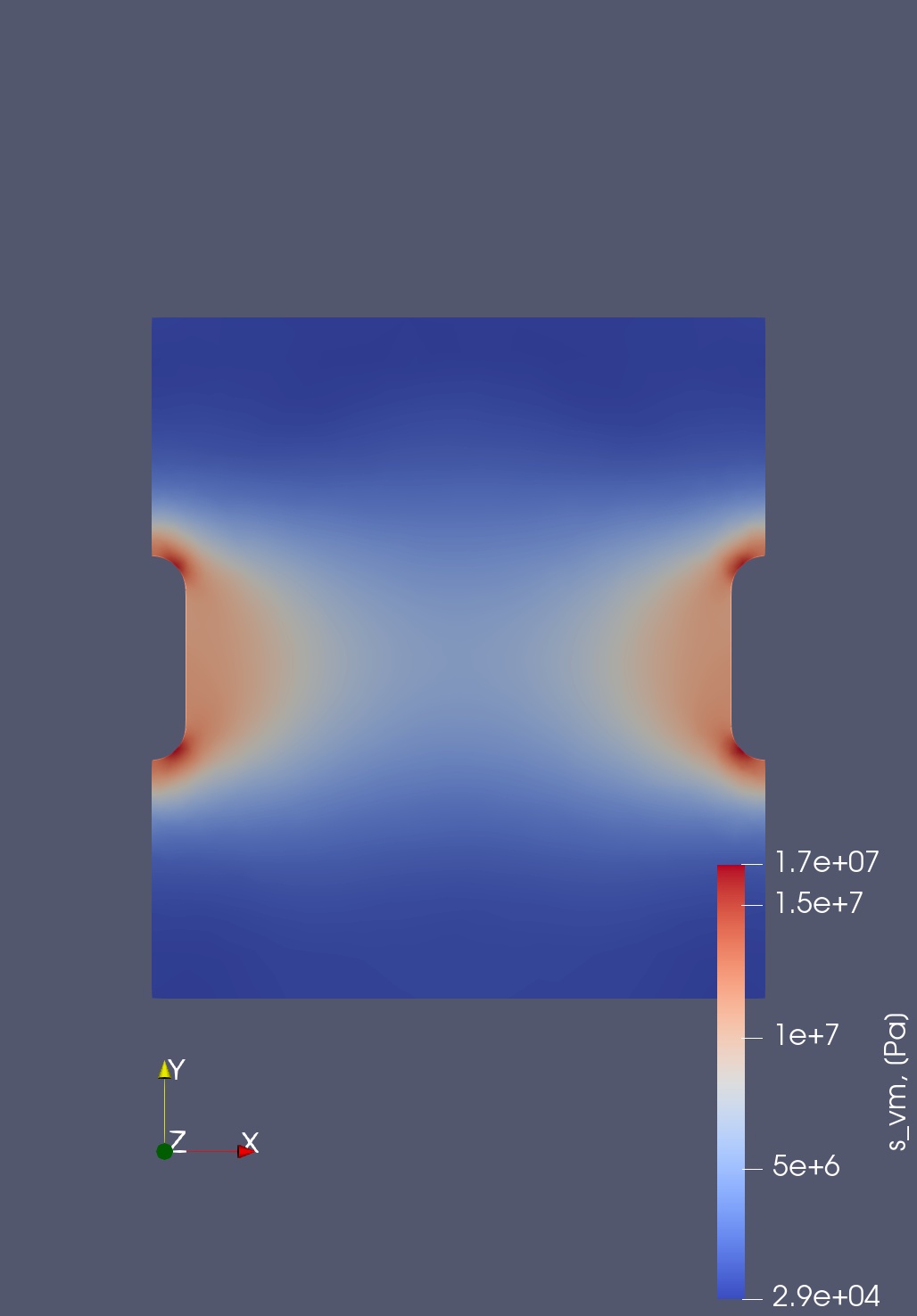}}
	\subfigure[]{
		\label{fig_multi_002}
		\includegraphics[width=0.21\linewidth]{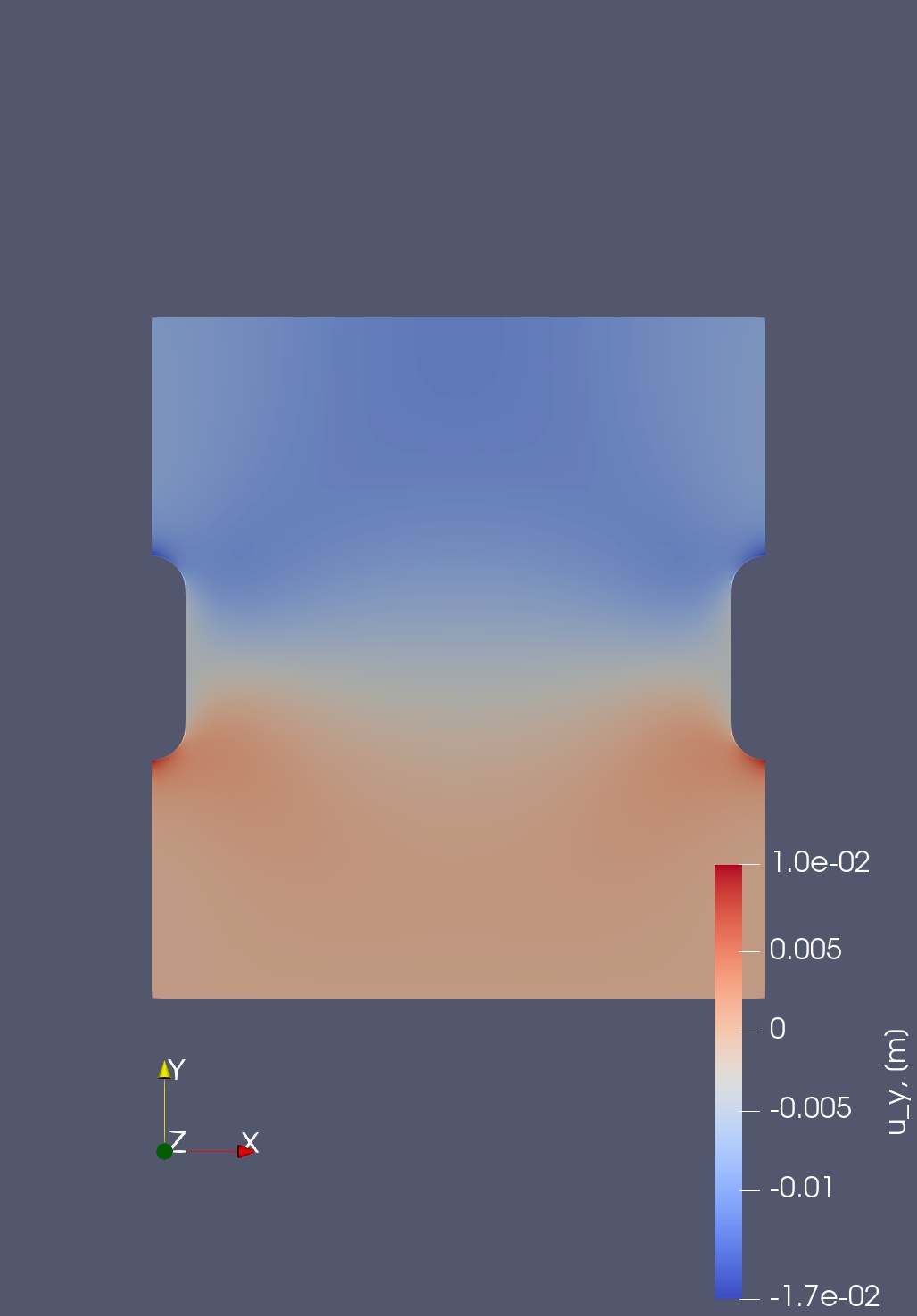}}
	\subfigure[]{
		\label{fig_multi_003}
		\includegraphics[width=0.21\linewidth]{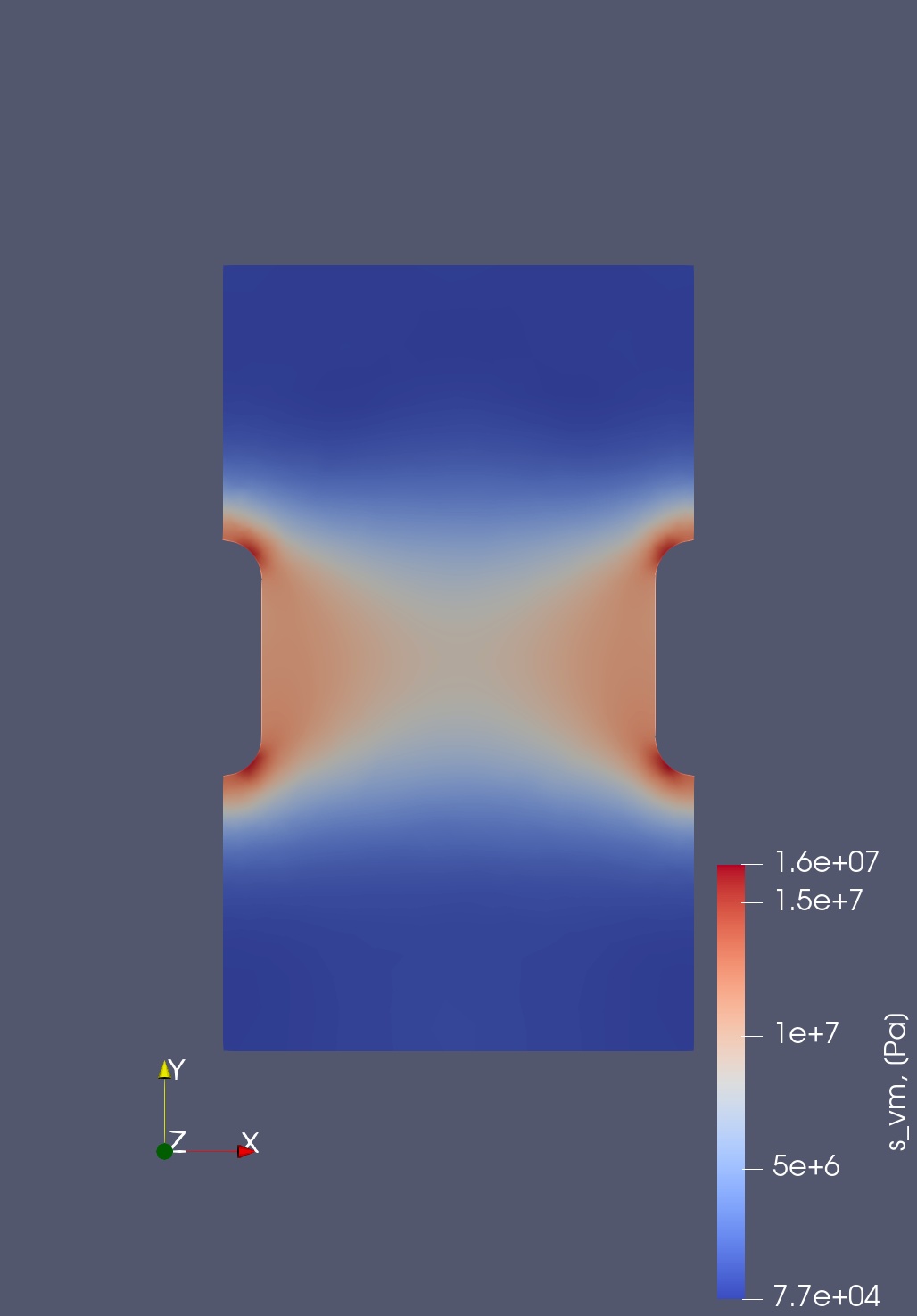}}
	\subfigure[]{
		\label{fig_multi_004}
		\includegraphics[width=0.21\linewidth]{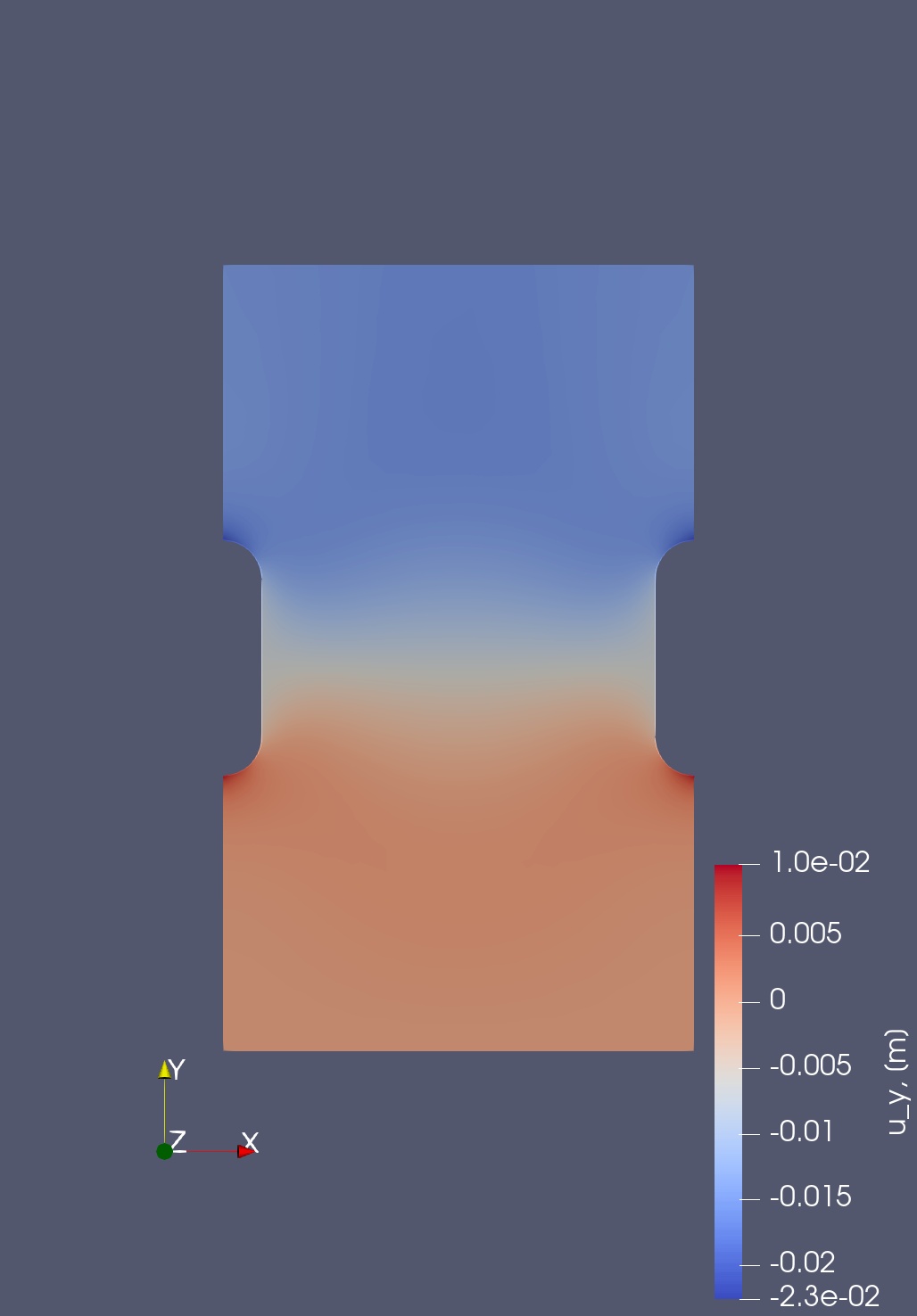}}
	\caption{\textbf{Multi-cavern simulation}: The above contours show the vertical displacement and von Mises stress for regular cylindrical shaped cavern after 275 days. Fig. \ref{fig_multi_001} and Fig. \ref{fig_multi_002} show the von Mises stress and vertical deformation for $CTC = 320 $m respectively. Fig. \ref{fig_multi_003} and Fig. \ref{fig_multi_004} show the von Mises stress and vertical deformation for $CTC = 200 $m respectively.
	}
	\label{fig_multiregucreepgraphs}
\end{figure*}

\begin{figure*}[h]
	\centering   		
	\subfigure[]{
		\label{fig_multiirreg_001}
		\includegraphics[width=0.21\linewidth]{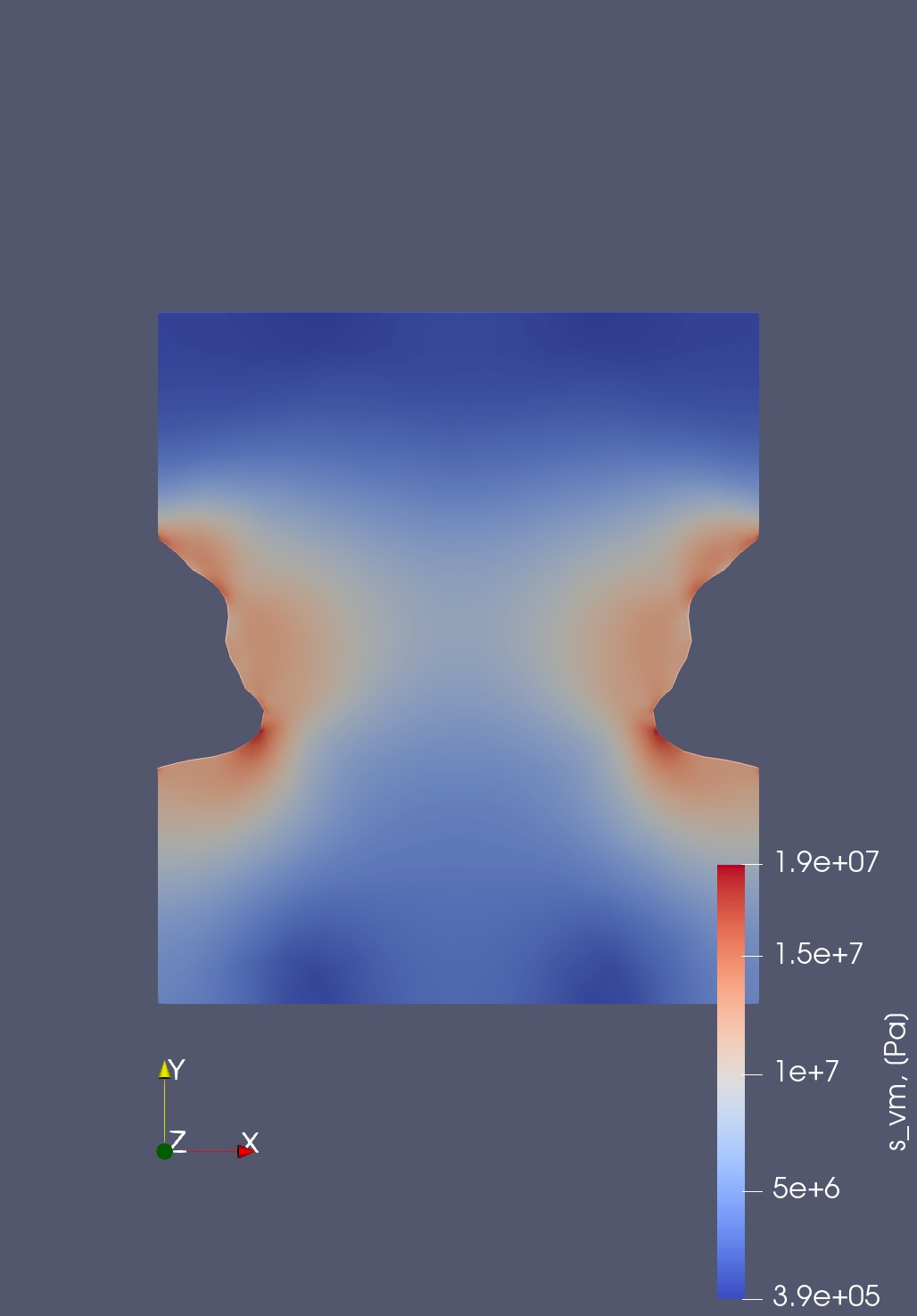}}
	\subfigure[]{
		\label{fig_multiirreg_002}
		\includegraphics[width=0.21\linewidth]{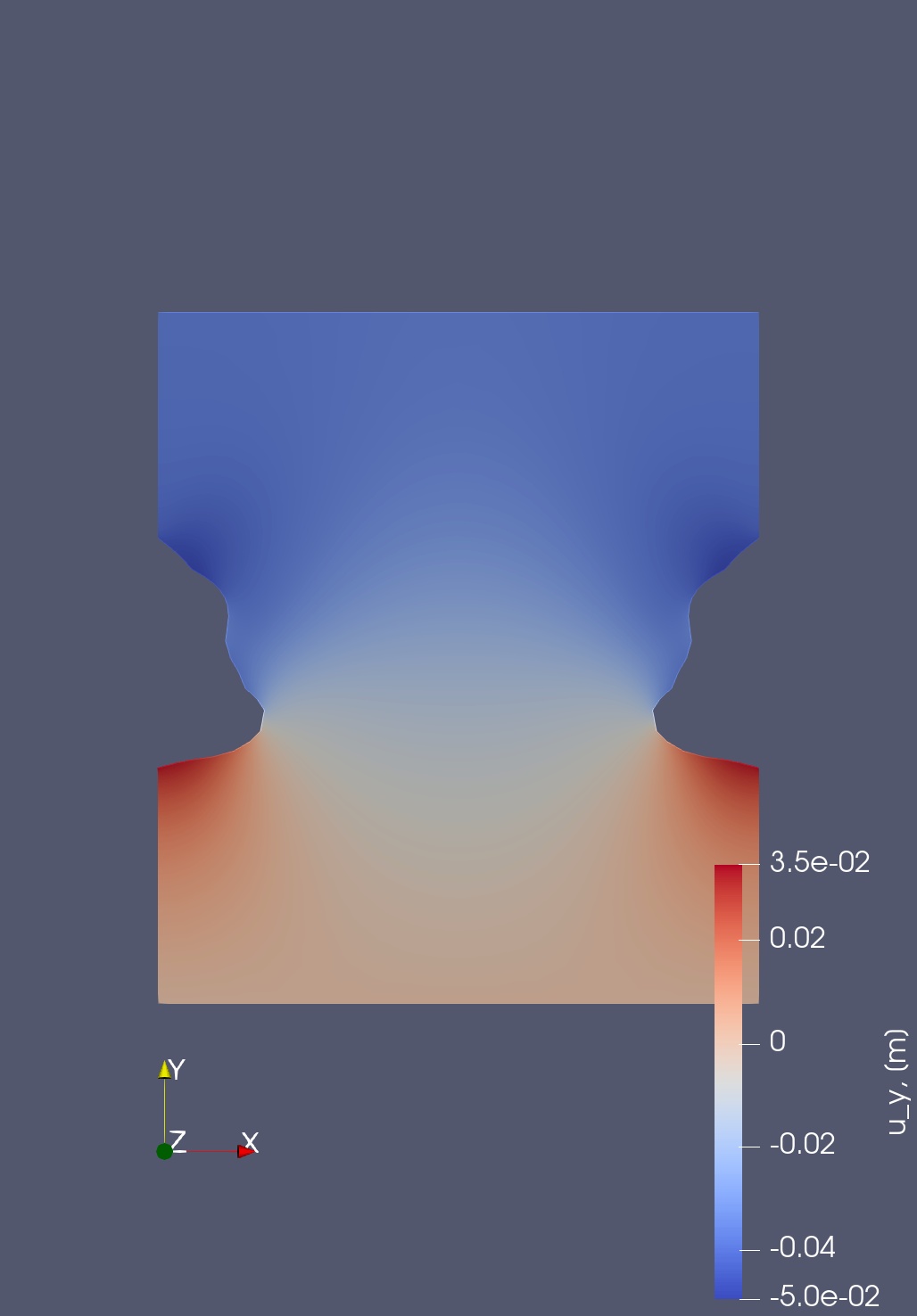}}
	\subfigure[]{
		\label{fig_multiirreg_003}
		\includegraphics[width=0.21\linewidth]{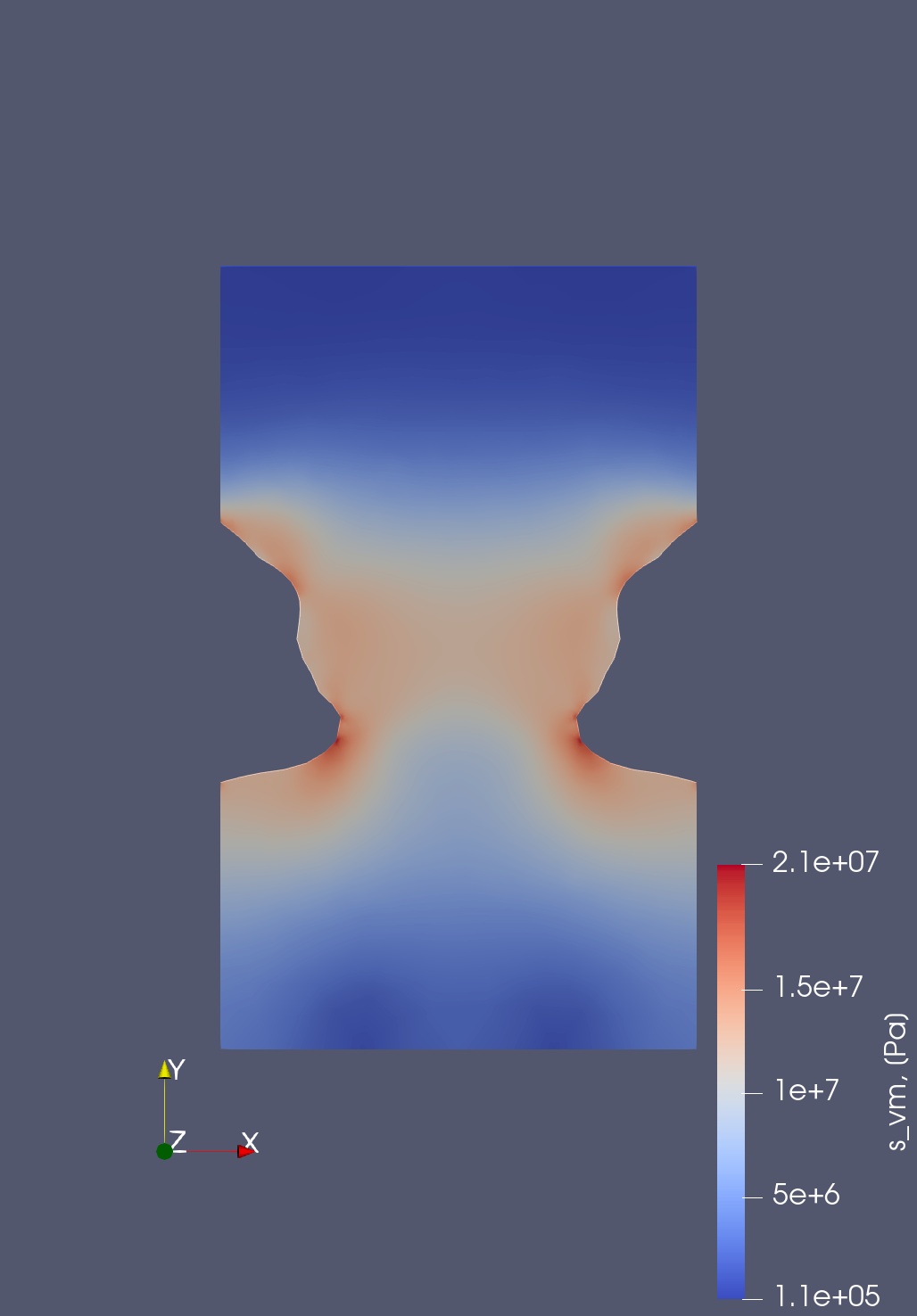}}
	\subfigure[]{
		\label{fig_multiirreg_004}
		\includegraphics[width=0.21\linewidth]{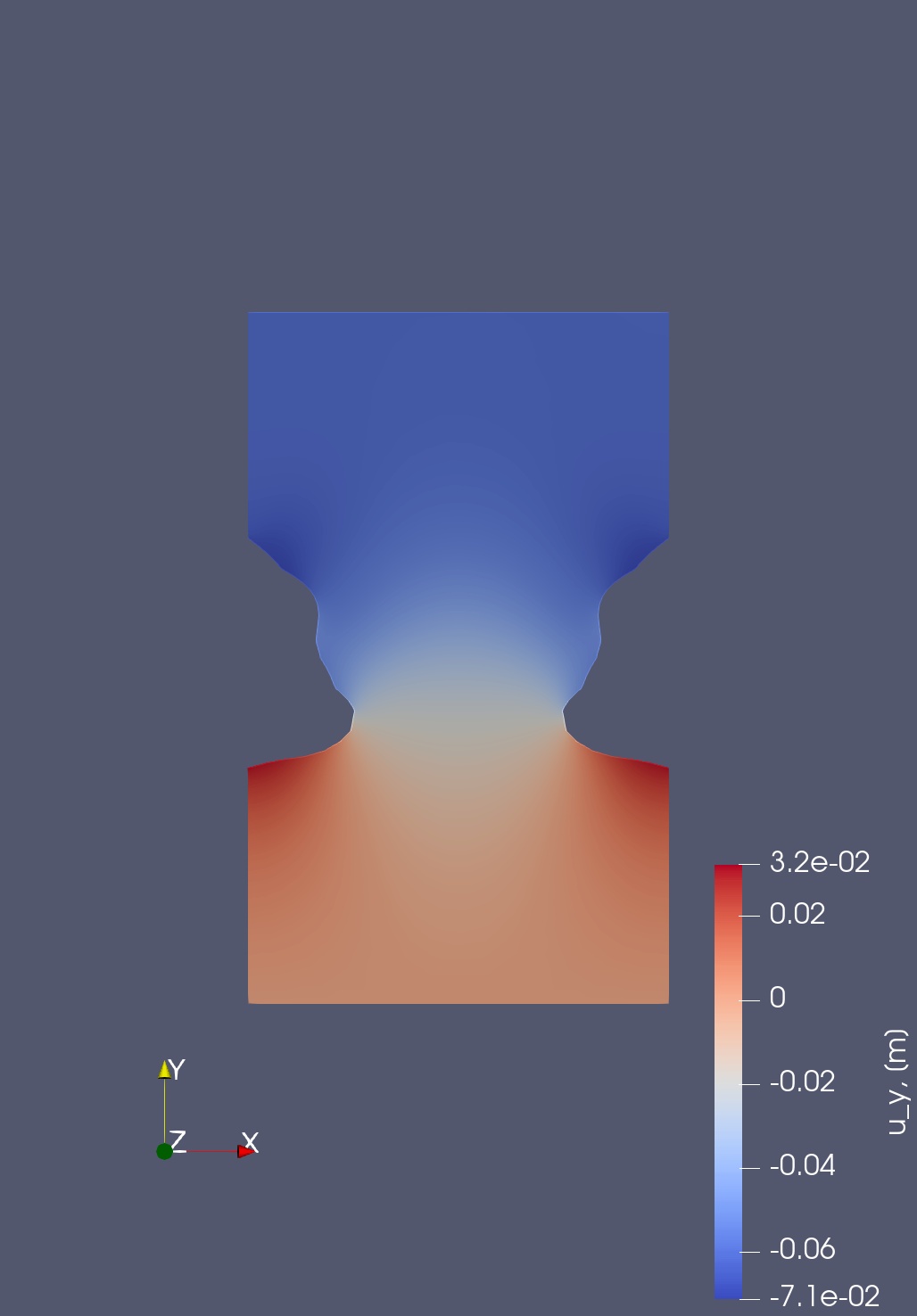}}
	\caption{\textbf{Multi-cavern simulation}: Illustration of the vertical displacement and von Mises stress for irregular shaped cavern after 275 days. Fig. \ref{fig_multiirreg_001} and Fig. \ref{fig_multiirreg_002} show the von Mises stress and vertical deformation for $CTC = 200 $m respectively. Fig. \ref{fig_multiirreg_003} and Fig. \ref{fig_multiirreg_004} show the von Mises stress and vertical deformation for $CTC = 140 $m, respectively.}
	\label{fig_multiirregcreepgraphs}
\end{figure*}

\begin{figure*}[h]
	\centering   	
	\subfigure[]{
		\label{ctcschematic}
		\includegraphics[width=0.2\linewidth]{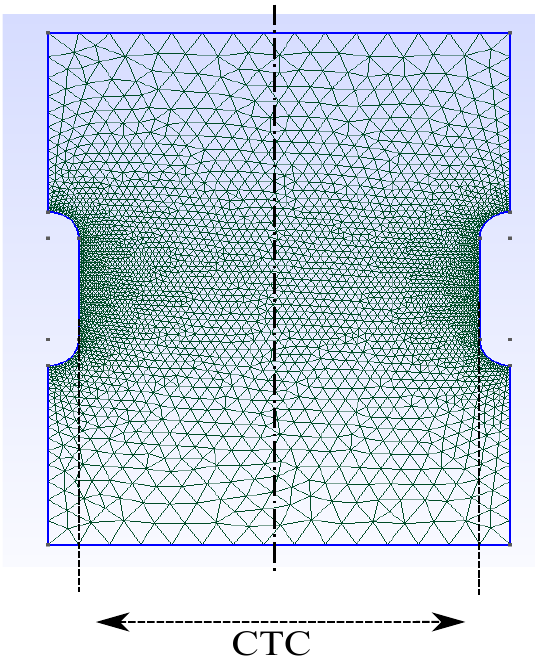}}	
	\subfigure[]{
		\label{svmdistanceregular}
		\includegraphics[scale = 0.17]{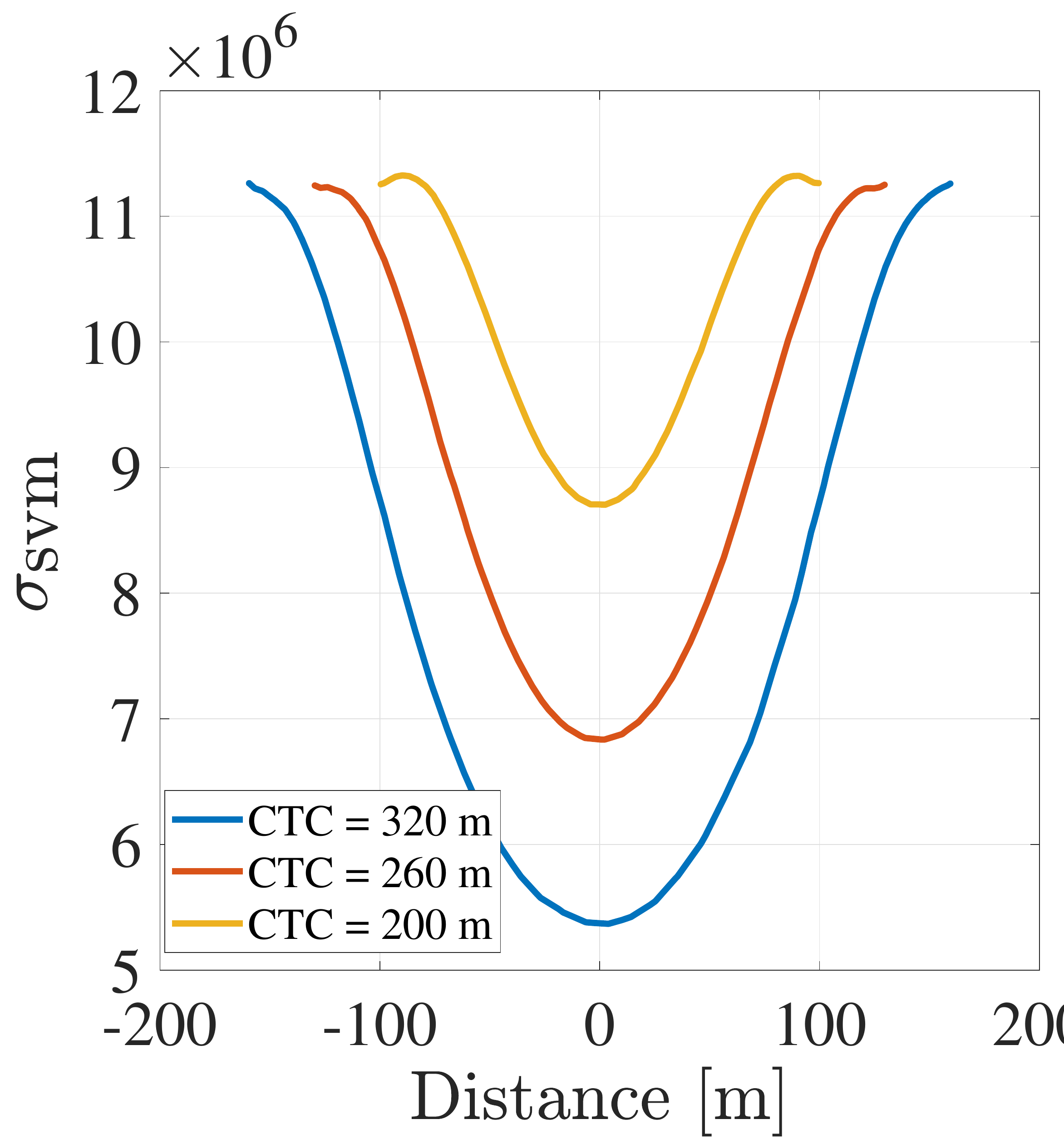}}
	\subfigure[]{
		\label{svmdistanceirregular}
		\includegraphics[scale = 0.18]{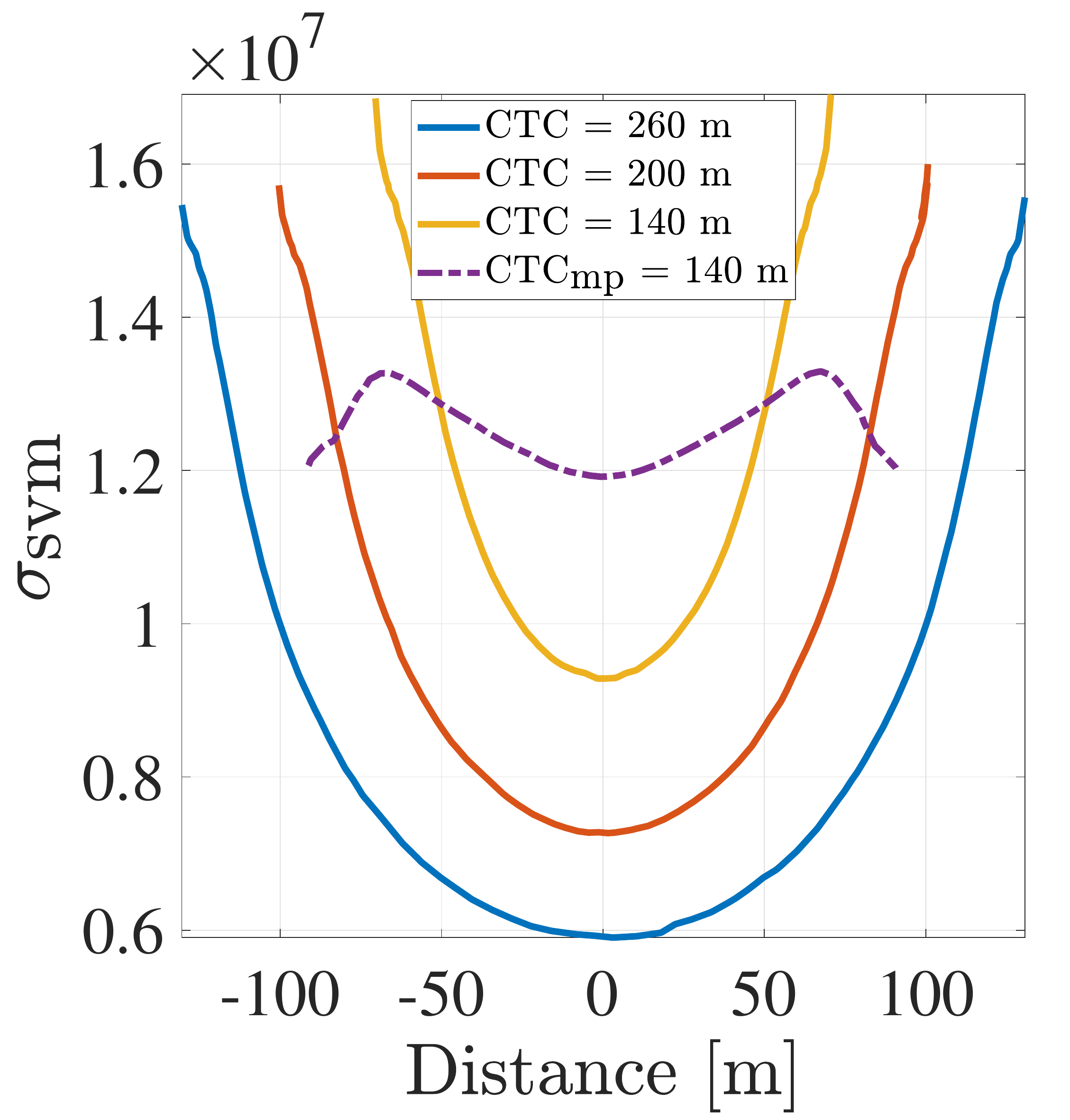}}
	\caption{\textbf{Multi-cavern simulation}: Fig. \ref{ctcschematic} shows the example of computational mesh showing cavern to cavern distance (CTC) for regular shaped cavern. Fig. \ref{svmdistanceregular} shows the variation of von Mises stress with stress for regular shaped caverns at different CTC. Fig. \ref{svmdistanceirregular} shows the variation of von Mises stress with stress for irregular shaped caverns at different CTC. 
	}
	\label{fig_multiplots}
\end{figure*}

Multi-cavern simulations are helpful when more than one cavern can be used close to each other in the geological domain for subsurface energy storage.  Cavern to cavern distance (CTC) between the caverns is critical to understand the influence of each cavern on the other. In this study, multi-cavern simulation studies were conducted for regular and irregular-shaped caverns. CTC is the minimum cavern to cavern distance possible in the geological domain. CTC for regular cylindrical-shaped cavern is shown in Figure \autoref{ctcschematic}. Roller boundary conditions were imposed along the caverns and traction-free boundary conditions on the top plane. \\
Multi cavern simulations for regular-shaped caverns showing von Mises stress and deformation ($u_y$) are shown in \autoref{fig_multiregucreepgraphs}. \autoref{fig_multiirregcreepgraphs} shows the same for irregular-shaped caverns. It can be seen that as the distance between the caverns reduces, causing an increase in Von-mises stress. These simulations can be used when there is heterogeneity in the geological domain which can handle less stress. Accordingly, Drucker Prager or Von Mises failure criteria could be used to determine how critical the distance is between the caverns \cite{Li2020}.  \\
Vertical deformation ($u_y$) is compared here since traction-free boundary conditions are applied on the top face. It can be seen that the closer the caverns it higher the vertical deformation is observed. When the caverns have complex geometry, the magnitude of the Von- Mises stress $s_{vm}$ and vertical deformation is higher. This could lead to subsidence \cite{Marketos2016} in the geological domain or could amplify seismicity. \\
\autoref{fig_multiplots} shows the variation of von Mises stress along the horizontal distance in the mid-plane section for regularly shaped caverns and the minimum distance (CTC plane) for complex shaped caverns. $CTC_{mp}$ test case shows the variation of von Mises stress at the mid-plane section of the irregular-shaped cavern. Here from Figure \autoref{svmdistanceregular}  it can be seen that the von Mises stress distribution is more uniform similar to the Gaussian distribution. \textcolor{black}{As CTC reduces, the minimum stress in the center of the homogeneous geological domain increases. Regularly-shaped caverns have minimum surface area due to their smooth surfaces. Hence, there is no higher stress observed near the cavern walls. The stress reduces slowly from peak stress closer to the cavern and drops to a minimum value near the central vertical plane of the geological domain. }  \\
From Fig. \autoref{svmdistanceirregular}, it can be seen that the stress distribution at the center is more spread out, and closer to the cavern, it peaks out. This is due to the irregular shape. From the curve representing $CTC_{mp}$ it can be seen that the stress increases at first and then reduces again. This is because this section is not the closest distance with the adjacent cavern and not the surface with the most curvature. 

\begin{figure*}[h]
	\centering   		
	\subfigure[]{
		\label{senseAB}
		\includegraphics[scale=0.34]{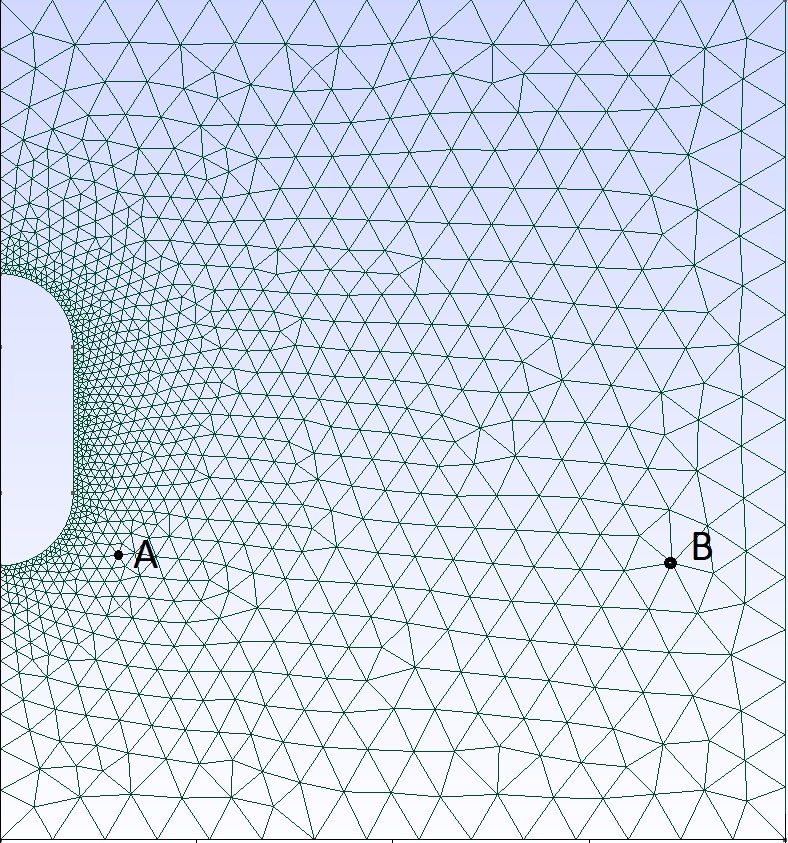}}
	\subfigure[]{
		\label{senseE}
		\includegraphics[width=0.3451\linewidth]{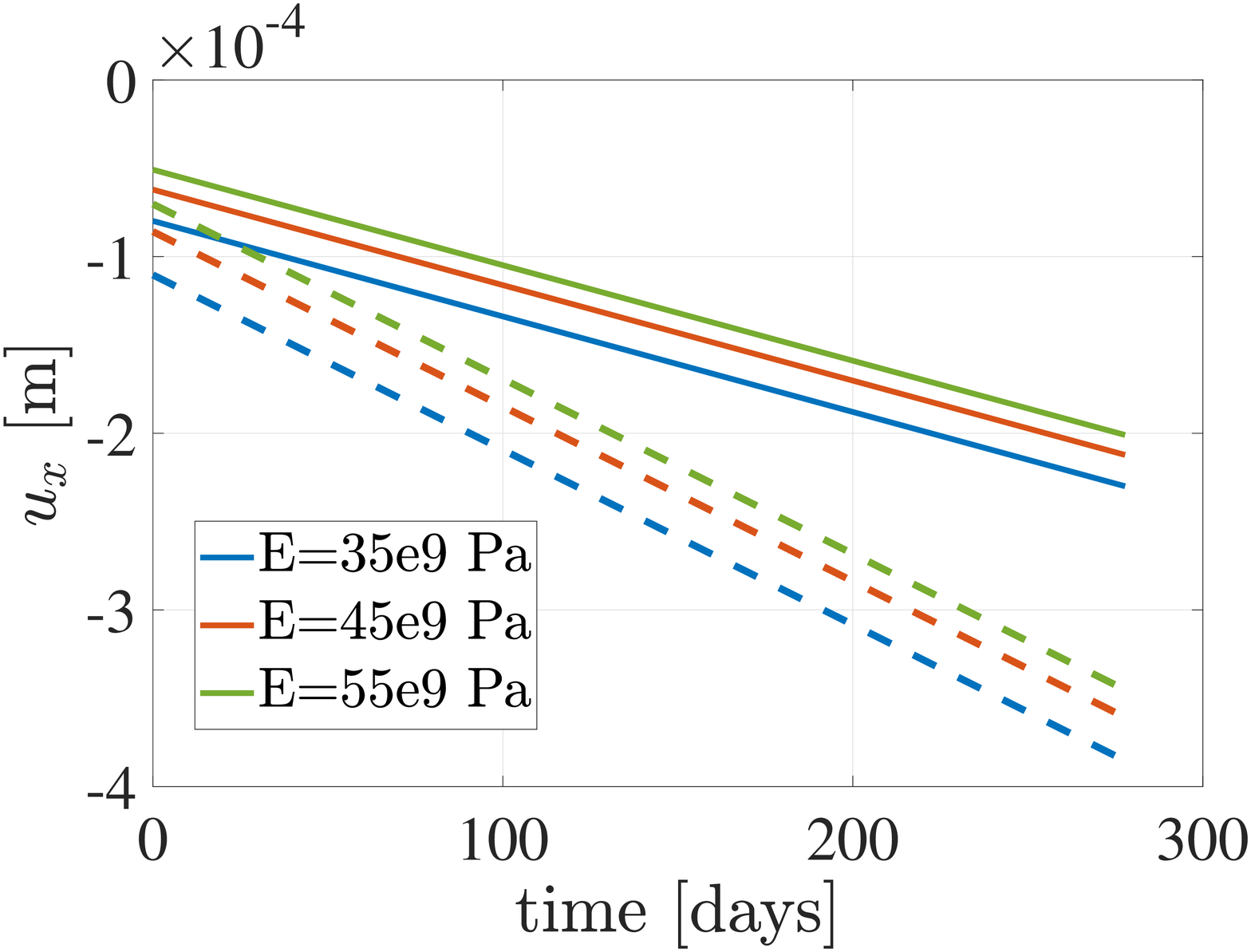}}
	\subfigure[]{
		\label{sensea}
		\includegraphics[width=0.3451\linewidth]{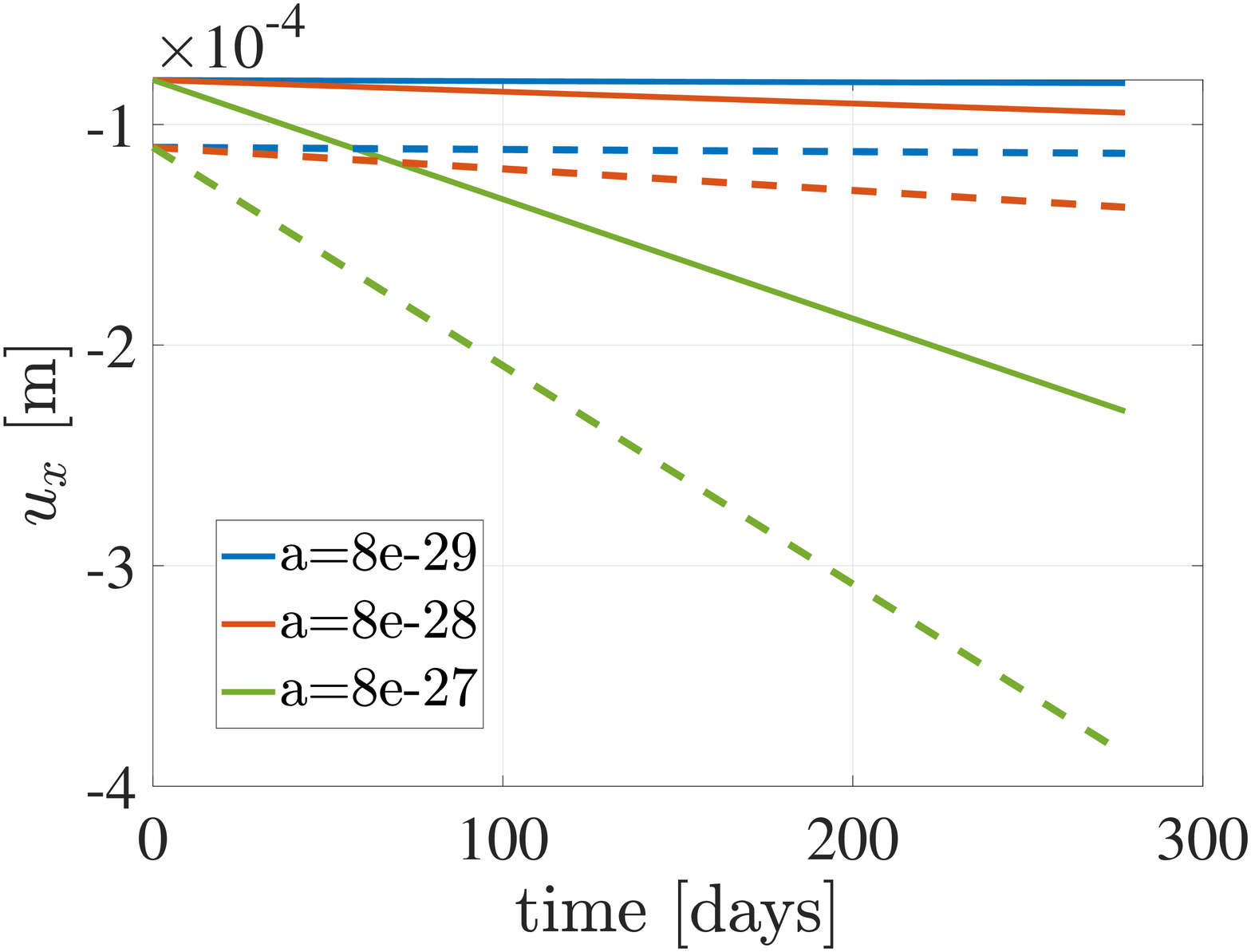}}
	\subfigure[]{
		\label{sensetemp}
		\includegraphics[width=0.32451\linewidth]{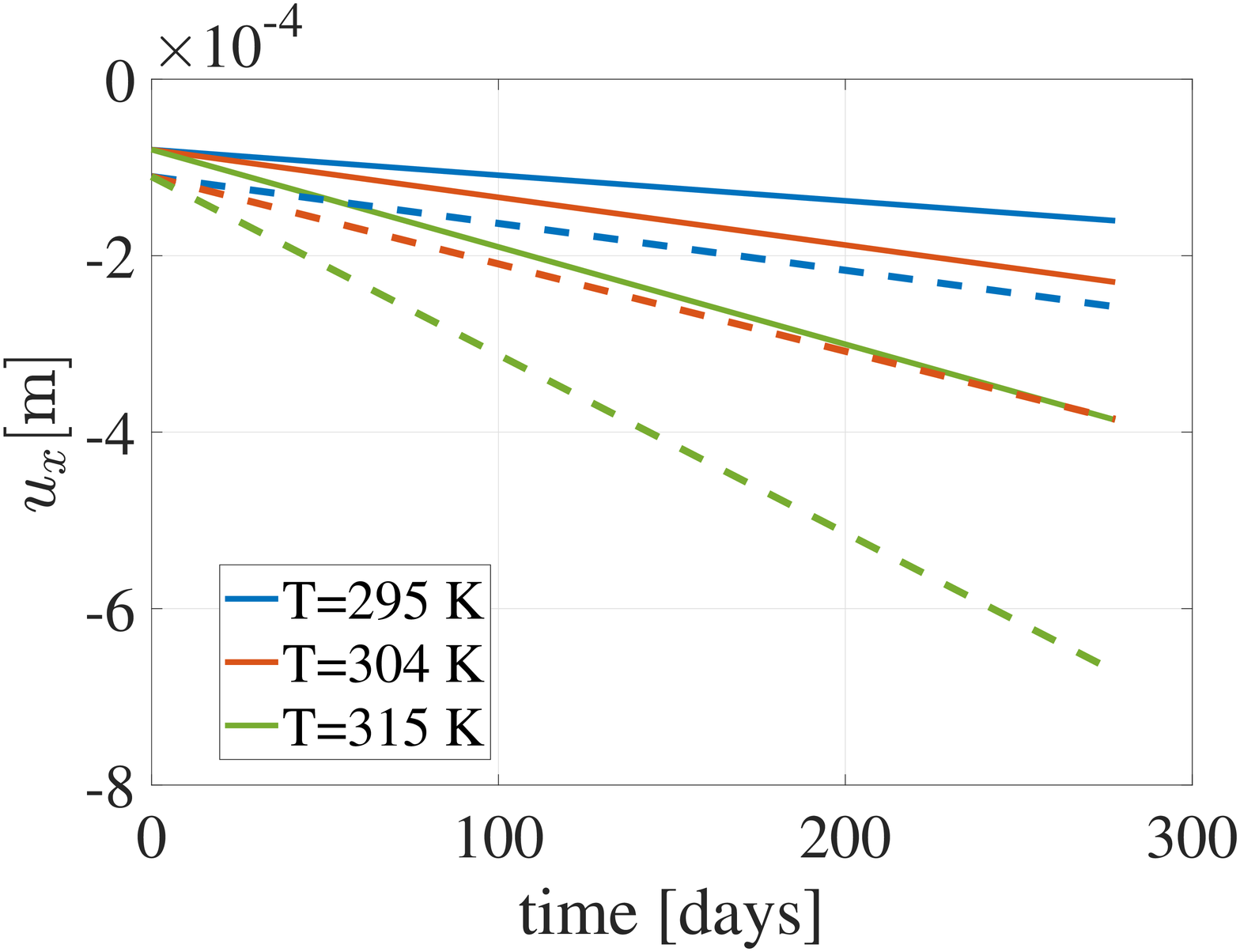}}
	\subfigure[]{
		\label{senseABn}
		\includegraphics[width=0.32451\linewidth]{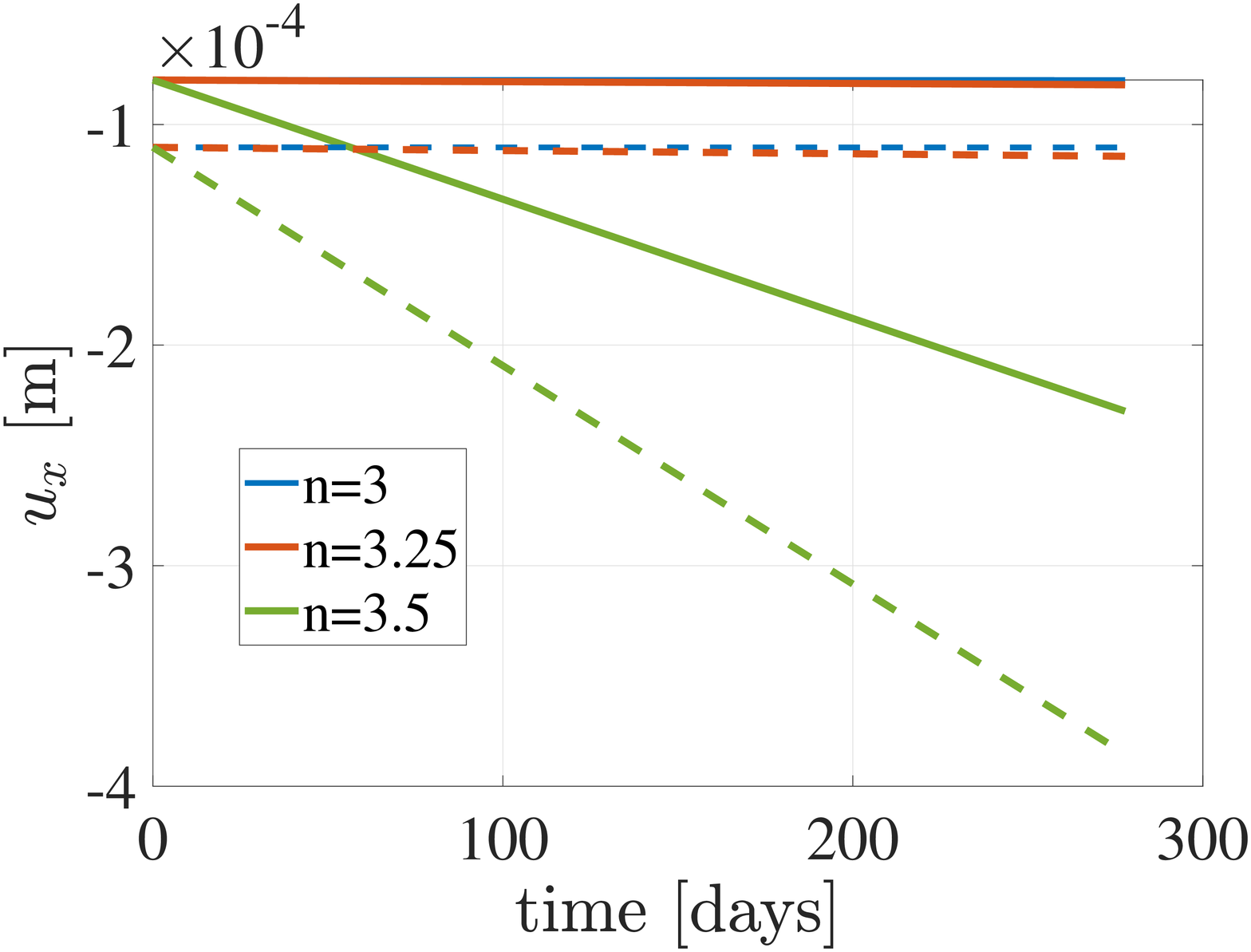}}
	\subfigure[]{
		\label{senseABdepth}
		\includegraphics[width=0.32451\linewidth]{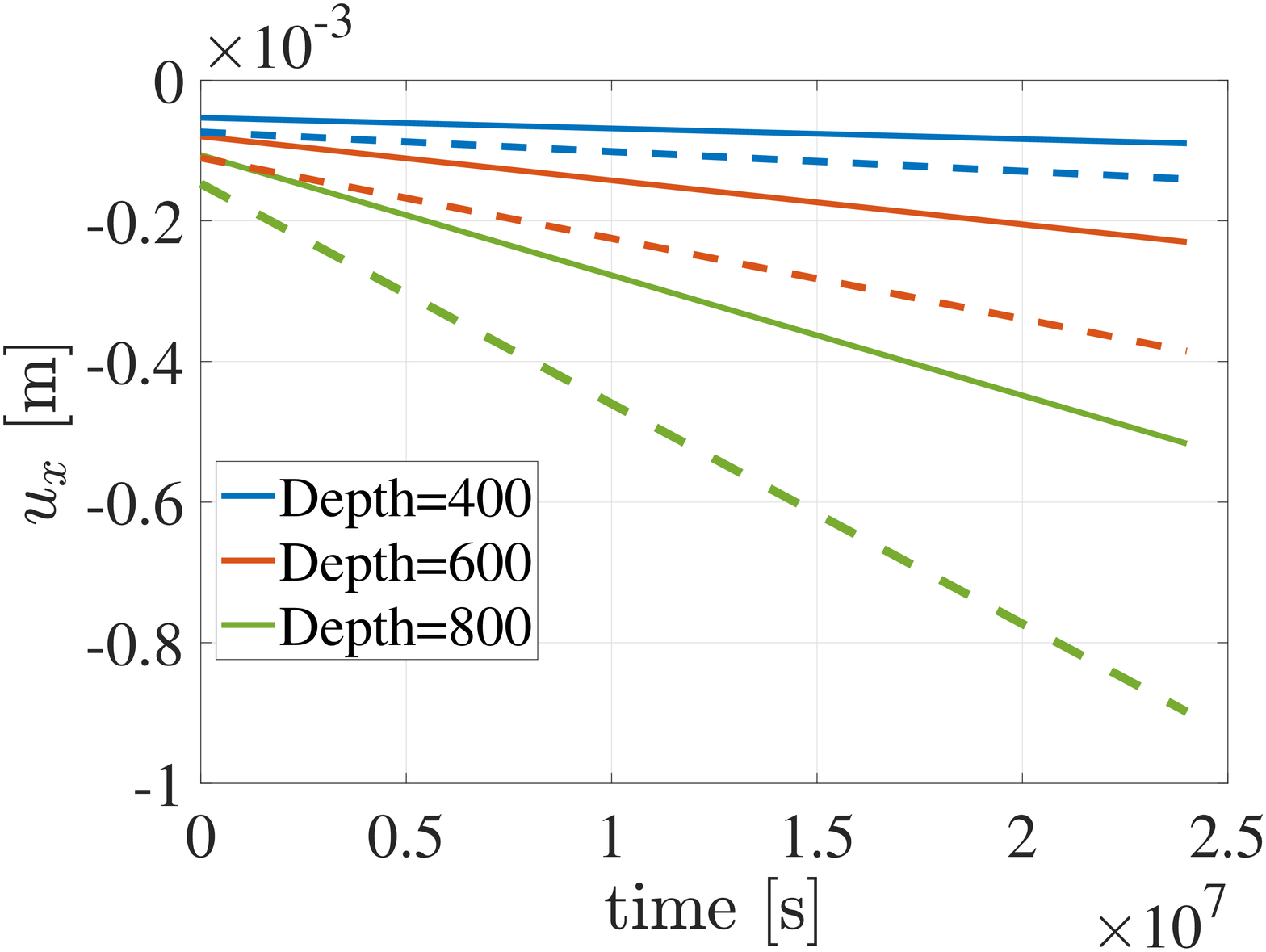}}
	\caption{The above plots show the variation of horizontal deformation at two different points A and B with time for different Young's modulus ($E$),  creep constant ($a$), temperature ($T$), creep exponent $n$ and depth of the cavern respectively. The dotted lines (- - -) show the variation for the node closer to the cavern (Point A) and the full lines (---) show the variation of deformation (Point B) in the far field of the cavern.}
	\label{fig_sensitivityfull}
\end{figure*} 

\subsection{Sensitivity analysis}
\label{Sensitivityanalysis}
Creep is a complex phenomenon that involves various parameters that can influence the deformation and stresses over time. To quantify the impact of the critical parameters, sensitivity analysis needs to be conducted. The parameters chosen are cavern depth, creep constant,  \textcolor{black}{ Young's modulus of halite rock}, temperature, and creep exponent. Energy storage technology could involve different operating conditions and heterogeneous properties of rock salt. Due to this, the above parameters are chosen to study their influence on the time-dependent deformation. Fig. \autoref{senseAB} shows the two points (A and B) in the domain where sensitivity analysis is conducted. The influence of Young's modulus on horizontal deformation is shown in Fig. \autoref{senseE}. Change in Young's modulus changes the elastic deformation. \textcolor{black}{Higher the Young's modulus, lower is the deformation}. The salt cavern's far-field shows a lower slope than the node closer to the cavern because of higher stresses closer to the cavern. The sensitivity of deformation towards shear modulus is not shown here since it can be expected that the deformation is not sensitive towards it. Fig. \ref{sensea} shows the variation of horizontal deformation with different creep multipliers. Higher the creep multiplier, the higher the rate of change in deformation with time, leading to a higher accumulation of plastic deformation. The increase in creep constant is non-linearly increasing the deformation after a certain period.      \\
The influence of temperature on deformation is seen in Fig. \ref{sensetemp}. The higher the temperature, the higher is the accumulated creep deformation. Again, a similar trend is observed as the last plot where the temperature change is not linearly changing the change in deformation. The impact of creep exponent $n$ is shown in Fig. \ref{senseABn}. The higher the creep exponent, the higher is the deformation. Compared to creep constant and temperature, creep exponent is very sensitive to the accumulated deformation. For $n=3$ and $n=3.25$ the variation of deformation with time is almost horizontal, however for creep exponent, $n=3.5$, the rate of change in deformation is very high. Comparing the sensitivity plots for the creep parameters, it can be seen that the influence of creep exponent is biggest, followed by creep constant, and lastly, it is creep temperature.  Fig \ref{senseABdepth} shows the variation of horizontal deformation with time for different depths of the cavern. In the geological domain with a lot of heterogeneity and interlayers, it is important to construct the cavern in the domain with minimum cracks and heterogeneity. Depth of the cavern is a parameter that can allow this argument. The higher the depth, the higher is the lithostatic pressure leading to higher creep deformation.

\iffalse 
\begin{figure*}
	\centering   		
	\subfigure[]{
		\label{2asense}
		\includegraphics[height=0.32\textwidth]{images/point_A/domain_A2.png}}
	\subfigure[]{
		\label{2csense}
		\includegraphics[height=0.32\textwidth]{images/sens/plot1.pdf}}
	\caption{Sensitivity analysis: Two points  A and B are considered in the domain to understand how the horizontal deformation varies with time Fig. \ref{2asense}. Fig. \ref{2csense} shows the variation of deformation with time for points A and B.}
	\label{fig_sensitivity}
\end{figure*} 
\fi

\iffalse 
\begin{figure*}
	\centering   		
	\subfigure[]{
		\label{fig_depth_sens}
		\includegraphics[width=0.31\linewidth]{images/sens/plot2.pdf}}
	\subfigure[]{
		\label{fig_temp_sens}
		\includegraphics[width=0.31\linewidth]{images/sens/plot5.pdf}}
	\subfigure[]{
		\label{fig_bulk_sens}
		\includegraphics[width=0.31\linewidth]{images/sens/plot4.pdf}}
	\subfigure[]{
		\label{fig_shear_sens}
		\includegraphics[width=0.31\linewidth]{images/sens/plot6.pdf}}
	\subfigure[]{
		\label{fig_hetero_sens}
		\includegraphics[width=0.31\linewidth]{images/sens/plot3.pdf}}
	\caption{Sensitivity analyses for different parameters, as indicated in their captions. In Fig. \ref{fig_hetero_sens}, A1 and B1 belong to the homogeneous domain, while A2 and B2 to the heterogeneous domain.}
	\label{fig_sens00111}
\end{figure*} 
\fi 

\section{Conclusion}

In this work, the influence of complex shapes and material heterogeneity in the geological domain on salt caverns employed for energy storage technology is studied using a 2D finite element solver. The secondary (i.e., steady-state) creep behavior is introduced in the mathematical model based on power law, and the power-law parameters were taken from the work of Carter et al. \cite{Carter1993}. The numerical methodology and the chosen creep constitutive constants are compared with the experimental data. Tertiary creep is introduced by utilizing the damage evolution parameter, allowing to predicting the material failure \cite{KonstantinNaumenko2007}. The developed simulator allowed for both explicit and implicit time integration schemes. After the consistency check and benchmarking with the experimental data, Various conclusions can be derived from this work. \\
1) It is evident from the above results that creep is a slow phenomenon of insignificant magnitude compared to the elastic component for year-long operations. This justifies the present strategy of omitting the low-stress creep mechanism of pressure solution from the current analysis. However, on a scale of several years with a significant amount of heterogeneity, the effect of creep strains on the deformation and stresses can become evident, which would eventually cause the material to reach the tertiary stage leading to failure at zones of critical stress intensity. Less than five \% volume changes in the cavern were observed considering cylindrical homogeneous caverns after 275 days.\\
2) The role of heterogeneity by considering only different elastic parameters is showed for irregular-shaped caverns. Depending on the type of impurity (potash, halite), the location of impurity in the domain, and the type of distribution imposed in the geological model, stress and deformation distributions vary.  \\
3) When heterogeneous interlayers are incorporated in the simulation model with the right constitutive models for the composition, a much higher increase in deformation is observed locally and around the cavern than the homogeneous test case and the heterogeneous test case without considering creep properties (test case 5). In general, bischofite showed higher creep deformations compared to carnallite. When the interlayers are located near the high curvature region, that region has the highest potential to be a failure zone.  \\
4) The influence of curvature of the caverns was studied in this work for both with and without heterogeneity. Interlayers near the high curvature showed at least five times more deformation than interlayers located near the minimum curvature. Potential failure zones by computing permeability were also identified to be in the regions around high curvature. 
5) Tertiary creep was also incorporated in this model with the assumption creep rate depends on the damage state and stress of the system. The underlying physics could be included however the constants involved in the damage law had to be assumed due to unavailable literature for more extended time scales.    \\
6) A detailed sensitivity analysis of various parameters shows their influence on the deformation, which can gauge the deformation in real field test cases.\\
7)  The developed open-source simulator was extensively tested on various test cases, as shown in the paper. The developed model is publicly available as an open-source simulator in the TU Delft repository of the ADMIRE project located at \\ https://gitlab.tudelft.nl/ADMIRE\_Public/Salt\_Cavern. The developed code allows the user to use different features simultaneously to understand the underlying field-relevant physics quickly.  \\
8) Multi-cavern simulations were critical to understanding the significance of CTC, irregular shape of the caverns on the stress distributions in the geological domain.\\  

%It is evident from the above results that creep is a slow phenomenon of insignificant magnitude compared to the elastic component, for year-long operations. This justifies the present strategy of omitting the low stress creep mechanism of pressure solution from the present analysis. However, on a scale of several years, the effect of creep strains on the deformation and stresses can become evident, which would eventually cause the material to reach the tertiary stage leading to failure at zones of critical stress intensity. A detailed sensitivity analysis of various parameters shows their influence on the deformation which can be used to gauge the deformation in real field test cases. In future studies, the role of additional creep mechanisms such as pressure solution and uncertainties in dislocation creep parameters will be considered further, as will effects of hydrogen permeation on damage development in the cavern wall. 

The role of additional creep mechanisms such as pressure solution and uncertainties in dislocation creep parameters will be considered in the future, also the effects of gas (hydrogen or green methane) permeation on damage development in the cavern wall. Future work involves including thermal and (visco)plastic strains to allow for more reliable simulations and sensitivity analyses. Since the field test cases are large-scale of the order (km), efficient formulations such as multi-scale formulation would be developed to reduce the computational costs. Further research would be required to benchmark the tertiary creep model with lab and field data. Other heterogeneous interlayers such as shale rocks need to be studied considering the right creep constitutive laws and underlying physics are chosen.

\bibliography{bibliography}

\begin{thebibliography}{10}
\urlstyle{rm}
\expandafter\ifx\csname url\endcsname\relax
  \def\url#1{\texttt{#1}}\fi
\expandafter\ifx\csname urlprefix\endcsname\relax\def\urlprefix{URL }\fi
\expandafter\ifx\csname doiprefix\endcsname\relax\def\doiprefix{DOI: }\fi
\providecommand{\bibinfo}[2]{#2}
\providecommand{\eprint}[2][]{\url{#2}}

\bibitem{DONADEI2016113}
\bibinfo{author}{Donadei, S.} \& \bibinfo{author}{Schneider, G.-S.}
\newblock \bibinfo{title}{Chapter 6 - compressed air energy storage in
  underground formations}.
\newblock In \bibinfo{editor}{Letcher, T.~M.} (ed.)
  \emph{\bibinfo{booktitle}{Storing Energy}}, \bibinfo{pages}{113--133},
  \doiprefix\url{https://doi.org/10.1016/B978-0-12-803440-8.00006-3}
  (\bibinfo{publisher}{Elsevier}, \bibinfo{address}{Oxford},
  \bibinfo{year}{2016}).

\bibitem{Laban_2020}
\bibinfo{author}{Laban, M.}
\newblock \bibinfo{journal}{\bibinfo{title}{Hydrogen storage in salt caverns:
  Chemical modelling and analysis of large-scale hydrogen storage in
  underground salt caverns}}.
\newblock
  {\emph{\JournalTitle{http://resolver.tudelft.nl/uuid:d647e9a5-cb5c-47a4-b02f-10bc48398af4}}}
   (\bibinfo{year}{2020}).

\bibitem{Lord2014}
\bibinfo{author}{Lord, A.~S.}, \bibinfo{author}{Kobos, P.~H.} \&
  \bibinfo{author}{Borns, D.~J.}
\newblock \bibinfo{journal}{\bibinfo{title}{{Geologic storage of hydrogen:
  Scaling up to meet city transportation demands}}}.
\newblock {\emph{\JournalTitle{International Journal of Hydrogen Energy}}}
  \textbf{\bibinfo{volume}{39}}, \bibinfo{pages}{15570--15582},
  \doiprefix\url{10.1016/j.ijhydene.2014.07.121} (\bibinfo{year}{2014}).

\bibitem{Caglayan2020}
\bibinfo{author}{Caglayan, D.~G.} \emph{et~al.}
\newblock \bibinfo{journal}{\bibinfo{title}{{Technical potential of salt
  caverns for hydrogen storage in Europe}}}.
\newblock {\emph{\JournalTitle{International Journal of Hydrogen Energy}}}
  \textbf{\bibinfo{volume}{45}}, \bibinfo{pages}{6793--6805},
  \doiprefix\url{10.1016/j.ijhydene.2019.12.161} (\bibinfo{year}{2020}).

\bibitem{tno_2012}
\bibinfo{author}{TNO}.
\newblock \bibinfo{title}{Informatiebladen zoutwinning}.
\newblock \bibinfo{type}{Tech. Rep.}, \bibinfo{institution}{Innovation for
  Life} (\bibinfo{year}{2012}).

\bibitem{Jackson1986}
\bibinfo{author}{Jackson, M.} \& \bibinfo{author}{Talbot, C.}
\newblock \bibinfo{journal}{\bibinfo{title}{External shapes, strain rates, and
  dynamics of salt structures}}.
\newblock {\emph{\JournalTitle{Geological Society of America Bulletin - GEOL
  SOC AMER BULL}}} \textbf{\bibinfo{volume}{97}},
  \doiprefix\url{10.1130/0016-7606(1986)97<305:ESSRAD>2.0.CO;2}
  (\bibinfo{year}{1986}).

\bibitem{Jeremic1994}
\bibinfo{author}{Jeremic, M.}
\newblock \emph{\bibinfo{title}{{Rock Mechanics in Salt Mining}}}
  (\bibinfo{publisher}{CRC Press}, \bibinfo{year}{1994}).

\bibitem{KonstantinNaumenko2007}
\bibinfo{author}{Naumenko, K.} \& \bibinfo{author}{Altenbach, H.}
\newblock \emph{\bibinfo{title}{Modeling of Creep for Structural Analysis}}
  (\bibinfo{publisher}{Springer-Verlag GmbH}, \bibinfo{year}{2007}).

\bibitem{Betten2008}
\bibinfo{author}{Betten, J.}
\newblock \emph{\bibinfo{title}{{Creep mechanics (Third Edition)}}}
  (\bibinfo{publisher}{Springer Berlin Heidelberg}, \bibinfo{year}{2008}).

\bibitem{Carter1993}
\bibinfo{author}{Carter, N.}, \bibinfo{author}{Horseman, S.},
  \bibinfo{author}{Russell, J.} \& \bibinfo{author}{Handin, J.}
\newblock \bibinfo{journal}{\bibinfo{title}{Rheology of rocksalt}}.
\newblock {\emph{\JournalTitle{Journal of Structural Geology}}}
  \textbf{\bibinfo{volume}{15}}, \bibinfo{pages}{1257--1271},
  \doiprefix\url{10.1016/0191-8141(93)90168-a} (\bibinfo{year}{1993}).

\bibitem{Spiers1990}
\bibinfo{author}{Spiers, C.} \emph{et~al.}
\newblock \bibinfo{journal}{\bibinfo{title}{Experimental determination of
  constitutive parameters governing creep of rocksalt by pressure solution}}.
\newblock {\emph{\JournalTitle{Geological Society Special Publication}}}
  \textbf{\bibinfo{volume}{54}}, \bibinfo{pages}{215--227},
  \doiprefix\url{10.1144/GSL.SP.1990.054.01.21} (\bibinfo{year}{1990}).

\bibitem{Heege2005}
\bibinfo{author}{Heege, J.~T.}, \bibinfo{author}{Bresser, J.~D.} \&
  \bibinfo{author}{Spiers, C.}
\newblock \bibinfo{journal}{\bibinfo{title}{Dynamic recrystallization of wet
  synthetic polycrystalline halite: dependence of grain size distribution on
  flow stress, temperature and strain}}.
\newblock {\emph{\JournalTitle{Tectonophysics}}}
  \textbf{\bibinfo{volume}{396}}, \bibinfo{pages}{35--57},
  \doiprefix\url{10.1016/j.tecto.2004.10.002} (\bibinfo{year}{2005}).

\bibitem{Hunsche1999}
\bibinfo{author}{Hunsche, U.} \& \bibinfo{author}{Hampel, A.}
\newblock \bibinfo{journal}{\bibinfo{title}{Rock salt {\textemdash} the
  mechanical properties of the host rock material for a radioactive waste
  repository}}.
\newblock {\emph{\JournalTitle{Engineering Geology}}}
  \textbf{\bibinfo{volume}{52}}, \bibinfo{pages}{271--291},
  \doiprefix\url{10.1016/s0013-7952(99)00011-3} (\bibinfo{year}{1999}).

\bibitem{Hampel2017}
\bibinfo{author}{Hampel, A.} \& \bibinfo{author}{Schulze, O.}
\newblock \bibinfo{title}{The composite dilatancy model: A constitutive model
  for the mechanical behavior of rock salt}.
\newblock In \emph{\bibinfo{booktitle}{The Mechanical Behavior of Salt
  {\textendash} Understanding of {THMC} Processes in Salt}},
  \bibinfo{pages}{99--107}, \doiprefix\url{10.1201/9781315106502-12}
  (\bibinfo{publisher}{{CRC} Press}, \bibinfo{year}{2017}).

\bibitem{Xing2014}
\bibinfo{author}{Xing, W.} \emph{et~al.}
\newblock \bibinfo{journal}{\bibinfo{title}{{Experimental study of mechanical
  and hydraulic properties of bedded rock salt from the Jintan location}}}.
\newblock {\emph{\JournalTitle{Acta Geotechnica}}}
  \textbf{\bibinfo{volume}{9}}, \bibinfo{pages}{145--151},
  \doiprefix\url{10.1007/s11440-013-0231-x} (\bibinfo{year}{2014}).

\bibitem{Park2018}
\bibinfo{author}{Park, B.~Y.}, \bibinfo{author}{Sobolik, S.~R.} \&
  \bibinfo{author}{Herrick, C.~G.}
\newblock \bibinfo{journal}{\bibinfo{title}{{Geomechanical Model Calibration
  Using Field Measurements for a Petroleum Reserve}}}.
\newblock {\emph{\JournalTitle{Rock Mechanics and Rock Engineering}}}
  \textbf{\bibinfo{volume}{51}}, \bibinfo{pages}{925--943},
  \doiprefix\url{10.1007/s00603-017-1370-4} (\bibinfo{year}{2018}).

\bibitem{Munson1982}
\bibinfo{author}{Munson, D.~E.} \& \bibinfo{author}{Dawson, P.~R.}
\newblock \bibinfo{journal}{\bibinfo{title}{{A transient creep model for salt
  during stress loading and unloading}}}.
\newblock {\emph{\JournalTitle{SAND82-0962. Sandia National
  Laboratories,Albuquerque}}}  (\bibinfo{year}{1982}).

\bibitem{Yang2015}
\bibinfo{author}{Yang, S.~Q.}, \bibinfo{author}{Xu, P.} \&
  \bibinfo{author}{Ranjith, P.~G.}
\newblock \bibinfo{journal}{\bibinfo{title}{{Damage model of coal under creep
  and triaxial compression}}}.
\newblock {\emph{\JournalTitle{International Journal of Rock Mechanics and
  Mining Sciences}}} \textbf{\bibinfo{volume}{80}}, \bibinfo{pages}{337--345},
  \doiprefix\url{10.1016/j.ijrmms.2015.10.006} (\bibinfo{year}{2015}).

\bibitem{Ma2013ab}
\bibinfo{author}{Ma, L.~J.} \emph{et~al.}
\newblock \bibinfo{journal}{\bibinfo{title}{{A new elasto-viscoplastic damage
  model combined with the generalized hoek-brown failure criterion for bedded
  rock salt and its application}}}.
\newblock {\emph{\JournalTitle{Rock Mechanics and Rock Engineering}}}
  \textbf{\bibinfo{volume}{46}}, \bibinfo{pages}{53--66},
  \doiprefix\url{10.1007/s00603-012-0256-8} (\bibinfo{year}{2013}).

\bibitem{Ma2017}
\bibinfo{author}{Ma, L.}, \bibinfo{author}{Wang, M.}, \bibinfo{author}{Zhang,
  N.}, \bibinfo{author}{Fan, P.} \& \bibinfo{author}{Li, J.}
\newblock \bibinfo{journal}{\bibinfo{title}{{A Variable-Parameter Creep Damage
  Model Incorporating the Effects of Loading Frequency for Rock Salt and Its
  Application in a Bedded Storage Cavern}}}.
\newblock {\emph{\JournalTitle{Rock Mechanics and Rock Engineering}}}
  \textbf{\bibinfo{volume}{50}}, \bibinfo{pages}{2495--2509},
  \doiprefix\url{10.1007/s00603-017-1236-9} (\bibinfo{year}{2017}).

\bibitem{betten_2014}
\bibinfo{author}{Betten, J.}
\newblock \emph{\bibinfo{title}{Creep mechanics}}
  (\bibinfo{publisher}{Springer}, \bibinfo{year}{2014}).

\bibitem{Li2016}
\bibinfo{author}{Li, S.} \& \bibinfo{author}{Urai, J.~L.}
\newblock \bibinfo{journal}{\bibinfo{title}{Rheology of rock salt for salt
  tectonics modeling}}.
\newblock {\emph{\JournalTitle{Petroleum Science}}}
  \textbf{\bibinfo{volume}{13}}, \bibinfo{pages}{712--724},
  \doiprefix\url{10.1007/s12182-016-0121-6} (\bibinfo{year}{2016}).

\bibitem{Berest2019}
\bibinfo{author}{B{\'{e}}rest, P.} \emph{et~al.}
\newblock \bibinfo{journal}{\bibinfo{title}{{Very Slow Creep Tests on Salt
  Samples}}}.
\newblock {\emph{\JournalTitle{Rock Mechanics and Rock Engineering}}}
  \textbf{\bibinfo{volume}{52}}, \bibinfo{pages}{2917--2934},
  \doiprefix\url{10.1007/s00603-019-01778-9} (\bibinfo{year}{2019}).

\bibitem{Urai2017}
\bibinfo{author}{Urai, J.} \& \bibinfo{author}{Spiers, C.}
\newblock \bibinfo{title}{The effect of grain boundary water on deformation
  mechanisms and rheology of rocksalt during long-term deformation}.
\newblock In \emph{\bibinfo{booktitle}{The Mechanical Behavior of Salt -
  Understanding of THMC Processes in Salt}}, \bibinfo{pages}{149--158},
  \doiprefix\url{10.1201/9781315106502-17} (\bibinfo{publisher}{{CRC} Press},
  \bibinfo{year}{2017}).

\bibitem{Berest2012}
\bibinfo{author}{B{\'{e}}rest, P.} \emph{et~al.}
\newblock \bibinfo{title}{Very slow creep tests on rock samples}.
\newblock In \emph{\bibinfo{booktitle}{Mechanical Behaviour of Salt {VII}}},
  \doiprefix\url{10.1201/b12041-12} (\bibinfo{publisher}{{CRC} Press},
  \bibinfo{year}{2012}).

\bibitem{Peach2001}
\bibinfo{author}{Peach, C.~J.}, \bibinfo{author}{Spiers, C.~J.} \&
  \bibinfo{author}{Trimby, P.~W.}
\newblock \bibinfo{journal}{\bibinfo{title}{Effect of confining pressure on
  dilatation, recrystallization, and flow of rock salt at 150c}}.
\newblock {\emph{\JournalTitle{Journal of Geophysical Research: Solid Earth}}}
  \textbf{\bibinfo{volume}{106}}, \bibinfo{pages}{13315--13328},
  \doiprefix\url{10.1029/2000jb900300} (\bibinfo{year}{2001}).

\bibitem{TerHeege2005}
\bibinfo{author}{{Ter Heege}, J.~H.}, \bibinfo{author}{{De Bresser}, J.~H.} \&
  \bibinfo{author}{Spiers, C.~J.}
\newblock \bibinfo{journal}{\bibinfo{title}{{Dynamic recrystallization of wet
  synthetic polycrystalline halite: Dependence of grain size distribution on
  flow stress, temperature and strain}}}.
\newblock {\emph{\JournalTitle{Tectonophysics}}}
  \textbf{\bibinfo{volume}{396}}, \bibinfo{pages}{35--57},
  \doiprefix\url{10.1016/j.tecto.2004.10.002} (\bibinfo{year}{2005}).

\bibitem{TerHeege2005a}
\bibinfo{author}{{Ter Heege}, J.~H.}, \bibinfo{author}{{De Bresser}, J.~H.} \&
  \bibinfo{author}{Spiers, C.~J.}
\newblock \bibinfo{journal}{\bibinfo{title}{{Rheological behaviour of synthetic
  rocksalt: The interplay between water, dynamic recrystallization and
  deformation mechanisms}}}.
\newblock {\emph{\JournalTitle{Journal of Structural Geology}}}
  \textbf{\bibinfo{volume}{27}}, \bibinfo{pages}{948--963},
  \doiprefix\url{10.1016/j.jsg.2005.04.008} (\bibinfo{year}{2005}).

\bibitem{Marketos2016}
\bibinfo{author}{Marketos, G.}, \bibinfo{author}{Spiers, C.~J.} \&
  \bibinfo{author}{Govers, R.}
\newblock \bibinfo{journal}{\bibinfo{title}{Impact of rock salt creep law
  choice on subsidence calculations for hydrocarbon reservoirs overlain by
  evaporite caprocks}}.
\newblock {\emph{\JournalTitle{Journal of Geophysical Research: Solid Earth}}}
  \textbf{\bibinfo{volume}{121}}, \bibinfo{pages}{4249--4267},
  \doiprefix\url{10.1002/2016jb012892} (\bibinfo{year}{2016}).

\bibitem{Cornet2018}
\bibinfo{author}{Cornet, J.~S.}, \bibinfo{author}{Dabrowski, M.} \&
  \bibinfo{author}{Schmid, D.~W.}
\newblock \bibinfo{journal}{\bibinfo{title}{{Long term creep closure of salt
  cavities}}}.
\newblock {\emph{\JournalTitle{International Journal of Rock Mechanics and
  Mining Sciences}}} \textbf{\bibinfo{volume}{103}}, \bibinfo{pages}{96--106},
  \doiprefix\url{10.1016/J.IJRMMS.2018.01.025} (\bibinfo{year}{2018}).

\bibitem{Cornet2018a}
\bibinfo{author}{Cornet, J.~S.} \& \bibinfo{author}{Dabrowski, M.}
\newblock \bibinfo{journal}{\bibinfo{title}{{Nonlinear Viscoelastic Closure of
  Salt Cavities}}}.
\newblock {\emph{\JournalTitle{Rock Mechanics and Rock Engineering 2018
  51:10}}} \textbf{\bibinfo{volume}{51}}, \bibinfo{pages}{3091--3109},
  \doiprefix\url{10.1007/S00603-018-1506-1} (\bibinfo{year}{2018}).

\bibitem{Urai2008}
\bibinfo{author}{Urai, J.}, \bibinfo{author}{Schléder, Z.},
  \bibinfo{author}{Spiers, C.} \& \bibinfo{author}{Kukla, P.}
\newblock \emph{\bibinfo{title}{Flow and Transport Properties of Salt Rocks}},
  \bibinfo{pages}{277--290} (\bibinfo{publisher}{Springer, Berlin, Heidelberg},
  \bibinfo{year}{2008}).

\bibitem{Song2013}
\bibinfo{author}{Song, R.}, \bibinfo{author}{Yue-ming, B.},
  \bibinfo{author}{Jing-Peng, Z.}, \bibinfo{author}{De-yi, J.} \&
  \bibinfo{author}{Chun-he, Y.}
\newblock \bibinfo{journal}{\bibinfo{title}{{Experimental investigation of the
  fatigue properties of salt rock}}}.
\newblock {\emph{\JournalTitle{International Journal of Rock Mechanics and
  Mining Sciences}}} \textbf{\bibinfo{volume}{64}}, \bibinfo{pages}{68--72},
  \doiprefix\url{10.1016/j.ijrmms.2013.08.023} (\bibinfo{year}{2013}).

\bibitem{He2019}
\bibinfo{author}{He, M.}, \bibinfo{author}{Li, N.}, \bibinfo{author}{Zhu, C.},
  \bibinfo{author}{Chen, Y.} \& \bibinfo{author}{Wu, H.}
\newblock \bibinfo{journal}{\bibinfo{title}{{Experimental investigation and
  damage modeling of salt rock subjected to fatigue loading}}}.
\newblock {\emph{\JournalTitle{International Journal of Rock Mechanics and
  Mining Sciences}}} \textbf{\bibinfo{volume}{114}}, \bibinfo{pages}{17--23},
  \doiprefix\url{10.1016/j.ijrmms.2018.12.015} (\bibinfo{year}{2019}).

\bibitem{Liu2014}
\bibinfo{author}{Liu, J.}, \bibinfo{author}{Xie, H.}, \bibinfo{author}{Hou,
  Z.}, \bibinfo{author}{Yang, C.} \& \bibinfo{author}{Chen, L.}
\newblock \bibinfo{journal}{\bibinfo{title}{{Damage evolution of rock salt
  under cyclic loading in unixial tests}}}.
\newblock {\emph{\JournalTitle{Acta Geotechnica}}}
  \textbf{\bibinfo{volume}{9}}, \bibinfo{pages}{153--160},
  \doiprefix\url{10.1007/s11440-013-0236-5} (\bibinfo{year}{2014}).

\bibitem{Yin2019}
\bibinfo{author}{Yin, H.} \emph{et~al.}
\newblock \bibinfo{journal}{\bibinfo{title}{{Study on Damage and Repair
  Mechanical Characteristics of Rock Salt Under Uniaxial Compression}}}.
\newblock {\emph{\JournalTitle{Rock Mechanics and Rock Engineering}}}
  \textbf{\bibinfo{volume}{52}}, \bibinfo{pages}{659--671},
  \doiprefix\url{10.1007/s00603-018-1604-0} (\bibinfo{year}{2019}).

\bibitem{Wang2017}
\bibinfo{author}{Wang, T.} \emph{et~al.}
\newblock \bibinfo{journal}{\bibinfo{title}{{Failure Analysis of Overhanging
  Blocks in the Walls of a Gas Storage Salt Cavern: A Case Study}}}.
\newblock {\emph{\JournalTitle{Rock Mechanics and Rock Engineering}}}
  \textbf{\bibinfo{volume}{50}}, \bibinfo{pages}{125--137},
  \doiprefix\url{10.1007/s00603-016-1102-1} (\bibinfo{year}{2017}).

\bibitem{Kachanov1986}
\bibinfo{author}{Kachanov, L.}
\newblock \emph{\bibinfo{title}{{Introduction to continuum damage mechanics}}}
  (\bibinfo{publisher}{Springer Netherlands}, \bibinfo{year}{1986}).

\bibitem{Peach1996}
\bibinfo{author}{Peach, C.~J.} \& \bibinfo{author}{Spiers, C.~J.}
\newblock \bibinfo{journal}{\bibinfo{title}{{Influence of crystal plastic
  deformation on dilatancy and permeability development in synthetic salt
  rock}}}.
\newblock {\emph{\JournalTitle{Tectonophysics}}}
  \textbf{\bibinfo{volume}{256}}, \bibinfo{pages}{101--128},
  \doiprefix\url{10.1016/0040-1951(95)00170-0} (\bibinfo{year}{1996}).

\bibitem{Liu2017}
\bibinfo{author}{Liu, H.~Z.}, \bibinfo{author}{Xie, H.~Q.},
  \bibinfo{author}{He, J.~D.}, \bibinfo{author}{Xiao, M.~L.} \&
  \bibinfo{author}{Zhuo, L.}
\newblock \bibinfo{journal}{\bibinfo{title}{{Nonlinear creep damage
  constitutive model for soft rocks}}}.
\newblock {\emph{\JournalTitle{Mechanics of Time-Dependent Materials}}}
  \textbf{\bibinfo{volume}{21}}, \bibinfo{pages}{73--96},
  \doiprefix\url{10.1007/s11043-016-9319-7} (\bibinfo{year}{2017}).

\bibitem{Ma2013damage}
\bibinfo{author}{Ma, L.~J.} \emph{et~al.}
\newblock \bibinfo{journal}{\bibinfo{title}{{A new elasto-viscoplastic damage
  model combined with the generalized hoek-brown failure criterion for bedded
  rock salt and its application}}}.
\newblock {\emph{\JournalTitle{Rock Mechanics and Rock Engineering}}}
  \textbf{\bibinfo{volume}{46}}, \bibinfo{pages}{53--66},
  \doiprefix\url{10.1007/s00603-012-0256-8} (\bibinfo{year}{2013}).

\bibitem{Xu2018}
\bibinfo{author}{Xu, T.} \emph{et~al.}
\newblock \bibinfo{journal}{\bibinfo{title}{{The Modeling of Time-Dependent
  Deformation and Fracturing of Brittle Rocks Under Varying Confining and Pore
  Pressures}}}.
\newblock {\emph{\JournalTitle{Rock Mechanics and Rock Engineering}}}
  \textbf{\bibinfo{volume}{51}}, \bibinfo{pages}{3241--3263},
  \doiprefix\url{10.1007/s00603-018-1491-4} (\bibinfo{year}{2018}).

\bibitem{Hao2016}
\bibinfo{author}{Hao, T.~S.} \& \bibinfo{author}{Liang, W.~G.}
\newblock \bibinfo{journal}{\bibinfo{title}{{A New Improved Failure Criterion
  for Salt Rock Based on Energy Method}}}.
\newblock {\emph{\JournalTitle{Rock Mechanics and Rock Engineering}}}
  \textbf{\bibinfo{volume}{49}}, \bibinfo{pages}{1721--1731},
  \doiprefix\url{10.1007/s00603-015-0851-6} (\bibinfo{year}{2016}).

\bibitem{Zhao2021}
\bibinfo{author}{Zhao, K.} \emph{et~al.}
\newblock \bibinfo{journal}{\bibinfo{title}{{Damage Evolution and Deformation
  of Rock Salt Under Creep-Fatigue Loading}}}.
\newblock {\emph{\JournalTitle{Rock Mechanics and Rock Engineering}}}
  \textbf{\bibinfo{volume}{54}}, \bibinfo{pages}{1985--1997},
  \doiprefix\url{10.1007/s00603-020-02342-6} (\bibinfo{year}{2021}).

\bibitem{Chemia2008}
\bibinfo{author}{Chemia, Z.}, \bibinfo{author}{Koyi, H.} \&
  \bibinfo{author}{Schmeling, H.}
\newblock \bibinfo{journal}{\bibinfo{title}{Numerical modelling of rise and
  fall of a dense layer in salt diapirs}}.
\newblock {\emph{\JournalTitle{Geophysical Journal International}}}
  \textbf{\bibinfo{volume}{172}}, \bibinfo{pages}{798--816},
  \doiprefix\url{10.1111/j.1365-246x.2007.03661.x} (\bibinfo{year}{2008}).

\bibitem{Chemia2009}
\bibinfo{author}{Chemia, Z.}, \bibinfo{author}{Schmeling, H.} \&
  \bibinfo{author}{Koyi, H.}
\newblock \bibinfo{journal}{\bibinfo{title}{The effect of the salt viscosity on
  future evolution of the gorleben salt diapir, germany}}.
\newblock {\emph{\JournalTitle{Tectonophysics}}}
  \textbf{\bibinfo{volume}{473}}, \bibinfo{pages}{446--456},
  \doiprefix\url{10.1016/j.tecto.2009.03.027} (\bibinfo{year}{2009}).

\bibitem{Chemia2008a}
\bibinfo{author}{Chemia, Z.} \& \bibinfo{author}{Koyi, H.}
\newblock \bibinfo{journal}{\bibinfo{title}{The control of salt supply on
  entrainment of an anhydrite layer within a salt diapir}}.
\newblock {\emph{\JournalTitle{Journal of Structural Geology}}}
  \textbf{\bibinfo{volume}{30}}, \bibinfo{pages}{1192--1200},
  \doiprefix\url{10.1016/j.jsg.2008.06.004} (\bibinfo{year}{2008}).

\bibitem{Koyi1996}
\bibinfo{author}{Koyi, H.}
\newblock \bibinfo{journal}{\bibinfo{title}{Salt flow by aggrading and
  prograding overburdens}}.
\newblock {\emph{\JournalTitle{Geological Society, London, Special
  Publications}}} \textbf{\bibinfo{volume}{100}}, \bibinfo{pages}{243--258},
  \doiprefix\url{10.1144/gsl.sp.1996.100.01.15} (\bibinfo{year}{1996}).

\bibitem{Koyi1998}
\bibinfo{author}{Koyi, H.}
\newblock \bibinfo{journal}{\bibinfo{title}{The shaping of salt diapirs}}.
\newblock {\emph{\JournalTitle{Journal of Structural Geology}}}
  \textbf{\bibinfo{volume}{20}}, \bibinfo{pages}{321--338},
  \doiprefix\url{10.1016/s0191-8141(97)00092-8} (\bibinfo{year}{1998}).

\bibitem{Li2009}
\bibinfo{author}{Li, S.}, \bibinfo{author}{Feng, L.}, \bibinfo{author}{Tang,
  P.}, \bibinfo{author}{Rao, G.} \& \bibinfo{author}{Bao, Y.}
\newblock \bibinfo{journal}{\bibinfo{title}{Calculation of depth to detachment
  and its significance in the kuqa depression: A discussion}}.
\newblock {\emph{\JournalTitle{Petroleum Science}}}
  \textbf{\bibinfo{volume}{6}}, \bibinfo{pages}{17--20},
  \doiprefix\url{10.1007/s12182-009-0003-2} (\bibinfo{year}{2009}).

\bibitem{Li2012}
\bibinfo{author}{Li, S.} \emph{et~al.}
\newblock \bibinfo{journal}{\bibinfo{title}{Numerical modelling of the
  displacement and deformation of embedded rock bodies during salt tectonics: A
  case study from the south oman salt basin}}.
\newblock {\emph{\JournalTitle{Geological Society, London, Special
  Publications}}} \textbf{\bibinfo{volume}{363}}, \bibinfo{pages}{503--520},
  \doiprefix\url{10.1144/sp363.24} (\bibinfo{year}{2012}).

\bibitem{Poliakov1993}
\bibinfo{author}{Poliakov, A.}, \bibinfo{author}{Podladchikov, Y.} \&
  \bibinfo{author}{Talbot, C.}
\newblock \bibinfo{journal}{\bibinfo{title}{Initiation of salt diapirs with
  frictional overburdens: numerical experiments}}.
\newblock {\emph{\JournalTitle{Tectonophysics}}}
  \textbf{\bibinfo{volume}{228}}, \bibinfo{pages}{199--210},
  \doiprefix\url{10.1016/0040-1951(93)90341-g} (\bibinfo{year}{1993}).

\bibitem{Schultz1993}
\bibinfo{author}{Schultz-Ela, D.}
\newblock \bibinfo{journal}{\bibinfo{title}{Evolution of extensional fault
  systems linked with salt diapirism modeled with finite elements}}.
\newblock {\emph{\JournalTitle{{AAPG} Bulletin}}}
  \textbf{\bibinfo{volume}{77}},
  \doiprefix\url{10.1306/d9cb6353-1715-11d7-8645000102c1865d}
  (\bibinfo{year}{1993}).

\bibitem{Keken1993}
\bibinfo{author}{van Keken, P.}, \bibinfo{author}{Spiers, C.},
  \bibinfo{author}{van~den Berg, A.} \& \bibinfo{author}{Muyzert, E.}
\newblock \bibinfo{journal}{\bibinfo{title}{The effective viscosity of
  rocksalt: implementation of steady-state creep laws in numerical models of
  salt diapirism}}.
\newblock {\emph{\JournalTitle{Tectonophysics}}}
  \textbf{\bibinfo{volume}{225}}, \bibinfo{pages}{457--476},
  \doiprefix\url{10.1016/0040-1951(93)90310-g} (\bibinfo{year}{1993}).

\bibitem{Nicolai_MSFEM}
\bibinfo{author}{Castelletto, N.}, \bibinfo{author}{Hajibeygi, H.} \&
  \bibinfo{author}{Tchelepi, H.~A.}
\newblock \bibinfo{journal}{\bibinfo{title}{Multiscale finite-element method
  for linear elastic geomechanics}}.
\newblock {\emph{\JournalTitle{Journal of Computational Physics}}}
  \textbf{\bibinfo{volume}{331}}, \bibinfo{pages}{337--356},
  \doiprefix\url{10.1016/j.jcp.2016.11.044} (\bibinfo{year}{2017}).

\bibitem{Nicola_2019}
\bibinfo{author}{Castelletto, N.}, \bibinfo{author}{Klevtsov, S.},
  \bibinfo{author}{Hajibeygi, H.} \& \bibinfo{author}{Tchelepi, H.~A.}
\newblock \bibinfo{journal}{\bibinfo{title}{Multiscale two-stage solver for
  biot's poroelasticity equations in subsurface media}}.
\newblock {\emph{\JournalTitle{Computational Geosciences}}}
  \textbf{\bibinfo{volume}{23}}, \bibinfo{pages}{207--224},
  \doiprefix\url{10.1007/s10596-018-9791-z} (\bibinfo{year}{2019}).

\bibitem{Sokolova2019}
\bibinfo{author}{Sokolova, I.}, \bibinfo{author}{Bastisya, M.~G.} \&
  \bibinfo{author}{Hajibeygi, H.}
\newblock \bibinfo{journal}{\bibinfo{title}{Multiscale finite volume method for
  finite-volume-based simulation of poroelasticity}}.
\newblock {\emph{\JournalTitle{Journal of Computational Physics}}}
  \textbf{\bibinfo{volume}{379}}, \bibinfo{pages}{309--324},
  \doiprefix\url{10.1016/j.jcp.2018.11.039} (\bibinfo{year}{2019}).

\bibitem{Rameshkumar2020}
\bibinfo{author}{Kumar, K.~R.} \& \bibinfo{author}{Hajibeygi, H.}
\newblock \bibinfo{title}{Multi-scale nonlinear modeling of subsurface energy
  storage: cyclic loading with inelastic creep deformation}.
\newblock In \emph{\bibinfo{booktitle}{17th European Conference on the
  Mathematics of Oil Recovery, ECMOR}} (\bibinfo{publisher}{European
  Association of Geoscientists and Engineers, EAGE}, \bibinfo{year}{2020}).

\bibitem{Rameshkumar2021}
\bibinfo{author}{{Ramesh Kumar}, K.} \& \bibinfo{author}{Hajibeygi, H.}
\newblock \bibinfo{journal}{\bibinfo{title}{{Multiscale simulation of inelastic
  creep deformation for geological rocks}}}.
\newblock {\emph{\JournalTitle{Journal of Computational Physics}}}
  \textbf{\bibinfo{volume}{440}}, \bibinfo{pages}{110439},
  \doiprefix\url{10.1016/j.jcp.2021.110439} (\bibinfo{year}{2021}).

\bibitem{Ma2013}
\bibinfo{author}{Ma, H.}, \bibinfo{author}{Yang, C.}, \bibinfo{author}{Liu, J.}
  \& \bibinfo{author}{Chen, J.}
\newblock \bibinfo{title}{The influence of cyclic loading on deformation of
  rock salt}.
\newblock In \emph{\bibinfo{booktitle}{Rock Characterisation, Modelling and
  Engineering Design Methods}}, \bibinfo{pages}{63--68},
  \doiprefix\url{10.1201/b14917-10} (\bibinfo{publisher}{{CRC} Press},
  \bibinfo{year}{2013}).

\bibitem{Khaledi2014}
\bibinfo{author}{Khaledi, K.}, \bibinfo{author}{Mahmoudi, E.},
  \bibinfo{author}{Schanz, T.} \& \bibinfo{author}{Datcheva, M.}
\newblock \bibinfo{title}{Finite element modeling of the behavior of salt
  caverns under cyclic loading}.
\newblock In \emph{\bibinfo{booktitle}{Geomechanics from Micro to Macro}},
  \bibinfo{pages}{945--950}, \doiprefix\url{10.1201/b17395-169}
  (\bibinfo{publisher}{{CRC} Press}, \bibinfo{year}{2014}).

\bibitem{Firme2016}
\bibinfo{author}{Firme, P. A. L.~P.}, \bibinfo{author}{Roehl, D.} \&
  \bibinfo{author}{Romanel, C.}
\newblock \bibinfo{journal}{\bibinfo{title}{{An assessment of the creep
  behaviour of Brazilian salt rocks using the multi-mechanism deformation
  model}}}.
\newblock {\emph{\JournalTitle{Acta Geotechnica 2016 11:6}}}
  \textbf{\bibinfo{volume}{11}}, \bibinfo{pages}{1445--1463},
  \doiprefix\url{10.1007/S11440-016-0451-Y} (\bibinfo{year}{2016}).

\bibitem{Firme2019}
\bibinfo{author}{Firme, P.~A.}, \bibinfo{author}{Roehl, D.} \&
  \bibinfo{author}{Romanel, C.}
\newblock \bibinfo{journal}{\bibinfo{title}{{Salt caverns history and
  geomechanics towards future natural gas strategic storage in Brazil}}}.
\newblock {\emph{\JournalTitle{Journal of Natural Gas Science and
  Engineering}}} \textbf{\bibinfo{volume}{72}},
  \doiprefix\url{10.1016/J.JNGSE.2019.103006} (\bibinfo{year}{2019}).

\bibitem{Khaledi2016}
\bibinfo{author}{Khaledi, K.}, \bibinfo{author}{Mahmoudi, E.},
  \bibinfo{author}{Datcheva, M.} \& \bibinfo{author}{Schanz, T.}
\newblock \bibinfo{journal}{\bibinfo{title}{{Stability and serviceability of
  underground energy storage caverns in rock salt subjected to mechanical
  cyclic loading}}}.
\newblock {\emph{\JournalTitle{International Journal of Rock Mechanics and
  Mining Sciences}}} \textbf{\bibinfo{volume}{86}}, \bibinfo{pages}{115--131},
  \doiprefix\url{10.1016/J.IJRMMS.2016.04.010} (\bibinfo{year}{2016}).

\bibitem{Wang2015}
\bibinfo{author}{Wang, T.} \emph{et~al.}
\newblock \bibinfo{journal}{\bibinfo{title}{{Failure analysis of thick
  interlayer from leaching of bedded salt caverns}}}.
\newblock {\emph{\JournalTitle{International Journal of Rock Mechanics and
  Mining Sciences}}} \textbf{\bibinfo{volume}{73}}, \bibinfo{pages}{175--183},
  \doiprefix\url{10.1016/J.IJRMMS.2014.11.003} (\bibinfo{year}{2015}).

\bibitem{Liang2007}
\bibinfo{author}{Liang, W.}, \bibinfo{author}{Yang, C.}, \bibinfo{author}{Zhao,
  Y.}, \bibinfo{author}{Dusseault, M.~B.} \& \bibinfo{author}{Liu, J.}
\newblock \bibinfo{journal}{\bibinfo{title}{{Experimental investigation of
  mechanical properties of bedded salt rock}}}.
\newblock {\emph{\JournalTitle{International Journal of Rock Mechanics and
  Mining Sciences}}} \textbf{\bibinfo{volume}{44}}, \bibinfo{pages}{400--411},
  \doiprefix\url{10.1016/J.IJRMMS.2006.09.007} (\bibinfo{year}{2007}).

\bibitem{Zhang2017}
\bibinfo{author}{Zhang, N.} \emph{et~al.}
\newblock \bibinfo{journal}{\bibinfo{title}{{Stability and availability
  evaluation of underground strategic petroleum reserve (SPR) caverns in bedded
  rock salt of Jintan, China}}}.
\newblock {\emph{\JournalTitle{Energy}}} \textbf{\bibinfo{volume}{134}},
  \bibinfo{pages}{504--514}, \doiprefix\url{10.1016/J.ENERGY.2017.06.073}
  (\bibinfo{year}{2017}).

\bibitem{Liu2019}
\bibinfo{author}{Liu, W.} \emph{et~al.}
\newblock \bibinfo{journal}{\bibinfo{title}{{Physical simulation of
  construction and control of two butted-well horizontal cavern energy storage
  using large molded rock salt specimens}}}.
\newblock {\emph{\JournalTitle{Energy}}} \textbf{\bibinfo{volume}{185}},
  \bibinfo{pages}{682--694}, \doiprefix\url{10.1016/J.ENERGY.2019.07.014}
  (\bibinfo{year}{2019}).

\bibitem{Nawaz2015}
\bibinfo{author}{Nawaz, M.}
\newblock \emph{\bibinfo{title}{{Deformation and transport processes in salt
  rocks : An experimental study exploring effects of pressure and stress
  relaxation}}}.
\newblock Ph.D. thesis, \bibinfo{school}{Utrecht university}
  (\bibinfo{year}{2015}).

\bibitem{Li2018}
\bibinfo{author}{Li, J.} \emph{et~al.}
\newblock \bibinfo{journal}{\bibinfo{title}{{Mathematical model of salt cavern
  leaching for gas storage in high-insoluble salt formations}}}.
\newblock {\emph{\JournalTitle{Scientific Reports 2017 8:1}}}
  \textbf{\bibinfo{volume}{8}}, \bibinfo{pages}{1--12},
  \doiprefix\url{10.1038/s41598-017-18546-w} (\bibinfo{year}{2018}).

\bibitem{Luangthip}
\bibinfo{author}{Luangthip, A.}, \bibinfo{author}{Wilalak, N.},
  \bibinfo{author}{Thongprapha, T.} \& \bibinfo{author}{Fuenkajorn, K.}
\newblock \bibinfo{journal}{\bibinfo{title}{{Effects of carnallite content on
  mechanical properties of Maha Sarakham rock salt}}}.
\newblock {\emph{\JournalTitle{Arabian Journal of Geosciences}}}
  \textbf{\bibinfo{volume}{10}}, \bibinfo{pages}{1--14},
  \doiprefix\url{10.1007/s12517-017-2945-9} (\bibinfo{year}{2017}).

\bibitem{Kraus1980}
\bibinfo{author}{Kraus, H.}
\newblock \emph{\bibinfo{title}{{Creep analysis}}} (\bibinfo{publisher}{Wiley},
  \bibinfo{address}{New York SE}, \bibinfo{year}{1980}).

\bibitem{Rust2015}
\bibinfo{author}{Rust, W.}
\newblock \emph{\bibinfo{title}{{Non-linear finite element analysis in
  structural mechanics}}} (\bibinfo{publisher}{Springer International
  Publishing}, \bibinfo{year}{2015}).

\bibitem{Bonte2012}
\bibinfo{author}{Bont{\'{e}}, D.}, \bibinfo{author}{van Wees, J.-D.} \&
  \bibinfo{author}{Verweij, J.}
\newblock \bibinfo{journal}{\bibinfo{title}{Subsurface temperature of the
  onshore netherlands: new temperature dataset and modelling}}.
\newblock {\emph{\JournalTitle{Netherlands Journal of Geosciences - Geologie en
  Mijnbouw}}} \textbf{\bibinfo{volume}{91}}, \bibinfo{pages}{491--515},
  \doiprefix\url{10.1017/s0016774600000354} (\bibinfo{year}{2012}).

\bibitem{Berest2015}
\bibinfo{author}{B{\'{e}}rest, P.}, \bibinfo{author}{B{\'{e}}raud, J.~F.},
  \bibinfo{author}{Gharbi, H.}, \bibinfo{author}{Brouard, B.} \&
  \bibinfo{author}{DeVries, K.}
\newblock \bibinfo{journal}{\bibinfo{title}{{A Very Slow Creep Test on an Avery
  Island Salt Sample}}}.
\newblock {\emph{\JournalTitle{Rock Mechanics and Rock Engineering 2015 48:6}}}
  \textbf{\bibinfo{volume}{48}}, \bibinfo{pages}{2591--2602},
  \doiprefix\url{10.1007/S00603-015-0850-7} (\bibinfo{year}{2015}).

\bibitem{Peach1991}
\bibinfo{author}{Peach, C.~J.}
\newblock \emph{\bibinfo{title}{{Influence of deformation on the fluid
  transport properties of salt rocks}}}.
\newblock Ph.D. thesis, \bibinfo{school}{Utrecht university}
  (\bibinfo{year}{1991}).

\bibitem{Duncan1993}
\bibinfo{author}{Duncan, E. J.~S.} \& \bibinfo{author}{Lajtai, E.~Z.}
\newblock \bibinfo{journal}{\bibinfo{title}{{The creep of potash salt rocks
  from Saskatchewan}}}.
\newblock {\emph{\JournalTitle{Geotechnical and Geological Engineering}}}
  \textbf{\bibinfo{volume}{11}}, \bibinfo{pages}{159--184}
  (\bibinfo{year}{1993}).

\bibitem{XiaoXueying2011}
\bibinfo{author}{{Xiao Xueying}} \emph{et~al.}
\newblock \bibinfo{title}{{Coagulating salt for potassium salt solid mine
  backfilling and manufacturing method thereof}} (\bibinfo{year}{China patent,
  2011}).

\bibitem{Zhang2012}
\bibinfo{author}{Zhang, H.}, \bibinfo{author}{Wang, Z.},
  \bibinfo{author}{Zheng, Y.}, \bibinfo{author}{Duan, P.} \&
  \bibinfo{author}{Ding, S.}
\newblock \bibinfo{journal}{\bibinfo{title}{{Study on tri-axial creep
  experiment and constitutive relation of different rock salt}}}.
\newblock {\emph{\JournalTitle{Safety Science}}} \textbf{\bibinfo{volume}{50}},
  \bibinfo{pages}{801--805}, \doiprefix\url{10.1016/J.SSCI.2011.08.030}
  (\bibinfo{year}{2012}).

\bibitem{Zhang2020}
\bibinfo{author}{Zhang, N.}, \bibinfo{author}{Liu, W.}, \bibinfo{author}{Zhang,
  Y.}, \bibinfo{author}{Shan, P.} \& \bibinfo{author}{Shi, X.}
\newblock \bibinfo{journal}{\bibinfo{title}{{Microscopic Pore Structure of
  Surrounding Rock for Underground Strategic Petroleum Reserve (SPR) Caverns in
  Bedded Rock Salt}}}.
\newblock {\emph{\JournalTitle{Energies}}} \textbf{\bibinfo{volume}{13}},
  \bibinfo{pages}{1565}, \doiprefix\url{10.3390/EN13071565}
  (\bibinfo{year}{2020}).

\bibitem{Taheri2020}
\bibinfo{author}{Taheri, S.~R.}, \bibinfo{author}{Pak, A.},
  \bibinfo{author}{Shad, S.}, \bibinfo{author}{Mehrgini, B.} \&
  \bibinfo{author}{Razifar, M.}
\newblock \bibinfo{journal}{\bibinfo{title}{{Investigation of rock salt layer
  creep and its effects on casing collapse}}}.
\newblock {\emph{\JournalTitle{International Journal of Mining Science and
  Technology}}} \textbf{\bibinfo{volume}{30}}, \bibinfo{pages}{357--365},
  \doiprefix\url{10.1016/J.IJMST.2020.02.001} (\bibinfo{year}{2020}).

\bibitem{Li2020}
\bibinfo{author}{Li, W.}, \bibinfo{author}{Miao, X.} \& \bibinfo{author}{Yang,
  C.}
\newblock \bibinfo{journal}{\bibinfo{title}{{Failure analysis for gas storage
  salt cavern by thermo-mechanical modelling considering rock salt creep}}}.
\newblock {\emph{\JournalTitle{Journal of Energy Storage}}}
  \textbf{\bibinfo{volume}{32}}, \bibinfo{pages}{102004},
  \doiprefix\url{10.1016/j.est.2020.102004} (\bibinfo{year}{2020}).

\end{thebibliography}

\section*{Acknowledgements}

Hadi Hajibeygi and Kishan Ramesh Kumar were sponsored by the NWO (Dutch Research Council) under ViDi scheme, project "ADMIRE". The authors would like to thank the members of ADMIRE research group and user committee for the fruitful discussions during the development of this work. Also members of DARSim and Delft Subsurface Storage (DSS) Theme are acknowledged for their supports related to the FEM simulation toolbox development.

\section*{Author contributions statement}
K.R worked on methodology, software and writing. A.M worked on methodology, software and writing. C.S worked on conceptualization, reviewing and editing. H.H worked on conceptualization, reviewing, editing, supervision and funding acquisition.

\section*{Competing interest}

The authors declare no competing  interest.

\appendix

\end{document}